\documentclass[superscriptaddress,aps,prl,english,floatfix,twocolumn,10pt]{revtex4-2}
\usepackage[english]{babel}
\usepackage{graphicx}
\usepackage{float}
\usepackage{natbib}
\usepackage{amssymb}
\usepackage{amsmath}
\usepackage{mathtools}
\usepackage[utf8]{inputenc}
\usepackage{braket}
\usepackage{subfigure}
\usepackage{pgfplots}
\usepackage{csquotes}
\usepackage{hhline}
\usepackage{amssymb}
\usepackage{xr}

\usepackage{dcolumn}
\usepackage{tabularx}
\usepackage[colorlinks=true,linkcolor=blue,citecolor=blue,urlcolor=blue]{hyperref}
\usepackage{longtable}
\usepackage{listings}
\usepackage{color}
\usepackage[normalem]{ulem} 
\usepackage[export]{adjustbox}
\newcolumntype{C}{>{\centering\arraybackslash}X}

\begin{document}
	
	\title{\textbf{Emergence of non-trivial phases in interacting non-Hermitian quasiperiodic chains with power-law hopping}}
	\author{Aditi Chakrabarty}
	\email{aditichakrabarty030@gmail.com}
	\affiliation{Department of Physics and Astronomy, National Institute of Technology, Rourkela, Odisha-769008, India}
    \author{Sanchayan Banerjee}
\affiliation{School of Physical Sciences, National Institute of Science Education and Research, Jatni 752050, India}
\affiliation{Homi Bhabha National Institute, Training School Complex, Anushaktinagar, Mumbai 400094, India}
        \author{Tapan Mishra}
\email{mishratapan@niser.ac.in}
\affiliation{School of Physical Sciences, National Institute of Science Education and Research, Jatni 752050, India}
\affiliation{Homi Bhabha National Institute, Training School Complex, Anushaktinagar, Mumbai 400094, India}
	\author{Sanjoy Datta}
	\email{dattas@nitrkl.ac.in}
	\affiliation{Department of Physics and Astronomy, National Institute of Technology, Rourkela, Odisha-769008, India}
	
        \date{\today}
	
	\begin{abstract}
        In the last few years, several works have identified the concurrence of the spectral, delocalization-localization and topological phase transitions in non-Hermitian quasiperiodic systems in the presence of time-reversal symmetry (TRS), with or without interaction. In this work, we investigate one-dimensional interacting non-Hermitian quasiperiodic lattices with asymmetric power-law hopping and unveil that although the Hamiltonian respects the TRS, the reality of the eigenspectrum does not necessarily indicate a topologically trivial non-Hermitian many-body localization (NHMBL) regime. In fact, we reveal the emergence of a topologically trivial intermediate regime, where the states that are primarily multifractal in nature can also possess a fully real spectrum, thereby restoring the TRS before crossing over to the NHMBL phase. Moreover, in the entire intermediate regime, the interaction completely destroys the multifractal and mobility edges observed in the non-interacting counterpart. Besides, we unveil that due to the long-range nature of the hopping, the entire topologically non-trivial ergodic regime under the periodic boundary condition does not always give rise to boundary localized skin modes under the open boundary condition. Our findings thus advances and deepens the understanding about the emergence of non-trivial phases due to the interplay of interaction and long-range hopping in non-Hermitian quasiperiodic systems.
        \end{abstract}
	
	\maketitle
	\section{Introduction}\label{Sec:Introduction}
	 The quest to understand whether Anderson localization \cite{Anderson} persists in the presence of interactions has garnered a great deal of interest in the last two decades, leading to the theoretical prediction of many-body localization (MBL) \cite{Gornyi,Basko,Nandkishore,Zlatko,Maksym} in quantum systems with random disorder \cite{Pal,Frank,Bardarson,Sierant,Alet}, that is also verified experimentally \cite{Kondov,Choi,Smith,Hess}.
    The MBL transition drew considerable attention of researchers as it breaks the predicted eigenstate thermalization hypothesis \cite{Deutsch,Srednicki} inherent in a closed and isolated quantum system. Due to this breakdown, the system in the MBL phase fails to thermalize and remains out of equilibrium indefinitely. This non-thermal phase is usually characterized by an ``area law" of entanglement entropy in contrast to the typical ``volume law" followed by the ergodic phase. Moreover, similar to the glassy phase, the entanglement entropy grows logarithmically on a divergent time scale \cite{Znidaric,Moore}, which is a key marker for identifying the phase possessing MBL. Another distinct feature MBL is the emergence of quasi-local integrals of motion (LIOM) \cite{Abanin,Scardicchio}, where the emergent integrability is stable to weak perturbations \cite{Organesyan}.\\
    \indent The notion of MBL has naturally been extended in closed interacting quantum lattices with an underlying quaisperiodic potential \cite{Huse_2013,Shylapnikov,DasSarma,Hong,Xu}, particularly in the Aubry-Andr\'{e}-Harper (AAH) type \cite{Aubry}, since in the non-interacting limit, such systems undergo a delocalization-localization transition even in one dimension.
    Interestingly, the MBL is reported to be more stable in the presence of AAH type potential since it suffers from less finite-size effects compared to its disordered counterpart \cite{Khemani}. Moreover, the universality class in quasiperiodic systems is also different as compared to the systems with random disorder \cite{Garg_2024}.
    Besides the theoretical interest in understanding the MBL in such quasiperiodic lattices, the optical lattices \cite{Schreiber,Henrik,Scherg,Bordia}, ultracold atomic systems \cite{Kohlert} and trapped ions \cite{Smith} have become attractive laboratories to study the interesting phenomena arising due to the interaction between quantum particles. In addition, the MBL in such systems have also been verified experimentally in open quantum systems \cite{Ehud}.\\
    \indent These developments naturally led to the following question: Does MBL persist in non-Hermitian systems, and if so, what unique features emerge due to interaction of the system with an external environment? In recent years, to address this question, particular attention has been devoted to the Hatano–Nelson model \cite{HatanoNelson1996, HatanoNelson1998}, characterized by asymmetric hopping amplitudes, in the presence of interaction. Several studies have explored this model under both random \cite{Kawabata, Banerjee} as well as quasiperiodic \cite{Huang, Dan-Wei-PRA21, Imura_2022, Imura_2023} potentials. The general conclusion from these studies is that, akin to the non-interacting scenario, the presence of time-reversal symmetry (TRS) leads to a complex-real spectral transition that is associated with the transition from the ergodic phase to the phase possessing non-Hermitian many-body localization (NHMBL). Moreover, such systems exhibit the many-body skin effect (MBSE) \cite{Lee_2020,Suthar,Numasawa,Li,Zhi} analogous to the non-interacting picture, where a change in the boundary condition is reflected in the bulk eigenstates (under periodic boundary condition (PBC)) becoming boundary-localized under open boundary condition (OBC) \cite{Lee}, albeit in the Fock space. Furthermore, the non-Hermitian many-body systems have also been probed from the point of view of topology, since both the non-interacting \cite{Zeuner,Yao,Gong,Kawabata_2019} as well as the Hermitian interacting \cite{Ryu,Hughes} counterparts exhibit novel topological phases. In the non-Hermitian systems, while the ergodic phase is topologically non-trivial, the NHMBL gives rise to a topologically trivial phase. On the other hand, over the years, several studies in the non-interacting systems have demonstrated that the existence of skin effect under the OBC is associated with a non-trivial topology of the energy spectrum in the ergodic phase under PBC \cite{Fang,Sato,Claes}. In addition, in the presence of interactions, previous works have illustrated that under the TRS, the ergodic-NHMBL and the complex-real spectral transitions under PBC are concurrent to a simultaneous change in the topology \cite{Kawabata,Huang}.\\
    \indent Furthermore, similar to the non-interacting Hermitian lattices, when the interaction is confined between the nearest-neighbors only, long-range hopping ($J\sim r^{-\beta}$) tends to enhance delocalization \cite{Khatami,Shlyapnikov,Sabyasachi}. Moreover, in the presence of interaction, earlier works suggest that MBL cannot exist for $\beta<2$ \cite{Burin,Sabyasachi} in systems with either random or quasiperiodic potential.
    Interestingly, in the presence of such long-range hopping and an incommensurate potential, the many-body mobility edges arise \cite{Garg} similar to single-particle mobility edges \cite{DasSarma}. On the other hand, recently, the idea of long-range hopping has been extended to non-Hermitian quasiperiodic systems without interaction \cite{Chen,Juan} that manifest identical phases as the Hermitian ones. Specifically, such systems display intermediate phases with multifractal or mobility edges in the single particle eigenstates depending on the strength of $\beta$ and the quasiperiodic potential. In addition, an increase in the range of hopping has been shown to be detrimental to the degree of localization of the skin modes \cite{Xianlong}. It is important to note that, in the limit of $\beta \rightarrow \infty$, one retrieves the behavior identical to the scenario with only the nearest-neighbor hopping.\\
    \indent Since the inclusion of asymmetric long-range hopping does not break the TRS of the Hamiltonian, it is then natural to wonder whether the correspondence between the spectral, ergodic-NHMBL, topology and MBSE still persists in such 1D non-Hermitian interacting systems. Motivated by the above question, in this work, we extensively study the 1D non-Hermitian quasiperiodic system with nearest-neighbor interaction and long-range hopping. We first search for the intermediate regimes with multifractal/mobility edges exhibited in the non-interacting counterparts and find that such edges disappear in the presence of interaction and rather consists of an intermediate regime. Moreover, we find evidence of a NHMBL regime and clearly demonstrate that the reality of eigenvalues does not necessarily imply NHMBL, which has often been used as an indicator to mark this phase. In particular, a fraction of the multifractal states in the intermediate regime can also have a completely real eigenspectrum. This in turn gives rise to a regime that is topologically trivial but does not belong to the NHMBL phase. Therefore, our results suggests that the usual one-to-one correspondence between the spectra and the topology with the underlying phases that typically appears in the $\beta \rightarrow \infty$ limit breaks down due to the interplay of long-range hopping and interaction. This also suggests that the restoration of TRS in such a system occurs prior to the NHMBL regime. Finally, we demonstrate that the ergodicity in such many-body systems, wherein the states are also topological in nature, does not necessarily give rise to the MBSE under OBC.\\
    \indent The organization of the remaining article is the following: In Sec.~\ref{Sec:Model}, we first discuss the interacting version of the non-Hermitian model with power-law hopping. This is followed by a brief explanation of the characterization techniques for identifying the phases under PBC in the static and dynamic scenarios in Sec.~\ref{Sec:Numerical_methods_PBC}, Sec.~\ref{Sec:Entropy} and Sec.~\ref{Sec:Winding_number}. We then present the relevant discussion about the recognition of MBSE under OBC in Sec.~\ref{Sec:SE_OBC}. We highlight the results on the effect of the interplay of long-range hopping and interaction using the different numerical techniques under both PBC and OBC in Sec.~\ref{Sec:Results}. Finally, we gather all our important findings and summarize them in Sec.\ref{Sec:Conclusions}. A few results required for completeness of this work are relegated to the Appendices.
    
	\section{The non-Hermitian power-law model with interaction}\label{Sec:Model}
        We introduce the non-Hermitian Hamiltonian considered in this work by recalling the non-interacting version of the model with a deterministic quasiperiodic potential and a power-law modified hopping as considered in Ref.~\cite{Xianlong}. Accordingly, in the presence of interaction, the minimal Hamiltonian then reads as,\\
        \begin{eqnarray}
	       \mathcal{H}=  \displaystyle\sum_{n,n', n' > n} \Big(\frac{J_R}{|n'-n|^\beta} c^\dag_{n'} c_{n} + \frac{J_L}{|n'-n|^\beta}c^\dag_{n} c_{n'} \Big)~~~~~~ \nonumber \\
           + \sum_{n} V \text{cos} (2\pi\alpha n+\phi) c^\dag_{n} c_{n}+ \displaystyle\sum_{n} U \hat{{\mathbf{n}}}_n  \hat{{\mathbf{n}}}_{n+1}.~~~~~~~
	\label{Eq:Hamiltonian}
	\end{eqnarray}

        Here, $c^\dag_{n}$ ($c_{n}$) create (annihilate) a spinless fermion at the $n$th site, and $\hat{\mathbf{n}}_{n}$ is the associated fermion number operator. $L=na$ denotes the length of the one-dimensional chain. $J_L$ and $J_R$ provide the rate of fermionic hopping towards the left and right respectively, modified by a power-law exponent $\beta$. Throughout this study, without loss of generality, we consider $J_R=1.0$ and $J_L=0.5 J_R$. All other parameters in the Hamiltonian are scaled with respect to $J_R$. Then, the three different terms in the Hamiltonian corresponds to the long-range asymmetric hopping, the on-site quasiperiodic potential given by strength $V$ and the nearest-neighbor density-density interaction governed by magnitude $U$ respectively.  $\phi \in [0,2\pi)$ is an offset phase used in the quasiperiodic potential that will be required for sample averaging. We specify the inverse golden ratio $\alpha=(\sqrt{5}-1)/2$, which can be approximated as the ratio of two consecutive Fibonacci numbers. Although this approximation suggests restricting the lattice sizes to numbers in the Fibonacci series, such restriction becomes inconvenient due to the few lattice sizes that can be accessible within the computational limit in the presence of interaction. Furthermore, the situation worsens because the system sizes have to be even numbers due to the restriction of half-filling as considered in our work. However, the finite-size effects become less pronounced upon considering the value of $\alpha$ as mentioned in Ref.~\cite{Huse_2013}.\\
        \indent We note in passing that the Hamiltonian given in Eq.~\ref{Eq:Hamiltonian} preserves the time-reversal symmetry (TRS) since it remains intact under complex conjugation. Therefore, diagonalization of the interacting Hamiltonian $\mathcal{H}$ yields a set of complex eigenenergies $\{E_j\}$ and the conjugates $\{E^*_j\}$ along with corresponding left and right eigenstates $\psi^j_{L}$ and $\psi^j_{R}$ satisfying the biorthonormality condition $\braket{\psi^j_{L}|\psi^{j'}_{R}}=\delta_{jj'}$, which will be used in the subsequent sections for analysis.
        
        \section{Numerical identification of the distinct phases}\label{Sec:Numerical methods}
        \subsection{BIPR and the fractal dimension $D_2$}\label{Sec:Numerical_methods_PBC}
        One of the most commonly used probes to detect the ergodic and localized phases in the single particle case is the Inverse Participation Ratio (IPR). This quantity has also been used to capture such a phase transition in systems with interaction \cite{Huse_2013,Moessner}. In the configuration basis, one can describe the $j$th right eigenstate of a system as:
        \begin{eqnarray}
        \ket{\psi^j_R}=\displaystyle\sum_{m=1}^{\mathcal{D}}\psi^j_{mR}\ket{m},
	\label{Eq:Wave_function}
	\end{eqnarray}
        The contribution of the $m$th Fock basis configuration in this state is given by the amplitude $\braket{m | \psi^j_R} = \psi^j_{mR}$. Here, $\mathcal{D}$ is the dimension of the Hilbert space which scales exponentially with the size of the lattice. In addition, a similar expression can be obtained for the left eigenstate. Furthermore, since the non-Hermitian system has distinct left and right eigenstates satisfying the biorthonormality condition, one can introduce the configuration-basis biorthogonal-IPR (BIPR) \cite{Wang_2019} as follows:

        \begin{eqnarray}
		\text{BIPR}_j=\frac{\displaystyle\sum_{m=1}^{\mathcal{D}} |\psi_{mL}^{j}\psi_{mR}^{j}|^2}{\Big(\displaystyle\sum_{m=1}^{\mathcal{D}} |{\psi_{mL}^j\psi_{mR}^{j}}|\Big)^2}.  ~~~~~
		\label{Eq:BIPR}
	\end{eqnarray}

        To obtain an overall picture of all the eigenstates, we average this quantity over all the eigenstates as,

        \begin{eqnarray}
        \left \langle \text{BIPR} \right \rangle = \frac{1}{\mathcal{D}} \displaystyle \sum_{j=1}^{\mathcal{D}} \text{BIPR}_j. ~~~~
        \label{Eq:BIPR_avg}
        \end{eqnarray}
        
        The $\left \langle \text{BIPR} \right \rangle$ quantifies how the many-body wave-functions spread and is inversely proportional to the localization volume in the configuration space, going down to zero in the ergodic phase where all the configurations of the Hilbert space have finite contribution to the state. In contrary, the MBL states are confined to a smaller volume in the configuration space, leading to a finite value. These features thus mimic the $\left \langle \text{BIPR} \right \rangle$ used in single particle case, albeit in the Fock-space basis, i.e, $\left \langle \text{BIPR} \right \rangle \sim \mathcal{D}^{-1(0)}$ for the ergodic (many-body localized) states in the presence of interaction. 

        In addition, the fractal dimension $\text{D}_2$ is a well-used diagnostic to quantitatively comprehend the fractal behavior \cite{Shylapnikov,Joana} which is given for a particular eigenstate $j$ as,

        \begin{eqnarray}
		\text{D}_2^j=- \frac{\text{ln} ~\text{BIPR}_j}{\text{ln}~{\mathcal{D}}}.
		\label{Eq:D_2}
	\end{eqnarray}

        The above expression encodes the information about the fully ergodic and many-body localized eigenstates corresponding to $\text{D}_2^j=1 ~\text{and} ~0$ respectively. When $\text{D}_2^j$ lies well away from these limits, the states are fractal, i.e., they are extended yet non-ergodic in nature. The mean fractal dimension over all the Fock-space eigenstates is written as $\left \langle \text{D}_2 \right \rangle$. Furthermore, please note that while $\text{D}_2^j$ alone cannot suggest multifractality, we have verified the multifractal nature of the eigenstates separately for various moments $q$ in Appendix~\ref{App:D_q}. However, to avoid computational cost for creating the entire phase diagram, we have presented results pertaining to $\braket{\text{D}_2}$ only, and called them multifractal.

        \subsection{Dynamical probes in the interacting system: the von-Neumann entanglement entropy}\label{Sec:Entropy}
        The quantum evolution of a many-body system provides useful information on the system's intrinsic behavior at long times.  In this work, we prepare the interacting system  given in Eq.~\ref{Eq:Hamiltonian} at time $t=0$ in an initial state $\ket{\psi_0}$ which is an unentangled one, i.e., the N\'{e}el state, where $\ket{\psi_0}= \ket{1010....}$. The right-hand side gives the occupation at the $n$th site and the position of the sites are arranged from left to right in ascending order. The system undergoes a non-unitary dynamics due to the underlying non-Hermiticity in the system. After time $t$, the right eigenstate becomes,
        \begin{eqnarray}
            \ket{\psi_R(t)}=\frac{e^{-i\mathcal{H}t}\ket{\psi_0}}{||e^{-i\mathcal{H}t}\ket{\psi_0}||},
            \label{Eq:Evolution}
        \end{eqnarray}
        where the normalization helps to explicitly restore the unitary nature of the dynamics in non-Hermitian systems.\\
        \indent
        We now partition the entire system of length $L$ into two subsystems A and B symmetrically at $L/2$. Thus, subsystem A refers to the lattice sites $1,2,.....L/2$, and subsystem B consists of sites labelled from $L/2+1,....L$. Our interest lies in understanding how strongly the first subsystem A is entangled with the second subsystem B. Such a quantification of the entanglement between the two subsystems is achieved using the half-chain von-Neumann entanglement entropy for subsystem A corresponding to the right eigenstate at a particular instant described as,
        \begin{eqnarray}
            S_e(t)=\overline{-\text{Tr}[\rho_A(t) \text{ln} \rho_A(t)]},
            \label{Eq:Entanglement_entropy}
        \end{eqnarray}
        where we obtain the reduced density matrix $\rho_A(t)$ by tracing out all the degrees of freedom of the subsystem B, i.e., $\rho_A(t)\equiv \text{Tr}_B\{\rho(t)\} = \text{Tr}_{L/2} [\ket{\psi_{R}(t)}\bra{\psi_{R}(t)}/ \braket{\psi_{R}(t)|\psi_{R}(t)}]$. The overhead bar indicates the average over all disorder configurations, which is the phase $\phi$ in our case.
        It is well known that in the ergodic phase, $S_e(t) \sim t$ and saturates to a large value at a relatively long time, whereas in the MBL phase the entanglement entropy grows logarithmically in time, i.e., $S_e(t) \sim \text{log}(t)$ \cite{Moore}. The entanglement entropy therefore acts as a quantity to understand whether one part of the system acts as a good bath for the rest of the system to thermalize. In the ergodic phase, the system approaches thermalization, whereas the NHMBL states do not thermalize and remains at much lower values even at long times. In other works, it has been shown that under OBC, the skin states follow an area-law of entropy (but without NHMBL), unlike the scenario under PBC \cite{Numasawa}. We keep ourselves restricted to PBC since the investigation under OBC is not expected to reveal any additional information pertaining to our interest in verifying the nature of the phases under PBC.
        
        \subsection{The topological phase transition: winding number under PBC}\label{Sec:Winding_number}

        In non-Hermitian systems, the entire complex spectra is taken into account for the identification of the topological phases. The determination of the winding number in such a spectrum crucially depends on the distribution of eigenenergies encircling a given reference point. Generalization of the winding number in non-Hermitian systems with asymmetric hopping has been recently made to the interacting counterpart \cite{Huang}, using a parameter $\Phi$ physically interpreted as a synthetic magnetic flux through a ring of length $L$. Under the gauge transformation, $c_n\rightarrow e^{i(\Phi/L) n}c_n$, the transformed Hamiltonian in Eq.~\ref{Eq:Hamiltonian} becomes,

       \begin{eqnarray}
        \mathcal{H}(\Phi) = \displaystyle\sum_{n,n', n' > n} \Big(
        \frac{J_R}{|n' - n|^\beta} e^{-i(n' - n)\Phi/L} c^\dagger_{n'} c_n ~~~~~~ \nonumber \\
        + \frac{J_L}{|n' - n|^\beta} e^{i(n' - n)\Phi/L} c^\dagger_n c_{n'} 
        \Big) + \sum_n V \cos(2\pi \alpha n + \phi) \, c^\dagger_n c_n \nonumber \\
        + \sum_n U \hat{\mathbf{n}}_n \hat{\mathbf{n}}_{n+1}.~~~~~~~~~~~~~~~~~~~~~~~~~
        \label{Eq:Hamiltonian_phi}
        \end{eqnarray}

        The winding number estimated around the reference energy $E_R$ (which is not an eigenenergy of $\mathcal{H}(\Phi)$) is then given by \cite{Gong,Shiozaki}:\\
        \begin{eqnarray}
            w=\frac{1}{2\pi i}\int_0^{2\pi} \partial_\Phi \text{ln det} \{\mathcal{H}(\Phi)-E_R\} d\Phi.
            \label{Eq:Winding_number}
        \end{eqnarray}
        
        Without loss of generality, we assume $E_R=0$. A positive (negative) winding indicates edge states towards the left (right) boundary. In this work, we computed the disorder averaged winding number $\left \langle w \right \rangle$ where different realizations over $\phi$ are considered till its convergence upto the fourth decimal place. This disorder averaging is essential since the energy spectra changes with $\phi$, as also demonstrated in Ref.~\cite{Huang}. It is important to note that in many-body systems, the non-trivial topological phases are much more complicated and can bear values other than $\left \langle w \right \rangle = \pm 1$. We then characterize the topologically trivial phase where $\left \langle w \right \rangle \equiv 0$ and the non-trivial phase $\left \langle w \right \rangle \neq 0$. 
        
        \subsection{Determination of the many-body skin effect under OBC: the density imbalance}\label{Sec:SE_OBC}
        Contrary to the non-Hermitian systems with asymmetric hopping without interaction, the real challenge in the many-body counterpart is to identify the skin states under the open boundary condition (OBC) owing to the large Hilbert space dimension. The notion of the many-body density distribution given by the occupation $\left \langle \hat{\textbf{n}} \right \rangle$ of the basis states becomes useful in drawing an analogy between the many-body and the single-particle pictures. It has been demonstrated that although the ground state does not exhibit an exponential localization that marks the skin effect, the density profile can exhibit an asymmetry between two halves of the system, which can be identified using the imbalance as a definitive marker \cite{Alsallom}. The imbalance of the occupation densities between the left and right halves in the eigenstate labeled $j$ is then given by:

        \begin{eqnarray}
            \mathcal{I}^j =\displaystyle \sum_{n\in \text{right half}} \left \langle \hat{\textbf{n}}_n \right \rangle - \sum_{n\in \text{left half}} \left \langle \hat{\textbf{n}}_n \right \rangle.
            \label{Eq:Imbalance}
        \end{eqnarray}
        The average imbalance over all the eigenstates ($\left\langle \mathcal{I} \right \rangle$) in the ergodic regime is then expected to be near 0, whereas for skin states it should ideally approach the value of the filling-fraction. It is also clear that the MBSE towards right is given by a positive $\left\langle \mathcal{I} \right \rangle$, whereas a negative $\left\langle \mathcal{I} \right \rangle$ indicates the skin states localized towards the left of the lattice. At this point, it is crucial to note that, since not all the skin states have complete occupation towards the right/left only, the average imbalance $\left\langle \mathcal{I} \right \rangle$ is expected to deviate from $L/2$ in our case, but still remains much greater than 0.
        
        \section{Results and Discussions}\label{Sec:Results}
        To identify the nature of the states and capture the phase diagram, we use a couple of more measures. Firstly, we find the fraction of ergodic eigenstates ($\phi_e$) similar to Ref.~\cite{Xianlong_2023}, where the states with $\text{D}_2>0.7$ are considered to be ergodic. The NHMBL regime is chalked out where all the states have $\text{D}_2<0.2$. 
        The second useful measure is the fraction of imaginary eigenenergies which is defined as $f_{im}=\mathcal{D}_{im}/\mathcal{D}$, where $\mathcal{D}_{im}$ is the number of eigenenergies with non-zero imaginary part estimated from the machine precision ($10^{-13}$), and $\mathcal{D}$ is the total number of eigenenergies which equals the Hilbert space dimension.\\
        \indent Before proceeding to the main discussion, we emphasize that locating the exact phase boundary is extremely challenging in systems with interaction due to the limitations in accessing larger system-sizes. Therefore, we make rough estimates using different measures to mark the separation in phases throughout this work. Our main focus is not in marking the phase boundaries, but to extract new phases that might appear due to the interplay of interaction and long-range hopping. In our simulations, for convenience, we have set $L=14$, unless otherwise specified. Please note that the phase diagrams obtained in this work using the biorthogonal framework remains unaltered upon considering either the left or the right eigenstates individually. For marking the MBSE phase, we have restricted ourselves to the right eigenstates only. We have used the PBC in Secs.~\ref{Sec:V_beta}, \ref{Sec:U_V}, \ref{Sec:EE}, \ref{Sec:Topology} and the OBC for analyzing the MBSE in Sec.~\ref{Sec:Imbalance}.

	\begin{figure}
        \begin{tabular}{p{\linewidth}c}
        \centering
            \includegraphics[width=0.245\textwidth,height=0.2\textwidth]{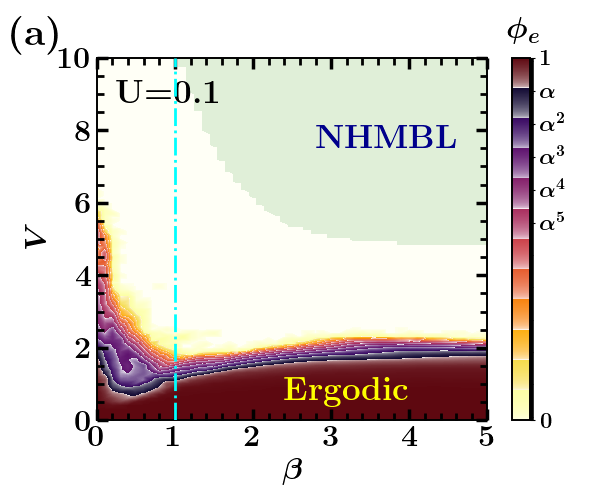}\hspace{-0.25cm}
            \includegraphics[width=0.245\textwidth,height=0.2\textwidth]{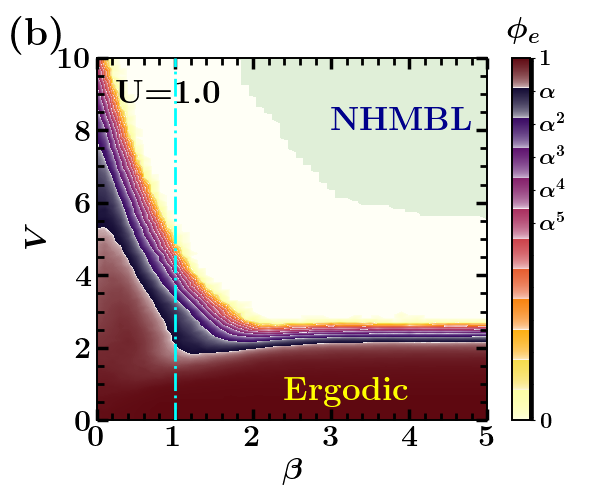}\\
            \includegraphics[width=0.245\textwidth,height=0.2\textwidth]{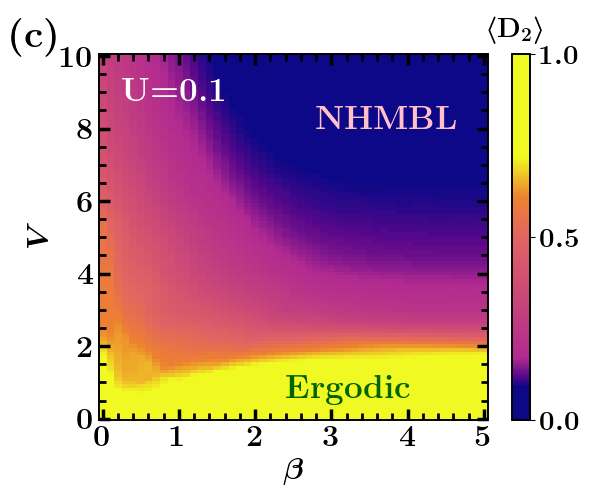}\hspace{-0.25cm}
            \includegraphics[width=0.245\textwidth,height=0.2\textwidth]{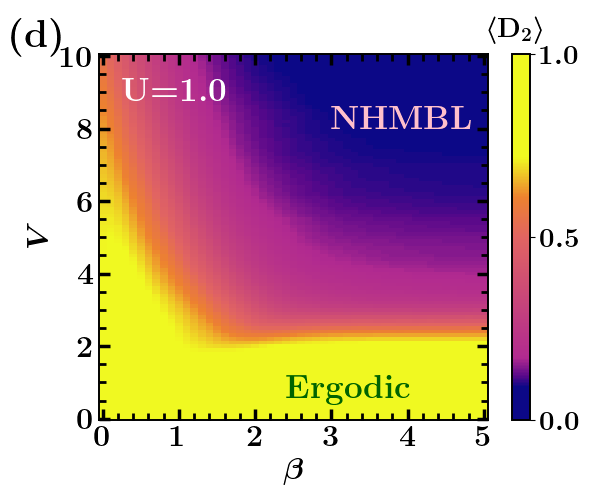}\\
            \includegraphics[width=0.245\textwidth,height=0.2\textwidth]{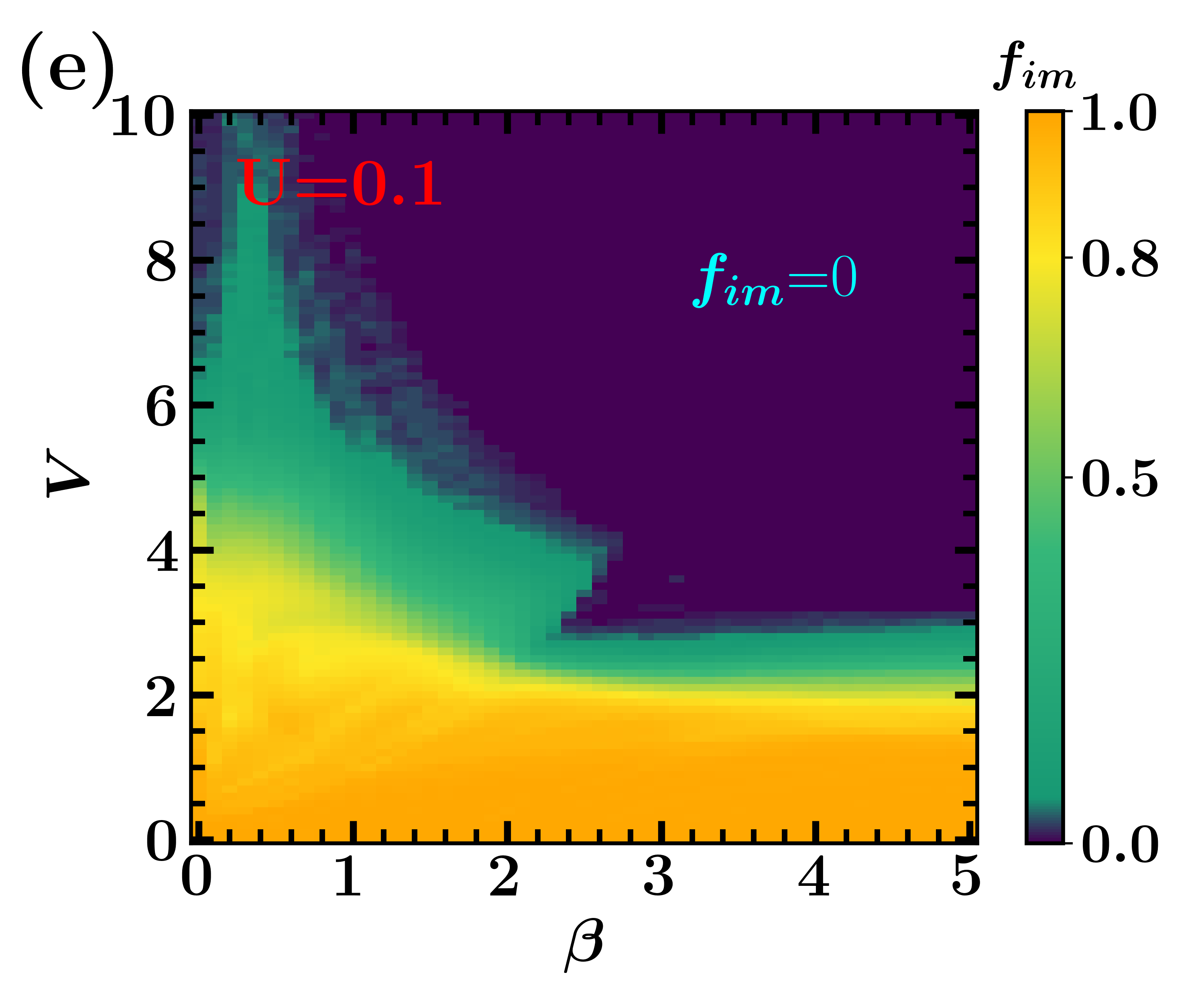}\hspace{-0.25cm}
            \includegraphics[width=0.245\textwidth,height=0.2\textwidth]{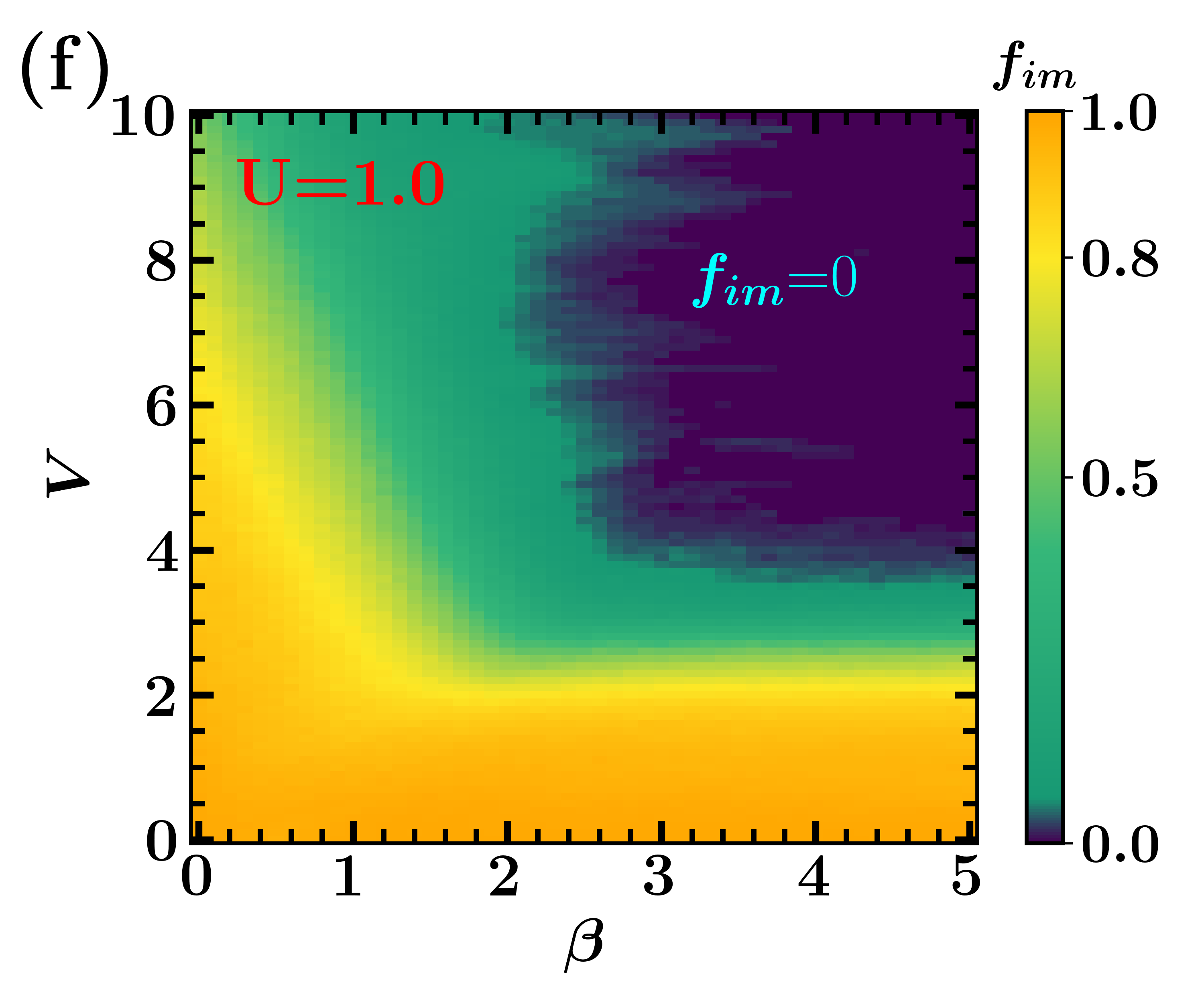}
        \caption{Different regimes characterized by the fraction of ergodic eigenstates ($\phi_e$) in terms of $\alpha$, in the parameter space spanning the strength of the quasiperiodic potential ($V$) and the power-law exponent ($\beta$) that controls the range of hopping. The interaction strength is set at: (a) $U=0.1$ and (b) $U=1.0$. (c) and (d) correspond to the same set of parameters as in (a) and (b) respectively, where the different phases are identified using the fractal dimension $\left \langle \text{D}_2 \right \rangle$. The $V-\beta$ phase diagrams for a broad recognition of the behavior of $f_{im}$ in the different regimes of the interacting system are shown at: (c) $U=0.1$ and (d) $U=1.0$, corresponding to the above figures.} 
        \label{Fig:Fig_1}
        \end{tabular}
        \end{figure}

        \begin{figure*}
            \includegraphics[width=0.245\textwidth,height=0.2\textwidth]{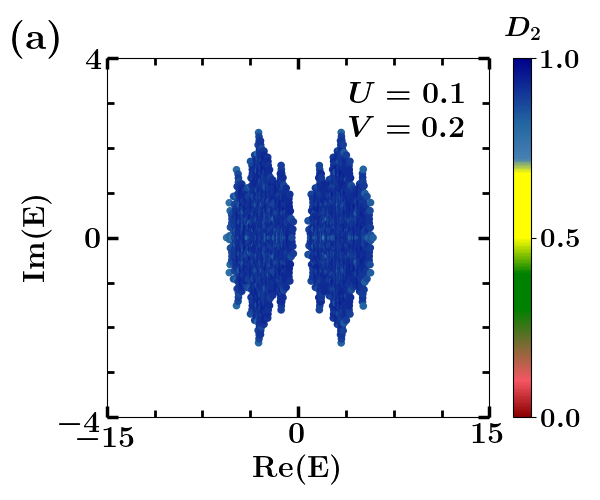}\hspace{-0.25cm}
            \includegraphics[width=0.245\textwidth,height=0.2\textwidth]{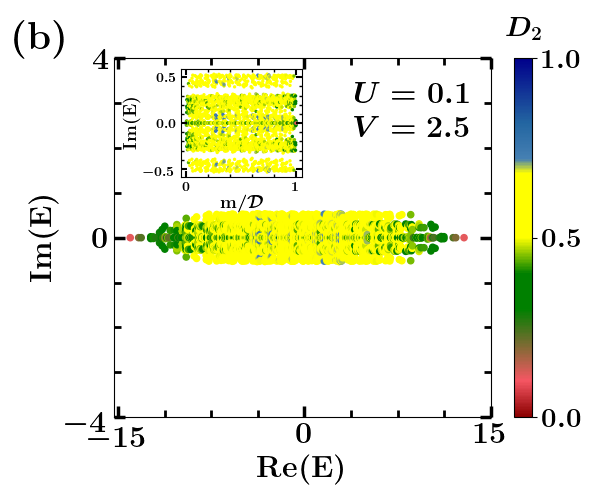}
            \includegraphics[width=0.245\textwidth,height=0.2\textwidth]{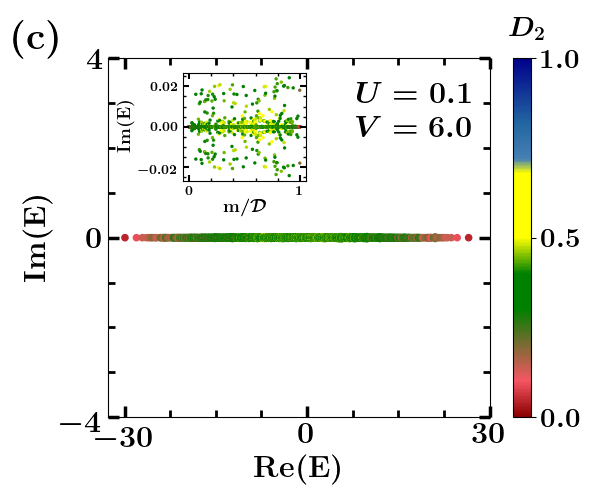}\\
            \includegraphics[width=0.245\textwidth,height=0.2\textwidth]{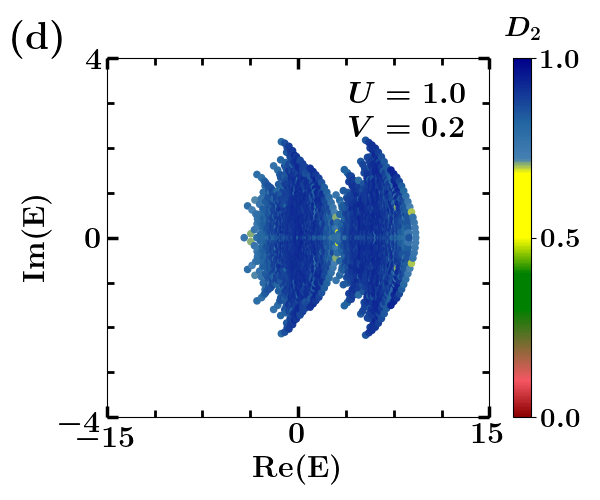}\hspace{-0.25cm}
            \includegraphics[width=0.245\textwidth,height=0.2\textwidth]{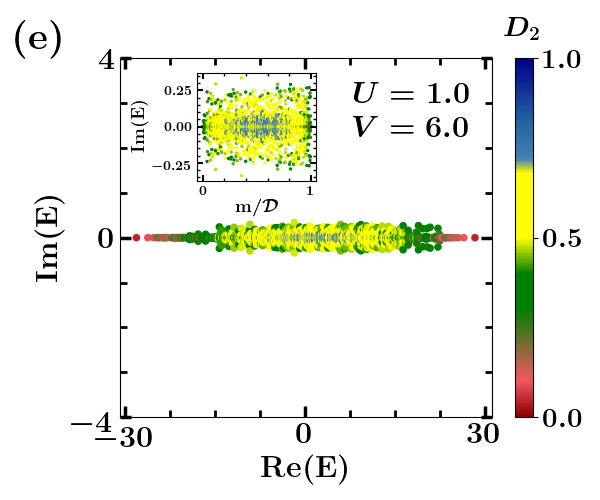}
            \includegraphics[width=0.245\textwidth,height=0.2\textwidth]{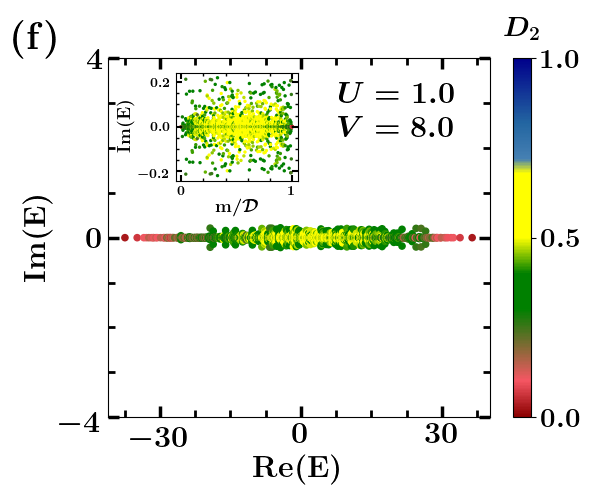}
        \caption{The energy spectrum in the complex plane at $\beta=0.5 (<1)$ and at different values of $U$ and $V$ chosen from Figs.~\ref{Fig:Fig_1}(a)-(f), wherein the multifractality of the different states are indicated using $\text{D}_2$. The other parameters are: (a) $U=0.1, V=0.2$ (ergodic regime), (b) $U=0.1, V=2.5$ (intermediate regime), (c) $U=0.1, V=8.0$ (intermediate regime), (d) $U=1.0, V=0.2$ (ergodic regime), (e) $U=1.0, V=6.0$ (intermediate regime), and (f) $U=1.0, V=8.0$ (intermediate regime). The intermediate regime clearly demonstrates an absence of multifractal/mobility edge in the system. The imaginary part of the eigenenergies with its fractal dimension $\text{D}_2$ as a function of the normalized eigenstate index in the many-particle Hilbert space where the intermediate regimes appear are shown in the figure insets.}
        \label{Fig:Fig_2}
        \end{figure*}

         \begin{figure*}
            \includegraphics[width=0.245\textwidth,height=0.2\textwidth]{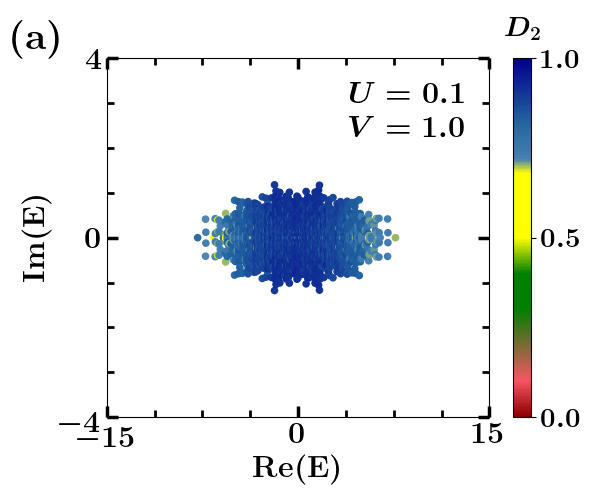}\hspace{-0.25cm}
            \includegraphics[width=0.245\textwidth,height=0.2\textwidth]{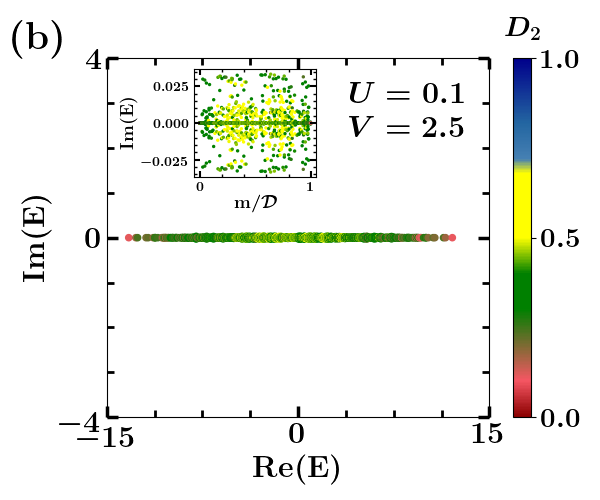}
            \includegraphics[width=0.245\textwidth,height=0.2\textwidth]{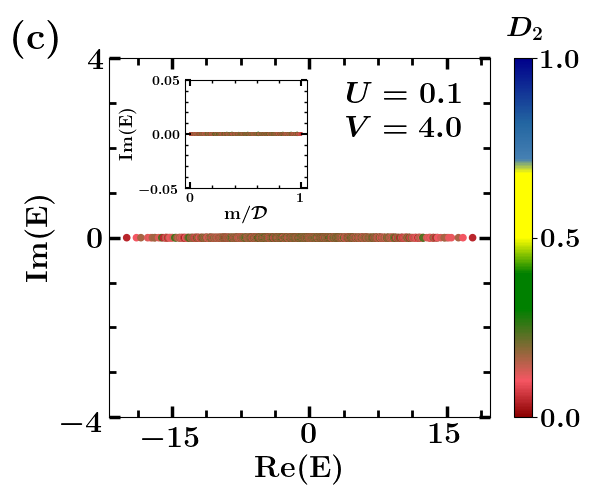}
            \includegraphics[width=0.245\textwidth,height=0.2\textwidth]{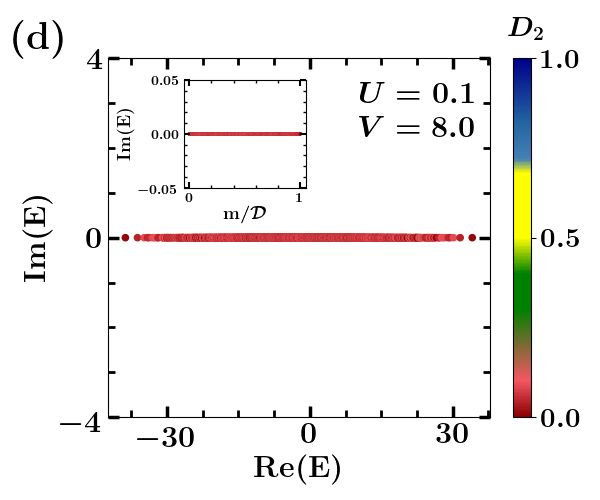}\\
            \includegraphics[width=0.245\textwidth,height=0.2\textwidth]{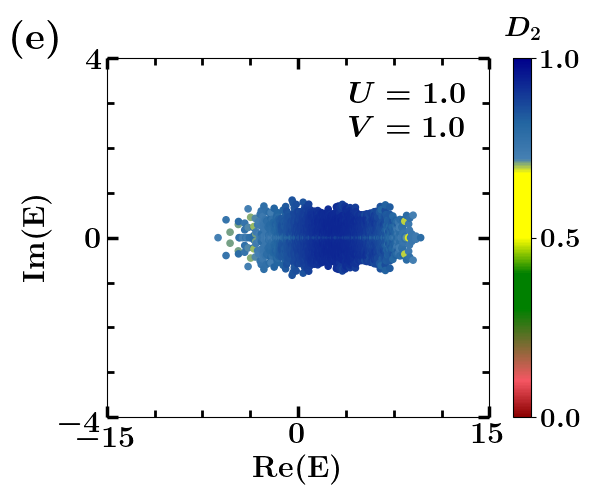}\hspace{-0.25cm}
            \includegraphics[width=0.245\textwidth,height=0.2\textwidth]{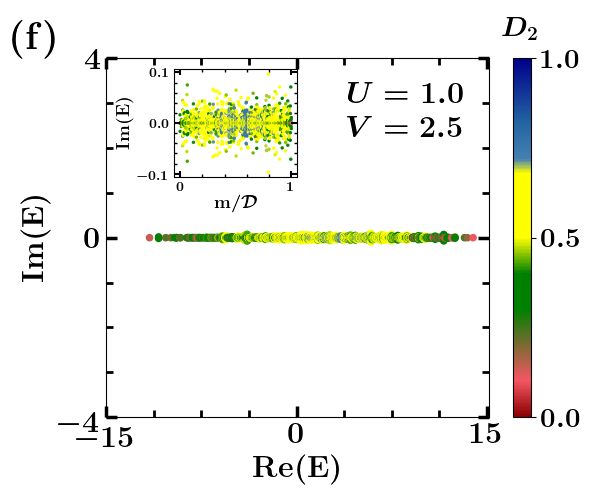}
            \includegraphics[width=0.245\textwidth,height=0.2\textwidth]{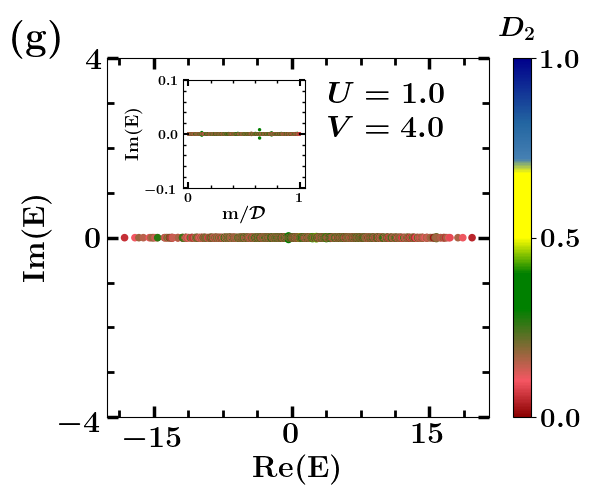}
            \includegraphics[width=0.245\textwidth,height=0.2\textwidth]{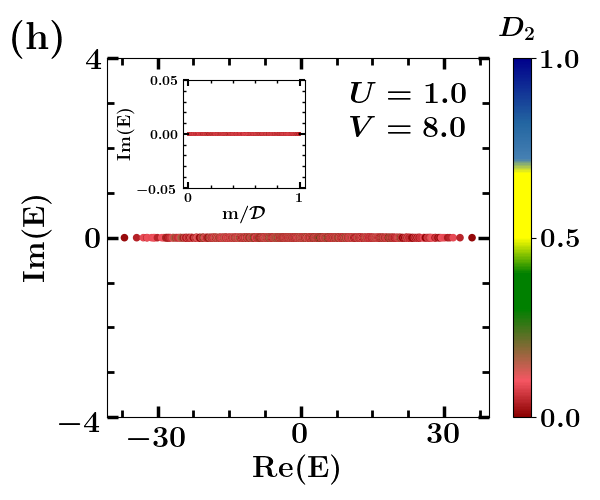}
        \caption{The energy spectrum in the complex plane at $\beta=3.0 (>1)$, similar to Fig.~\ref{Fig:Fig_2}. The parameters chosen in this case are: (a) $U=0.1, V=1.0$ (ergodic regime), (b) $U=0.1, V=2.5$ (intermediate regime), (c) $U=0.1, V=4.0$ (intermediate regime), and (d) $U=0.1, V=8.0$ (NHMBL regime). Figs. (e)-(h) in the lower panel are at identical values of the quasiperiodic potential as the ones in the above panel, but at $U=1.0$. Alike Fig.~\ref{Fig:Fig_2}, the imaginary part of the eigenenergies $vs.$ the normalized eigenstate index in the intermediate regimes are illustrated in the figure insets.}
        \label{Fig:Fig_3}
        \end{figure*}
        
        \subsection{Interaction-induced non-triviality: Phase dependence on $\beta$}\label{Sec:V_beta}
        In this section, we analyze the effect of the power-law exponent $\beta$ in the non-Hermitian chain at two strengths of interaction, i.e., $U=0.1$ (very low) and $U=1.0$ (moderate) respectively. It has been long known in the non-interacting counterparts that when the hopping is short-ranged, restricted to the nearest neighbor only, an ergodic-localization transition occurs at a critical strength of the quasiperiodic potential given by $V_c=2 \text{max}\{J_R,J_L\}$ \cite{Jiang}. In addition, the phase boundaries have also been drawn with power-law hopping \cite{Juan,Xianlong}. In such systems, a new dimension to the problem is added due to the presence of ergodic- multifractal (localized) edges at $\beta<1(>1)$ respectively. Such transitions between the two phases occurs at a normalized eigenstate-index $n/L=\alpha^s$, denoted as the $P_s$ regime dictated by $\phi_e = \alpha^s$. The $V-\beta$ phase diagram demonstrating the appearance of different $P_s$ regimes and the corresponding nature of $f_{im}$ in the non-interacting counterpart has been demonstrated in Fig.~\ref{Fig:Fig_A1} of Appendix~\ref{App:Non-interacting model}.

        \begin{figure}
        \begin{tabular}{p{\linewidth}c}
        \centering
            \includegraphics[width=0.22\textwidth,height=0.20\textwidth]{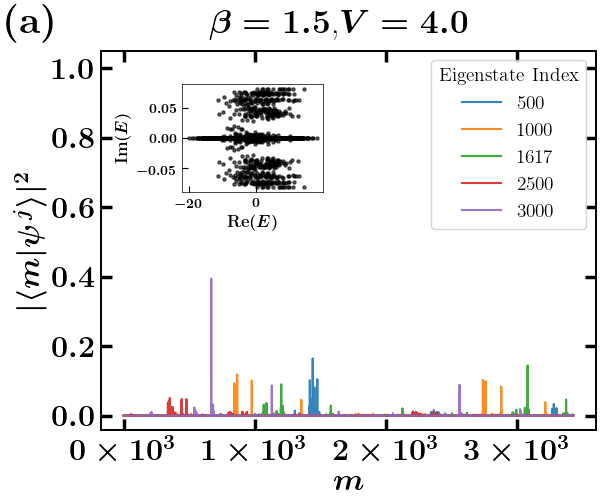}
            \includegraphics[width=0.22\textwidth,height=0.20\textwidth]{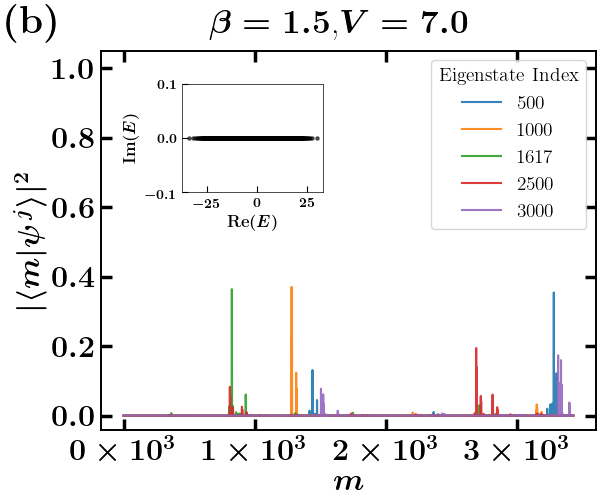}\\
            \includegraphics[width=0.22\textwidth,height=0.20\textwidth]{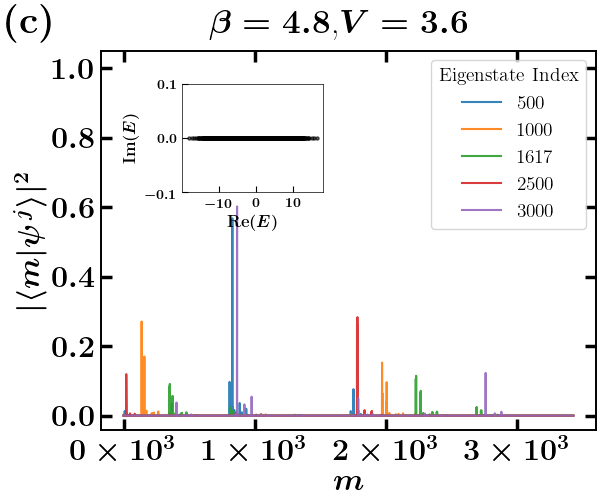}
            \includegraphics[width=0.22\textwidth,height=0.20\textwidth]{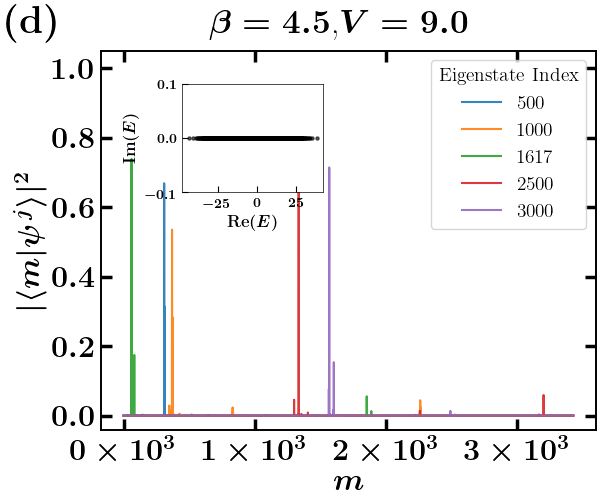}

        \caption{The probability distribution at $U=0.1$ of the various intermediate states are shown in (a)-(c), and the Fock-space NHMBL states are presented in (d). The insets demonstrate the nature of the energy spectrum in these three distinct phases. To clearly verify the nature of states in (a) to (d), we have used five representative states, i.e., $m=500,1000,1617,2500,3000$.}
        \label{Fig:Fig_4}
        \end{tabular}
        \end{figure}

        \indent Interestingly, in the presence of interaction, we find multiple emergent features as illustrated in Fig.~\ref{Fig:Fig_1}. At both low and moderate values of interaction (Figs.~\ref{Fig:Fig_1}(a) and (b) respectively), we find that the extent of the distinct $P_s$ regimes shrink significantly as compared to the non-interacting counterpart, as shown in Fig.~\ref{Fig:Fig_A1}(a). 
        Moreover, it is clear that a NHMBL regime appears when the hopping range is lowered and at large values of the quasiperiodic potential ($V$). There is an absence of one-to-one correspondence between the entire single-particle localized regime and the NHMBL one. In particular, there is no clear transition between the ergodic and NHMBL regimes, irrespective of the value of $\beta$ and $U$. It is important to note that, interestingly, with weak interaction, the NHMBL regime appears at $\beta<2$. In addition, at a low value of $\beta$, the ergodic regime is enhanced when the interaction is increased as demonstrated in Fig.~\ref{Fig:Fig_1}(b), while the NHMBL phase reduces slightly in extent. Moreover, on a closer inspection in the regimes on both sides of $\beta=1$, we find that the $P_s$ regimes lack a sharper boundary similar to Ref.~\cite{Xianlong_2023}, and in most of the parameter regimes other than the ergodic or NHMBL regimes, $\phi_e$ does not contain exactly $\alpha^s$ ergodic eigenstates, as one would typically expect from the non-interacting scenario (this is not demonstrated here explicitly, but will be clearer from the later discussion). In Figs.~\ref{Fig:Fig_2} and \ref{Fig:Fig_3}, we will explicitly show the absence of a clear multifractal/mobility edges in these regimes, in sharp contrast to the case without interaction. Therefore, we have refrained from labeling those as $P_s$ regimes and rather call them the `intermediate' ones wherever required. The different phases have also been verified using the level-statistics as demonstrated in Figs.~\ref{Fig:Fig_A3}(a)-(b) in Appendix~\ref{App:r_avg}. Furthermore, in Figs.~\ref{Fig:Fig_1}(c) and (d), we illustrate that in the intermediate regime, the $\left \langle \text{D}_2\right \rangle$ lies between 0.2 and 0.7, as the majority number of eigenstates are fractal in nature. We have verified the multifractal nature of these macroscopic number of eigenstates in the intermediate regime using average fractal dimension $\left \langle \text{D}_q \right \rangle$ for varying $q$ in Appendix~\ref{App:D_q}. Henceforth, to determine the multifractal nature of the states, only $\text{D}_2$ has been used to avoid additional numerical costs. \\
        \indent The next important question that immediately comes to the mind is that ``Does reality in the eigenspectrum always indicate the NHMBL?" To answer this, we plot the phase diagram using $f_{im}$ at $U=0.1$ and $U=1.0$ in Figs.~\ref{Fig:Fig_1}(e) and (f) respectively. We uncover that the system passes via an intermediate regime with real eigenvalues before crossing over to the fully NHMBL regime. This statement will be made clear in the subsequent discussion. Furthermore, such a difference in the energy spectrum where the intermediate eigenstates can have either complex or entirely real spectrum will have profound effects in the evolution of entanglement entropy, as discussed in Sec.~\ref{Sec:EE}. We now discuss the nature of the states in the intermediate regimes in more detail. In Figs.~\ref{Fig:Fig_2} and \ref{Fig:Fig_3}, we have plotted the complex energy spectrum along with the $\text{D}_2$ values for the individual Fock-space eigenstates at $\beta<1$ and $\beta>1$ respectively. We first consider the case when $\beta<1$. From Fig.~\ref{Fig:Fig_2}(a) and Fig.~\ref{Fig:Fig_2}(d), the appearance of ergodic eigenstates are evident at both $U=0.1$ and $U=1.0$ respectively. Interestingly, in the intermediate regime, we find a mixture of localized, multifractal and ergodic eigenstates as illustrated in Figs.~\ref{Fig:Fig_2}(b) and (e), both at low and moderate strengths of the nearest-neighbor interaction. However, upon increasing $V$, the ergodic eigenstates vanish entirely from the spectrum as demonstrated in Figs.~\ref{Fig:Fig_2}(c) and (f). Furthermore, from the insets Figs.~\ref{Fig:Fig_2}(b), (c), (e) and (f), it is evident that these states are not clearly separated at any particular energy, as is demonstrated in the figure insets. This is in sharp contrast to the ergodic-multifractal edge observed in the non-interacting picture.
        We now consider the scenario when $\beta>1$, 
        as shown in Fig.~\ref{Fig:Fig_3}. In the ergodic regime and intermediate ones at low $V$ as illustrated in Figs.~\ref{Fig:Fig_3}(a), (b), (e) and (f), the results are qualitatively similar to those obtained in Fig.~\ref{Fig:Fig_2} (a),(b), (d) and (e), i.e., the intermediate regimes demonstrate an absence of clear edge (in this case the mobility edge) in the Fock-space eigenstates. In this case, however, it is clear from Figs.~\ref{Fig:Fig_3}(c) and (g) that a part of the multifractal states in the intermediate regime can possess real eigenvalues, similar to the ones in the NHMBL regime. To conclude, it is quite evident from this preceding discussion that the ergodic-multifractal (localized) phase boundaries for $\beta<1$ and $\beta>1$ no longer exist in the presence of both low and moderate interaction strength and the intermediate regime rather consists of mostly multifractal states.\\
        \indent Another important aspect in such systems with asymmetric hopping is the existence of TRS in the Hamiltonian as already highlighted. This in turn leads to the existence of complex conjugate pairs of eigenvalues in the non-localized regime that breaks the TRS, whereas a completely real eigenspectrum following the restoration of TRS in the NHMBL regime at the transition when $\beta\rightarrow \infty$. In the presence of long-range hopping as demonstrated in Figs.~\ref{Fig:Fig_2} and \ref{Fig:Fig_3}, it is clearly evident that the restoration of TRS happens much before the emergence of the fully NHMBL regime. For further clarification regarding the existence of real spectrum even when the states are multifractal in nature, we plot the probability distribution of the Fock-states in distinct regimes at $U=0.1$ in Figs.~\ref{Fig:Fig_4}(a) to (d). From Figs.~\ref{Fig:Fig_4}(a)-(c), it is clear that in the intermediate regime, the weights are distributed and non-vanishing in several configurations over the Hilbert space. In Figs.~\ref{Fig:Fig_4}(a), we find a complex spectrum, whereas Figs.~\ref{Fig:Fig_4}(b) and (c) verifies the presence of real eigenspectrum (as shown in the figure insets). This is also true in the case of $U=1.0$ as demonstrated in Fig.~\ref{Fig:Fig_A5} of Appendix.~\ref{App:WF_analysis_U1.0}. Therefore, the reality of eigenspectrum as an indicator of the NHMBL regime as reported in several earlier works could be a misleading identifier of the phase.
        Moreover, it is also clear that not all the multifractal states in the intermediate regime are bound to possess real eigenvalues, which also contradicts the results in Ref.~\cite{Xianlong}. Lastly, in the NHMBL regime, the weights are distributed in different localization centers and some distant neighbors which have very small weights, similar to earlier findings, as demonstrated in Fig.~\ref{Fig:Fig_4}(d).
    
        \subsection{Effect of interaction in the short and long-range hopping limits: the $U-V$ phase diagram}\label{Sec:U_V}
\begin{figure}
        \begin{tabular}{p{\linewidth}c}
        \centering
            \includegraphics[width=0.245\textwidth,height=0.2\textwidth]{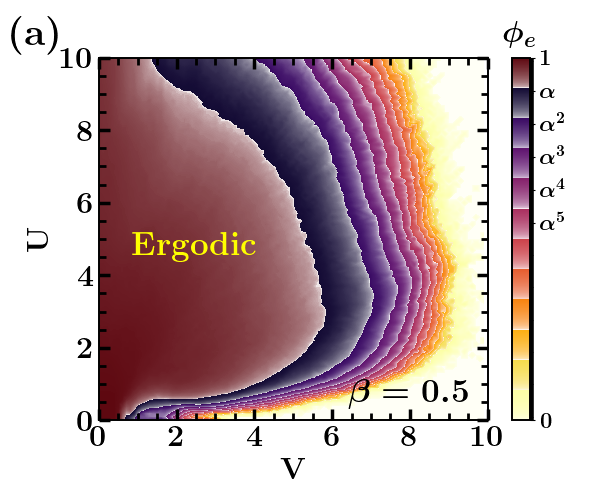}\hspace{-0.25cm}
            \includegraphics[width=0.245\textwidth,height=0.2\textwidth]{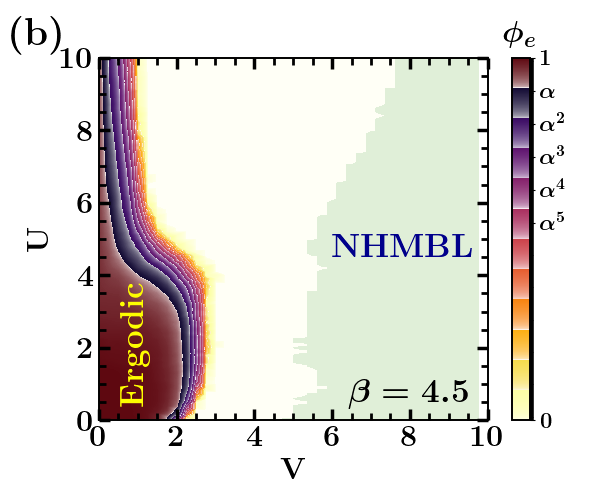}\\
            \includegraphics[width=0.245\textwidth,height=0.2\textwidth]{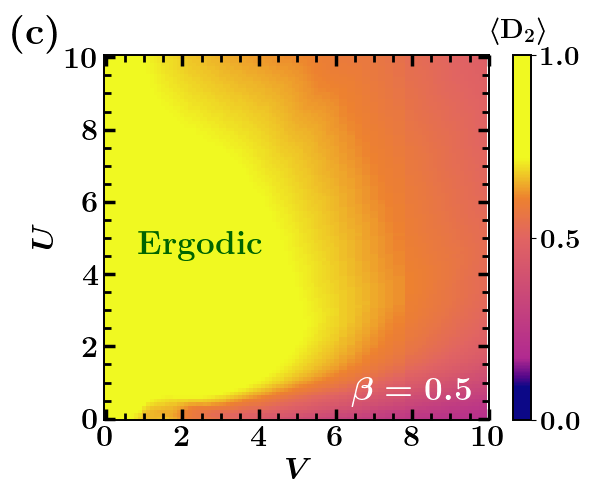}\hspace{-0.25cm}
            \includegraphics[width=0.245\textwidth,height=0.2\textwidth]{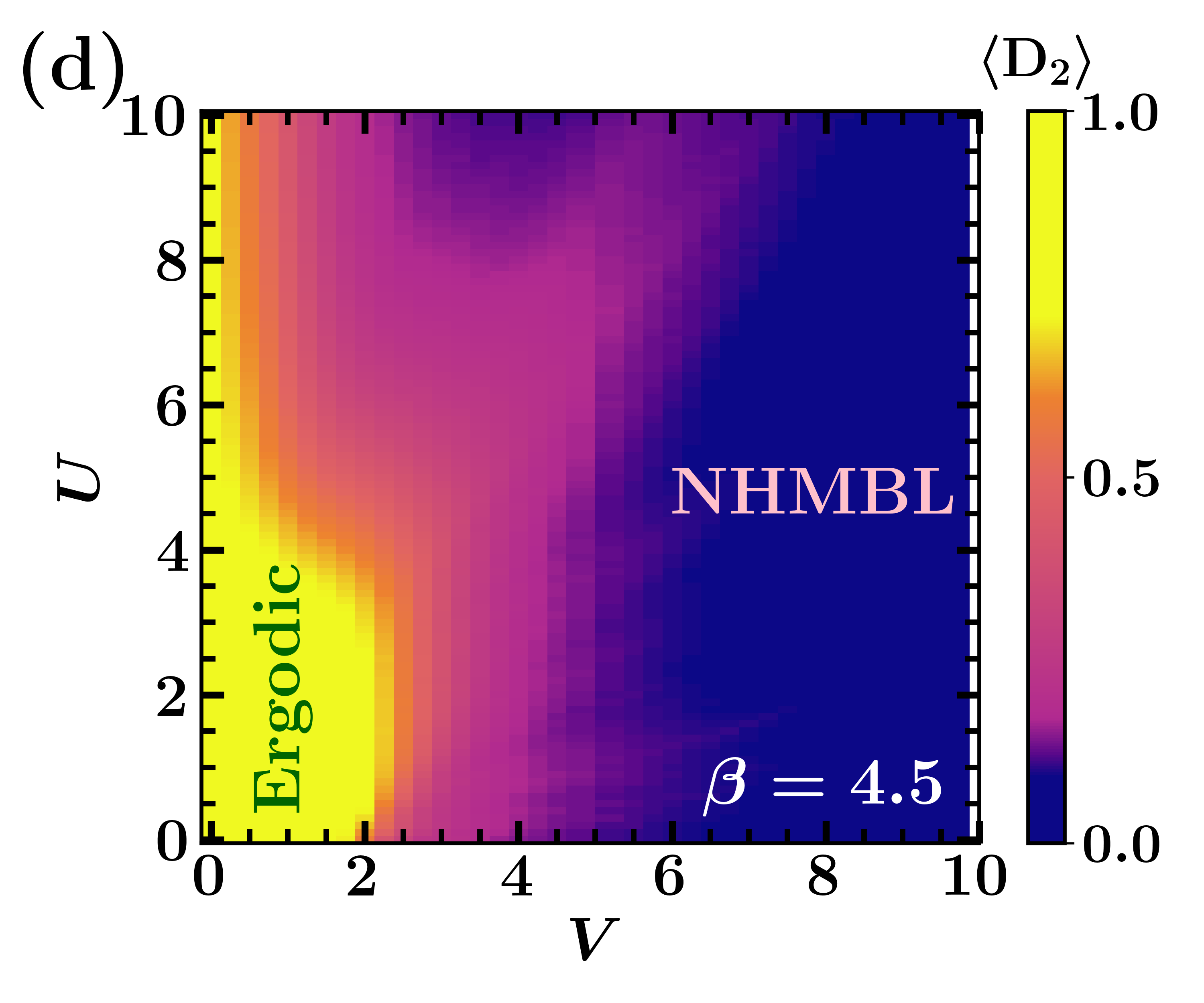}
        \caption{The phase diagram depicting (a),(b) $\phi_e$ and (c),(d) $\left \langle \text{D}_2 \right \rangle$, similar to Fig.~\ref{Fig:Fig_1}, at fixed values of $\beta= 0.5~ \text{and}~ \beta= 4.5$ respectively for varying interaction strength $U$ as a function of the quasiperiodic potential $V$.}
        \label{Fig:Fig_5}
        \end{tabular}
        \end{figure}

\begin{figure}
        \begin{tabular}{p{\linewidth}c}
        \centering
            \includegraphics[width=0.262\textwidth,height=0.225\textwidth]{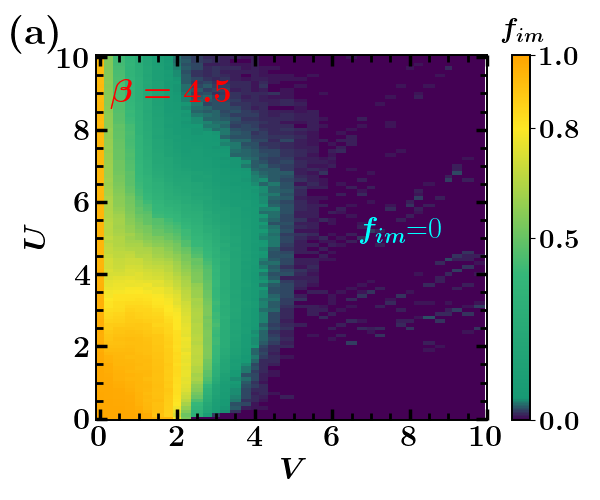}\hspace{-0.15cm}
            \includegraphics[width=0.22\textwidth,height=0.225\textwidth]{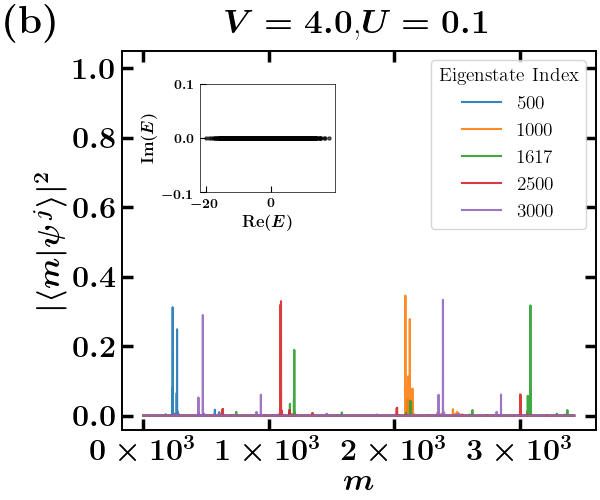}
        \caption{(a) $f_{im}$ in the $U-V$ phase diagram corresponding to Fig.~\ref{Fig:Fig_5}(b) at $\beta=4.5$. (b) A clear demonstration of the probability of the five selected intermediate eigenstates, wherein all the states are illustrated to have completely real eigenenergies as presented in the inset. }
        \label{Fig:Fig_6}
        \end{tabular}
        \end{figure}

        In this section, our aim is to understand the effect of interaction in both the short and long-range hopping limits. To do so, we construct the phase diagram using $\phi_e$ at $\beta=0.5$ and $\beta=4.5$ in Fig.~\ref{Fig:Fig_5}. Naively, one would expect the destruction of the ergodic phases upon adding interaction, at least at half-filling and when $\beta \rightarrow \infty$. Contrary to such expectations, a few works in the past have demonstrated a slight increase in the critical value of MBL transition with an increase in the strength of the interaction and the ergodic phase survives even at $U/J_R>1$ in the pure AAH model \cite{Huse_2013,Bera}. However, in such systems the ergodic regime vanishes beyond a certain strength of the interaction. In this work, we find that in the presence of long-range hopping, the ergodic regime extends quite substantially up to $U/J_R>>1$, as expected. The difference in the extent of the ergodic regime is clearly visible from Fig.~\ref{Fig:Fig_5}(a) and Fig.~\ref{Fig:Fig_5}(b) corresponding to the long and short-range hopping respectively. Furthermore, it is quite evident that with an increase in $\beta$, that both the ergodic and the intermediate regimes shrink, approaching the earlier reported results at $\beta \rightarrow \infty$ limit. In addition, it is evident that in the presence of long-range hopping, the NHMBL regime is completely absent, even when both the interaction and quasiperiodic potential are large enough.
        On the other hand, in the limit of short-range hopping, upon increasing the strength in quasiperiodic potential, the NHMBL regime appears. It is interesting to note that a greater disorder strength is required to approach the NHMBL regime when the interaction is strong enough. These phase diagrams are also consistent with the quantity $\braket{\text{D}_2}$ as illustrated in Figs.~\ref{Fig:Fig_5}(c) and Fig.~\ref{Fig:Fig_5}(d) in the long and short-range hopping limits respectively.\\
        \indent Besides, similar to Sec.~\ref{Sec:V_beta}, we plot the fraction of imaginary eigenenergies corresponding to different strengths of $U$ and $V$ at $\beta=4.5$ (such that one finds the NHMBL regime as well) in Fig.~\ref{Fig:Fig_6}(a). It is clear that some intermediate states with multifractality possess completely real eigenergies, identical to the previous observations in Fig.~\ref{Fig:Fig_4}. We further verify this in Fig.~\ref{Fig:Fig_6}(b), which is a clear indicative that the multifractal states with distributed probabilities in the Fock-space configurations exhibit absolutely real energies as illustrated in the inset.

        \begin{figure}
        \begin{tabular}{p{\linewidth}c}
        \centering
            \includegraphics[width=0.24\textwidth,height=0.24\textwidth]{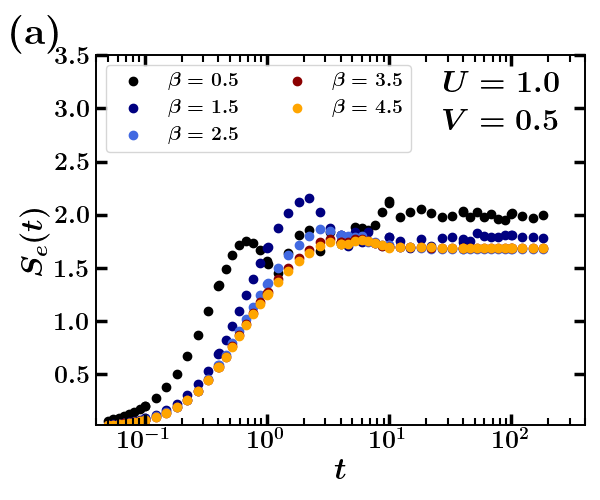}\hspace{-0.1cm}
            \includegraphics[width=0.24\textwidth,height=0.24\textwidth]{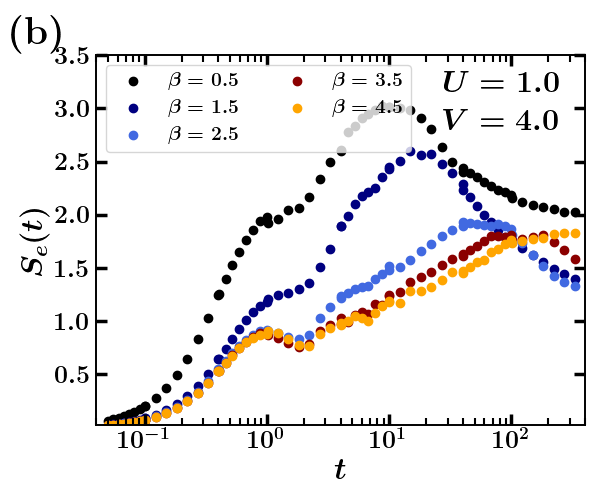}\hspace{-0.1cm}\\
            \includegraphics[width=0.24\textwidth,height=0.24\textwidth]{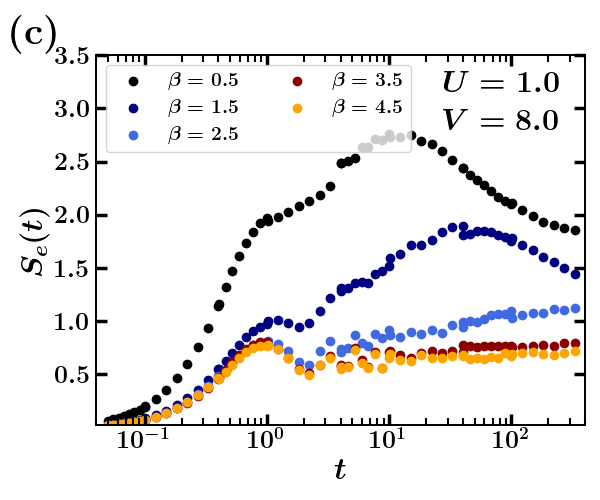}\hspace{-0.1cm}
            \includegraphics[width=0.24\textwidth,height=0.24\textwidth]{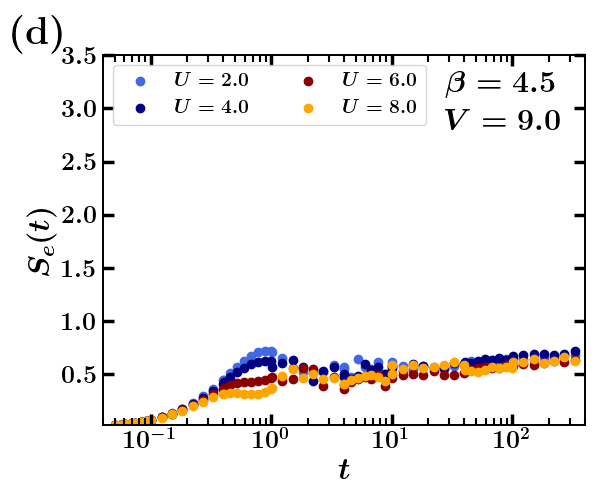}\hspace{-0.1cm}
        \caption{$S_e(t)$ vs. $t$ for varying power-law exponents $\beta$ at (a) $U=1.0,V=0.5$ (ergodic phases), (b) $U=1.0,V=4.0$ (intermediate phases), and (c) $U=1.0,V=8.0$ (intermediate and localized phases). The NHMBL regime at varying strengths of $U$ at $\beta=4.5, V=9.0$ is illustrated in (d). We have considered the average over 200 configurations of $\phi$ in all the figures.}
        \label{Fig:Fig_7}
        \end{tabular}
        \end{figure}

        \subsection{Behavior of entanglement entropy in the various phases}\label{Sec:EE}
        In previous works, the effect of long-range hopping in interacting Hermitian systems has been studied in the context of entanglement entropy and associated dynamics \cite{Roy,Nag}. In particular, in systems with power-law hopping and interaction, it was demonstrated that the von-Neumann entanglement entropy shows a faster growth in time for $\beta<1$ as compared to $\beta>1$ \cite{Nag}. Therefore, in order to have a complete understanding about the effect of $\beta$ in  non-Hermitian interacting systems, and also verify the type of phases discussed in the previous sections, we plot $S_e(t)$ vs. $t$ (as given in Eq.~\ref{Eq:Entanglement_entropy}) in the semi-logarithmic scale corresponding to the parameters in Fig.~\ref{Fig:Fig_1}(b). As illustrated in Fig.~\ref{Fig:Fig_7}(a), in the ergodic regime, we find an initial ballistic growth of the entanglement entropy that saturates to a large value at long times. In addition, similar to the observation in Ref.~\cite{Nag}, the growth of the entanglement entropy slows down with an increase in the value of $\beta$, indicating that in the presence of long-range hopping, the subsystem A entangles faster with subsystem B, spreading its information, as compared to the system with nearest-neighbor hopping. Furthermore, the saturating value for thermalization decreases with an increase in $\beta$. However, beyond a certain hopping-range, the saturated entropy remains constant with a further increase in $\beta$. Next, we select the value of $V$ such that the states are intermediate in nature as demonstrated in Fig.~\ref{Fig:Fig_7}(b). Similar to the previous finding, we find a faster growth of the entanglement when $\beta$ is low, but the behavior of the thermal value changes drastically. We notice that with an increase in $\beta$, initially the entanglement entropy at longer times decreases upto $\beta=2.5$. With a further increase in $\beta$, the long-time entropy starts increasing again. This is a characteristic feature of intermediate states in the non-Hermitian systems as reported in earlier works \cite{Imura_2023}. Another remarkable property is the non-monotonic behavior in entanglement entropy that arises from the non-unitary dynamics in the non-Hermitian systems as evident from Fig.~\ref{Fig:Fig_7}(b). As time progresses, the superposition governed by Eq.~\ref{Eq:Evolution} converges to a single or few eigenstates with maximum positive imaginary part in its eigenenergies. The initial growth of the entanglement entropy appears because of the dephasing mechanism, followed by the decay due to relaxation as discussed in Ref.~\cite{Banerjee}. 
        Moreover, in a similar non-Hermitian system, the authors claim that the maximum of the entanglement entropy occurs when the states have maximum ergodicity \cite{Imura_2022}. Surprisingly, we find that the maximum value of the entanglement entropy in the intermediate phase lies above that of the ergodic regime, but much below the Page value $S_{Page}=L/2~\text{ln}2-1/2$ \cite{Page}. We clearly find no such analogy in the system considered in our work.\\
        \indent We now pick another value of $V$ such that the NHMBL states are encountered. In Fig.~\ref{Fig:Fig_7}(c), at $V=8.0$, we find that in the intermediate regimes, the entanglement entropy behaves similar to Fig.~\ref{Fig:Fig_7}(b), while it saturates to a much smaller value, when $\beta$ is such that the states are NHMBL in nature. This is similar to the states deep in the NHMBL phase as demonstrated in Fig.~\ref{Fig:Fig_7}(d), where the linear growth terminates with a logarithmic growth for sufficiently long times, and a low saturation value of the entanglement entropy. 
        Additionally, for the sake of interested readers, we have demonstrated the system-size behavior of the entanglement entropy across the different phases in Appendix \ref{App:EE_system_sizes}.

        \begin{figure}
        \begin{tabular}{p{\linewidth}c}
        \centering
            \includegraphics[width=0.245\textwidth,height=0.2\textwidth]{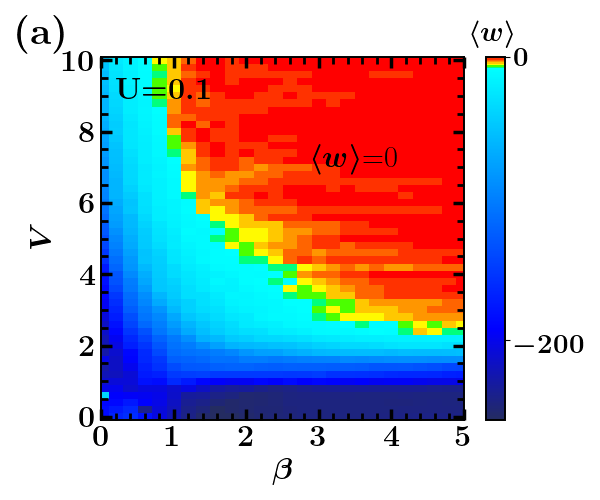}\hspace{-0.25cm}
            \includegraphics[width=0.245\textwidth,height=0.2\textwidth]{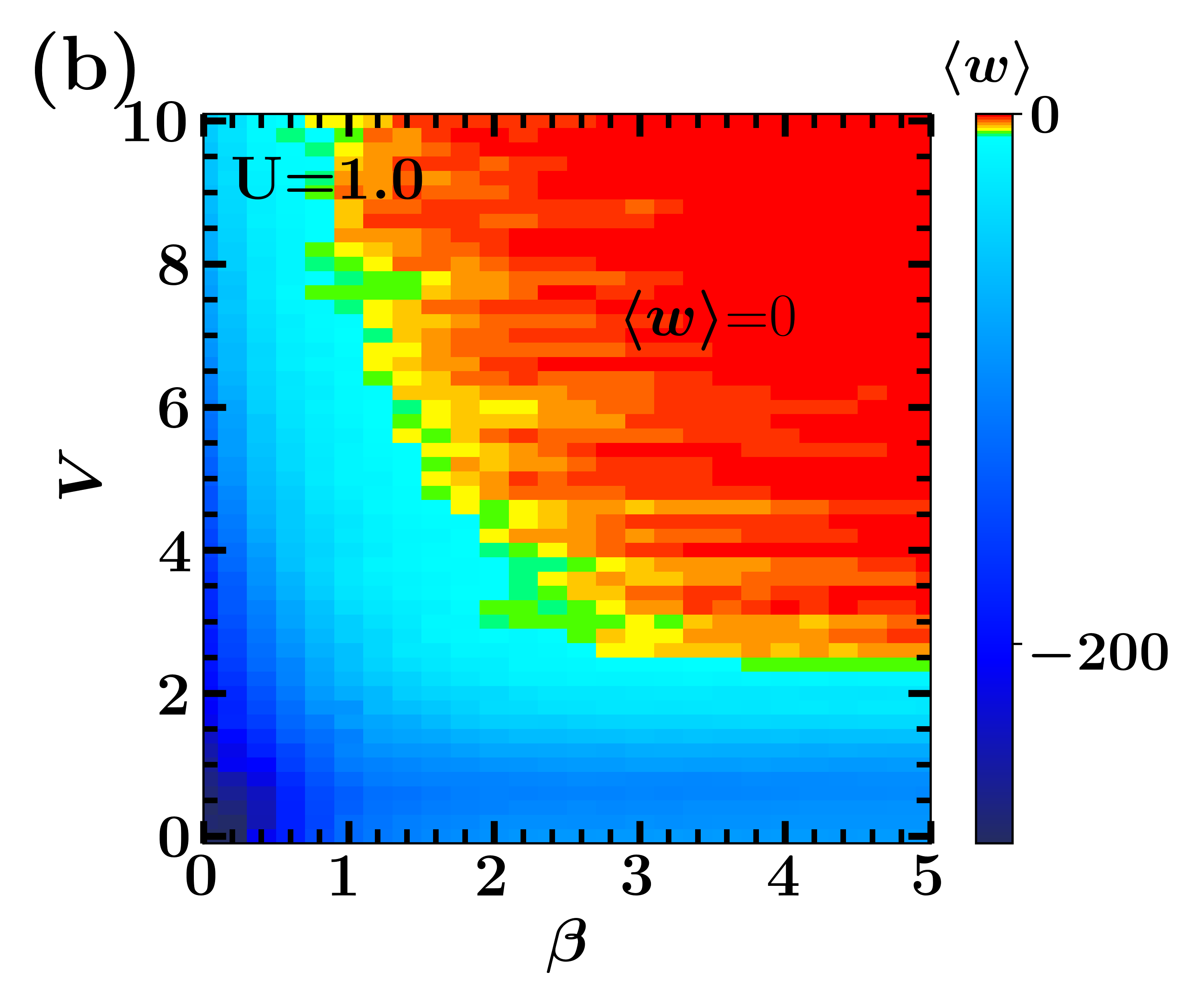}
        \caption{The many-body topological winding number averaged over several realizations $\left \langle w \right \rangle$ in the parameter space $V-\beta$ at (a) $U=0.1$ and (b) $U=1.0$ under PBC. Red indicates the topologically trivial phase, while other colors indicate the non-trivial phase.}
        \label{Fig:Fig_8}
        \end{tabular}
        \end{figure}

        \begin{figure}
        \begin{tabular}{p{\linewidth}c}
        \centering
            \includegraphics[width=0.245\textwidth,height=0.2\textwidth]{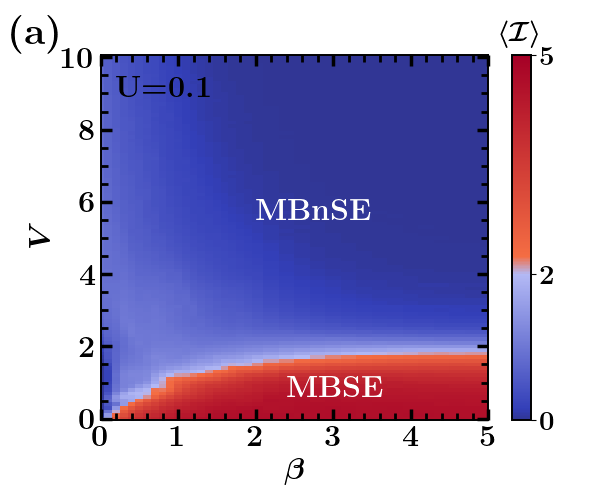}\hspace{-0.25cm}
            \includegraphics[width=0.245\textwidth,height=0.2\textwidth]{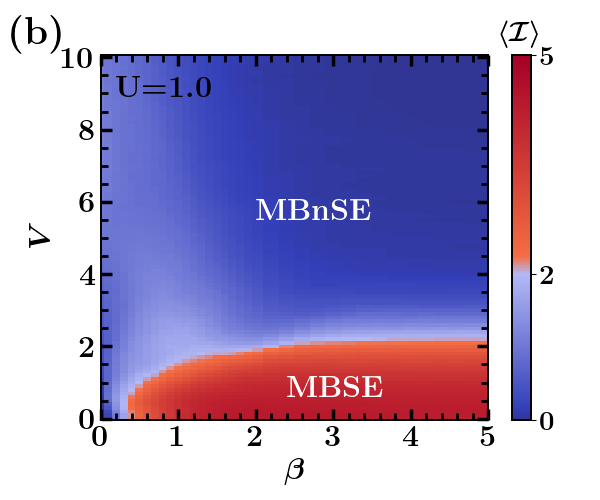}\\
            \includegraphics[width=0.245\textwidth,height=0.2\textwidth]{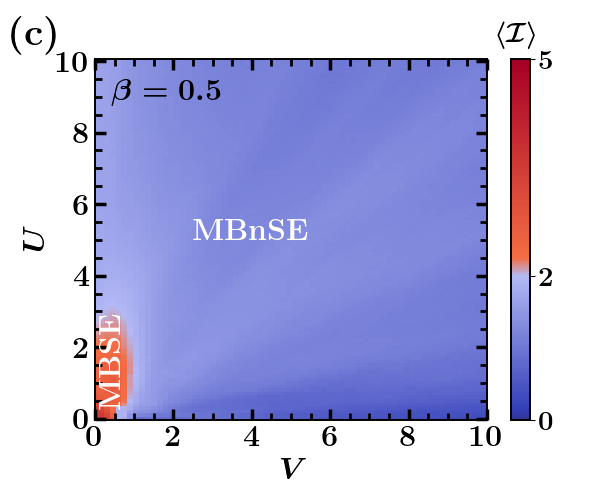}\hspace{-0.25cm}
            \includegraphics[width=0.245\textwidth,height=0.2\textwidth]{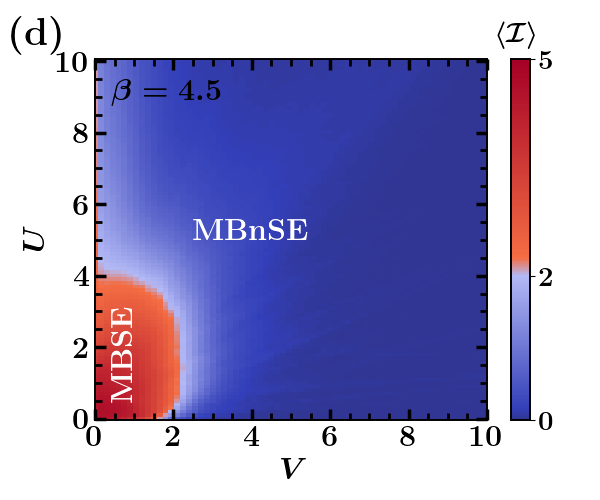}
        \caption{The average imbalance $\left \langle \mathcal{I} \right \rangle$ separating the MBSE (in red) and MBnSE (in blue) phases. In (a) and (b), we have demonstrated the behavior in the $V-\beta$ phase diagrams at $U=0.1 ~\text{and} ~U=1.0$ respectively, whereas in (c) and (d), the phase separations are shown at fixed values of $\beta=0.5~ \text{and} ~\beta=4.5$ respectively for varying $U$ and $V$.} 
        \label{Fig:Fig_9}
        \end{tabular}
        \end{figure}

        \begin{figure*}
        \centering
            \includegraphics[width=0.19\textwidth,height=0.16\textwidth]{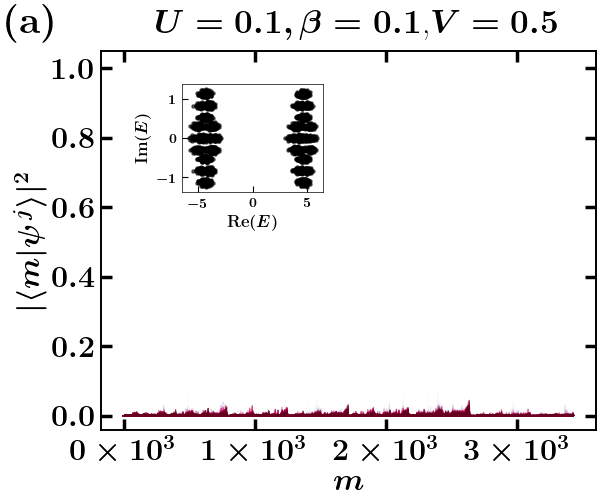}\hspace{-0.1cm}
            \includegraphics[width=0.19\textwidth,height=0.16\textwidth]{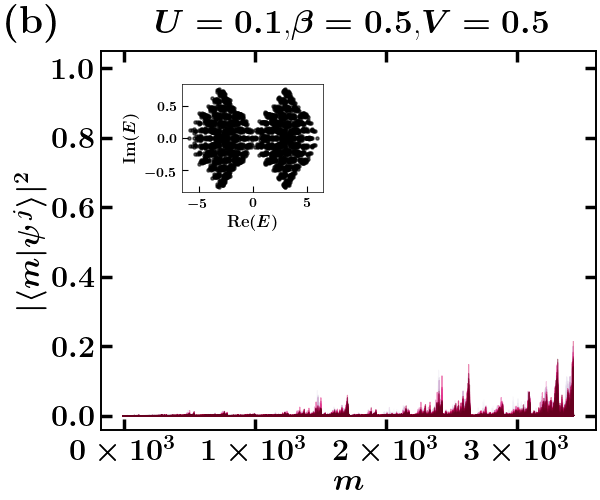}\hspace{-0.1cm}
            \includegraphics[width=0.19\textwidth,height=0.16\textwidth]{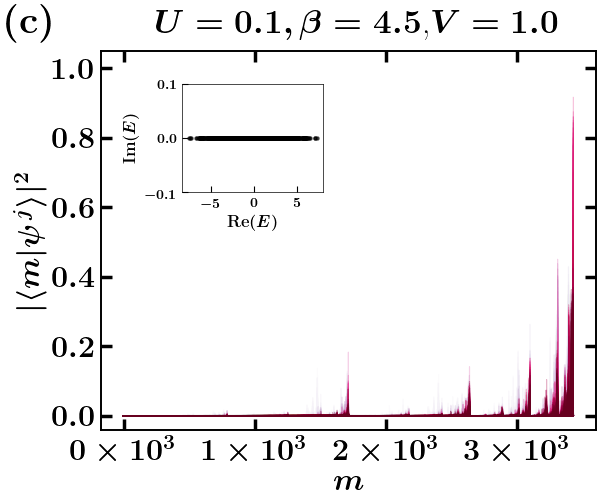}\hspace{-0.1cm}
            \includegraphics[width=0.19\textwidth,height=0.16\textwidth]{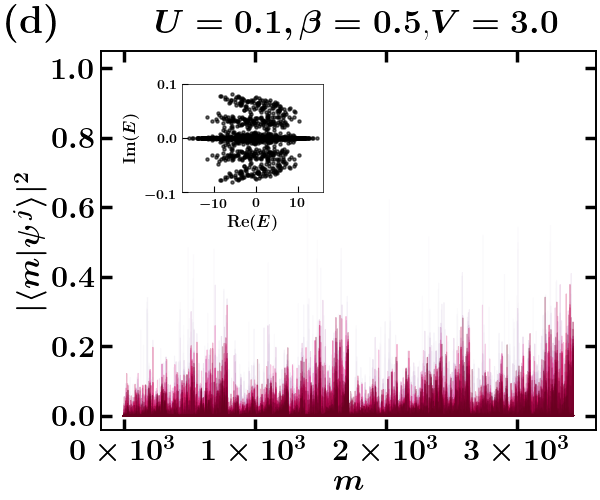}\hspace{-0.1cm}
            \includegraphics[width=0.19\textwidth,height=0.16\textwidth]{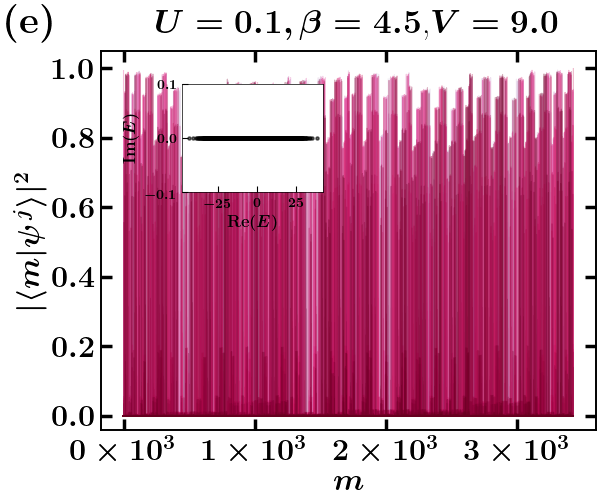}
        \caption{The wave-function probabilities under OBC for all the many-body eigenstates to demonstrate the existence or absence of the MBSE in accordance with the measure $\left \langle \mathcal{I} \right \rangle$, corresponding to Fig.~\ref{Fig:Fig_9}(a), i.e, at $U=0.1$. The other parameters are: (a) $\beta=0.1, V=0.5$, (b) $\beta=0.5, V=0.5$, (c) $\beta=4.5, V=1.0$, (d) $\beta=0.5, V=3.0$ and (e) $\beta=4.5, V=9.0$.}
        \label{Fig:Fig_10}
        \end{figure*}
        
        \subsection{The topologically distinct phases under PBC}\label{Sec:Topology}
        As already discussed, the topological transition has often been interlinked to the ergodic-NHMBL phase transition.
        In this work, as mentioned, the Hamiltonian is invariant under the TRS and also undergoes an ergodic-NHMBL phase transition. Furthermore, in Sec.~\ref{Sec:V_beta}, we have pointed out that in addition to the NHMBL phase, a part of the intermediate regime also restores the TRS leading to real eigenvalues. Naturally, we now wish to figure out the associated topology in these distinct phases before we can tie it to the NHMBL phase transition and the TRS of the Hamiltonian.\\
        \indent To do so, we estimate the average winding number $\left\langle w \right \rangle$ corresponding to Figs.~\ref{Fig:Fig_1}(a) and \ref{Fig:Fig_1}(b) over several disorder configurations as discussed in Sec.~\ref{Sec:Winding_number}. From Figs.~\ref{Fig:Fig_8}(a) and (b), it is evident that both at low and moderate strengths of interaction, the topological phases in the $V-\beta$ parameter space indicates topologically non-trivial phase corresponding to the entire ergodic and a part of the intermediate regime. On the other hand, in the NHMBL phase, the topology remains trivial as expected. However, in consistency with the previous finding of the multifractal states with real eigenvalues in the intermediate regime, we find topologically trivial phases before the appearance of NHMBL regime. It is crucial to note that the one-to-one correspondence between  Figs.~\ref{Fig:Fig_1}(a), \ref{Fig:Fig_1}(b) and Figs.~\ref{Fig:Fig_8}(a), \ref{Fig:Fig_8}(b) breaks down because we have considered disorder averaging in the latter to rule out chances of the conclusion being affected due to a particular potential configuration. Therefore, as we understand, the topological phase transition (associated to the TRS breaking) is not concurrent to the ergodic-NHMBL separation, due to a combined interplay of long-range hopping and interaction.

        \subsection{The Fock-space many-body skin effect}\label{Sec:Imbalance}
\begin{figure}
        \begin{tabular}{p{\linewidth}c}
        \centering
            \includegraphics[width=0.245\textwidth,height=0.2\textwidth]{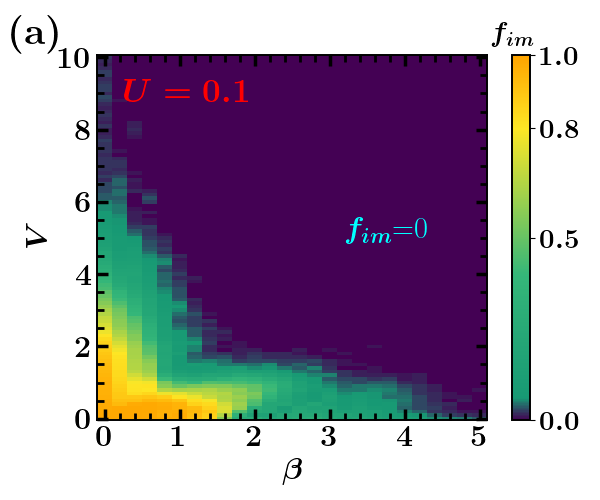}\hspace{-0.25cm}
            \includegraphics[width=0.245\textwidth,height=0.2\textwidth]{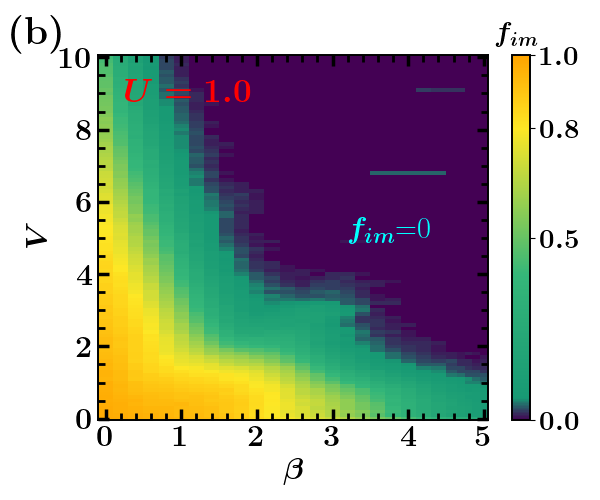}
        \caption{$f_{im}$ under OBC in the $V-\beta$ phase diagrams at the two distinct interaction strengths: (a) $U=0.1$ and (b) $U=1.0$ corresponding to Figs.~\ref{Fig:Fig_9}(a) and (b) respectively.}
        \label{Fig:Fig_11}
        \end{tabular}
        \end{figure}

        In the non-interacting systems, the topology has been associated to the skin effect in the past. In particular, it has been illustrated that skin modes under OBC appear in the topologically non-trivial ergodic regime under PBC. However, as already elaborately discussed in Secs.~\ref{Sec:V_beta}, \ref{Sec:U_V}, and \ref{Sec:Topology}, in systems with power-law hopping and interaction, topologically trivial intermediate states  emerge with completely real eigenenergies. On the other hand, in the ergodic phase and a part of the intermediate phase, the topology remains non-trivial. In this context, we investigate the appearance of skin modes by using imbalance of the occupation densities in the real-space as illustrated in Sec.~\ref{Sec:SE_OBC}. In Figs.~\ref{Fig:Fig_9}(a) and \ref{Fig:Fig_9}(b), we mark the MBSE phase in red and separate the phase where the skin states do not appear in blue. This phase will be called as the many-body no SE (MBnSE) phase from here on. As evident from these figures, the entire ergodic regime (as shown in Figs.~\ref{Fig:Fig_1}(a) and \ref{Fig:Fig_1}(b)) does not possess one-to-one correspondence with the MBSE regime as one would naively expect. In addition, we also verify this by varying the strength of the interaction at both short and long-range hopping limits in Figs.~\ref{Fig:Fig_9}(c) and \ref{Fig:Fig_9}(d) respectively. Interestingly, while this conclusion is valid in the long-range hopping limit, i.e., ergodic regime under PBC does not always correspond to MBSE phase under OBC, the correspondence between the ergodicity and skin states is re-established in the short-range hopping limit, as expected. Therefore, we conclude that the MBnSE ergodic phase is due to the combined effect of both nearest-neighbor interaction and the long-range power-law hopping in systems with asymmetric hopping.\\
        \indent Furthermore, in Fig.~\ref{Fig:Fig_10}, we verify the MBSE in the Fock space, to corroborate our results with the density imbalance. The parameters $U$, $V$, and $\beta$ are chosen such that we can clearly distinguish the behavior of the many-body eigenstates under OBC in the ergodic (MBnSE regime under OBC), MBSE, and NHMBL regimes. We have set $U=0.1$ (corresponding to Fig.~\ref{Fig:Fig_9}(a)) to demonstrate the nature in distinct regimes. However, we have also checked it for the other three phase diagrams (Figs.~\ref{Fig:Fig_9}(b)-(d)), but have not shown it explicitly. In Fig.~\ref{Fig:Fig_10}(a), we have selected the parameters inside the blue regime of Fig.~\ref{Fig:Fig_9}(a), that lies within the ergodic phase in Fig.~\ref{Fig:Fig_1}(a). The probability distribution clearly shows a MBnSE phase with complex eigenenergies as shown in the inset. Next, we proceed to demonstrate the MBSE behavior in Figs.~\ref{Fig:Fig_10}(b) and (c) picked up from red regime of Fig.~\ref{Fig:Fig_9}(a), which clearly indicates the tendency of the eigenstates in the Fock space towards the right. Interestingly, the energies in the MBSE can be both complex or real as illustrated. In Fig.~\ref{Fig:Fig_10}(d), we pick the eigenstates to be in the intermediate regime under PBC, which clearly shows the absence of MBSE as expected. We then demonstrate the NHMBL states which do not change its nature under a change in the boundary condition as depicted in Fig.~\ref{Fig:Fig_10}(e). Finally, to conclude the nature of eigenenergies of the MBSE phase under the OBC, we plot $f_{im}$ at $U=0.1$ and $U=1.0$ in the $V-\beta$ parameter space. On closely comparing Figs.~\ref{Fig:Fig_9}(a) and (b) with Figs.~\ref{Fig:Fig_11}(a) and (b), it is clearly evident that there exists skin states with both complex and real eigenenergies under the OBC, as reported earlier in systems without interaction \cite{Chakrabarty}. At last, we have compared the degree of localization of the MBSE states as a function of the power-law exponent $\beta$ in Appendix \ref{App:SE_beta}.
    
        \section{Conclusions}\label{Sec:Conclusions}
        In conclusions, the present work demonstrates that the interplay between power-law hopping and nearest-neighbor interaction in a non-Hermitian quasiperiodic system gives rise to rich and unconventional phases, markedly distinct from their non-interacting counterparts. Initially, we analyze how the single-particle $P_s$ regimes are altered by interactions in both short and long-range hopping limits. Notably, intermediate states that previously exhibited ergodic–multifractal (localized) edges without interaction evolve into an intermediate regime without a clear edge on both sides of $\beta = 1$. Surprisingly, the presence of real eigenenergies in the many-body spectrum does not necessarily imply NHMBL; intermediate phases can also host real spectra without being localized. This also means that the TRS of the Hamiltonian is restored much before it encounters the NHMBL regime. Furthermore, while we find a significant increase in the ergodic regime when the hopping is long-ranged, the ergodicity disappears at a strong interaction in case of short-range hopping, giving way to NHMBL phase at a strong quasiperiodic potential. In addition, we characterize the ergodic, intermediate, and NHMBL regimes through entanglement entropy, which reveals a non-monotonic behavior in the intermediate phase with complex eigenspectrum. On the other hand, when the intermediate states possess real eigenvalues, the entanglement entropy grows faster than the logarithmic evolution (that is typical of the NHMBL phase), but slower than the ballistic one in the ergodic phase, and the non-monotonic evolution over the time scale considered in our work vanishes. As expected, such an intermediate phase is also topologically trivial. Furthermore, we uncover that the topologically non-trivial ergodic regime under PBC does not always exhibit the MBSE under OBC, in the long-range hopping limit.  Besides, the resulting many-body skin states under OBC can bear either real or complex eigenenergies. This challenges our understanding of the skin-effect in systems where inter-particle interactions are present. Overall, we believe that our results deepen the understanding on how the interplay of hopping range and interaction reshapes the physics of non-Hermitian systems and call for a re-examination of the established expectations in such settings.
        
        \section*{Acknowledgments}\label{Sec:Acknowledgments}
        A.C. thanks the Council of Scientific $\&$ Industrial Research (CSIR)-HRDG, India, for external funding during the tenure via File No. 09/983(0047)/2020-EMR-I. A.C. and S.D. acknowledge the High Performance Computing (HPC) facility provided by SERB (DST), India, through Grant No. EMR/2015/001227, which was used for data computation in this work.
        All ED studies were performed with the QuSpin package \cite{Weinberg_Quspin_II}.

        \section*{Data availability}\label{Sec:Data}
        The data that support the above findings are not publicly available, but will be made available by the authors upon reasonable request.
    
	\appendix

        \section*{Appendices}
        
	\section{The spinless fermionic model without interaction}\label{App:Non-interacting model}
        To understand the effect of interaction in a non-Hermitian quasiperiodic Hamiltonian with power-law hopping, we first recall the various phases and attributes in the system without interaction. The non-interacting version of the Hamiltonian considered in our work is written as,
        \begin{eqnarray}
	       \mathcal{H}=  \displaystyle\sum_{n,n', n' > n} \Big(\frac{J_R}{|n'-n|^\beta} c^\dag_{n'} c_{n} + \frac{J_L}{|n'-n|^\beta}c^\dag_{n} c_{n'} \Big)~~~~~~ \nonumber \\
           + \sum_{n} V \text{cos} (2\pi\alpha n+\phi) c^\dag_{n} c_{n},~~~~~~~~~~~~~~~
	\label{Eq:Hamiltonian_no_interaction}
	\end{eqnarray}
        where the individual terms have been defined in Sec.~\ref{Sec:Model} of the main text. In the subsequent discussions, we investigate the different phases that we will be interested to look into throughout this work in the presence of interaction. It is important to note that the Hilbert space dimension in this case will be the same as number of lattices sites, i.e., $L$. We have considered a lattice with 987 sites for this purpose.
        
        \subsection{Effect of power-law hopping in the non-Hermitian system}\label{App:PL hopping}
\begin{figure}
		\begin{tabular}{p{\linewidth}c}
			\centering
			\includegraphics[width=0.249\textwidth,height=0.215\textwidth]{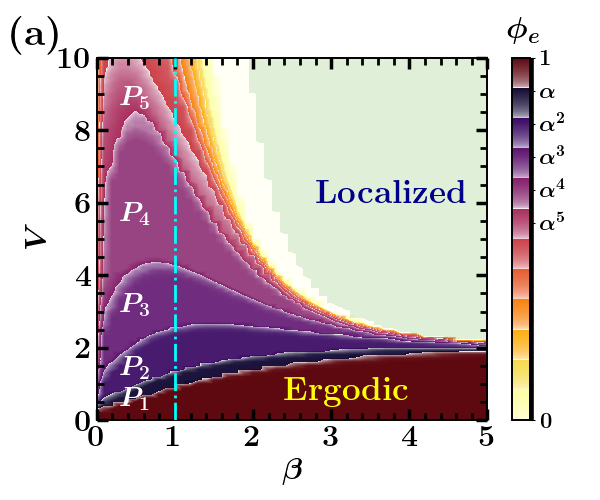}\hspace{-0.4cm}
           \includegraphics[width=0.249\textwidth,height=0.215\textwidth]{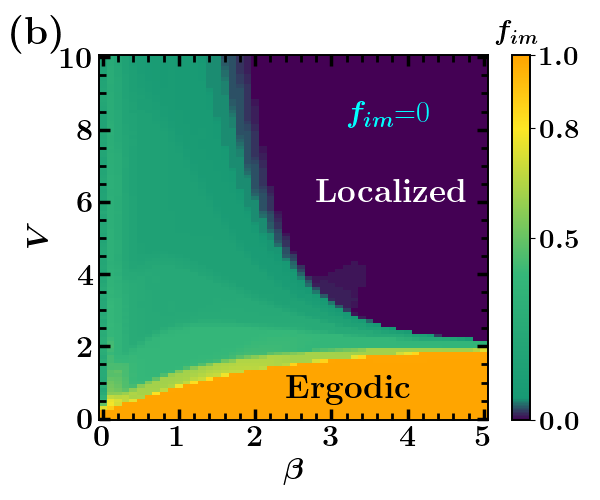}
		\caption{(a) The $V-\beta$ phase diagram illustrating the nature of $\phi_e$ in the non-interacting long-range chain with asymmetric hopping. The completely ergodic, single-particle localized and the $P_s$ regimes are indicated in different colors. b) $f_{im}$ for the identical parameters as in (a).}
		\label{Fig:Fig_A1}
		\end{tabular}
	\end{figure}

        In this section, we look into the effect of the power-law exponent $\beta$ on the phases in the non-Hermitian quasiperiodic system as described in Eq.~\ref{Eq:Hamiltonian_no_interaction}. In Fig.~\ref{Fig:Fig_A1}(a), we estimate $\phi_e$ as given in Sec.~\ref{Sec:Results}, and find that there exists a completely ergodic and a single-particle localized regime (in brown and light green colors respectively). We demonstrate the absence of localization for $\beta<2$. In addition, we find the emergence of the $P_s$ regimes, where each of the $P_s$ regime contains $\alpha^s$ ergodic eigenstates separated by a multifractal/localized phase at the normalized eigenstate index $n/L=\alpha^s$ (Please refer to the figures in Ref.~\cite{Xianlong} for more details). In particular, the $P_s$ regimes consist of ergodic-multifractal edges when $\beta<1$ and ergodic-localized edges when $\beta>1$ (not shown here for brevity). In addition, in Fig.~\ref{Fig:Fig_A1}(b), we verify that the localized regimes possess completely real eigenenergies, whereas the ergodic phase consists of a fully complex spectrum. $f_{im}$ in the $P_s$ regimes lies in between 0 and 1 due to the mixed nature of eigenenergies as expected.
        
        \subsection{SE in the non-Hermitian chain with power-law hopping}\label{App:SE}
\begin{figure}
		\begin{tabular}{p{\linewidth}c}
			\centering
			\includegraphics[width=0.245\textwidth,height=0.2\textwidth]{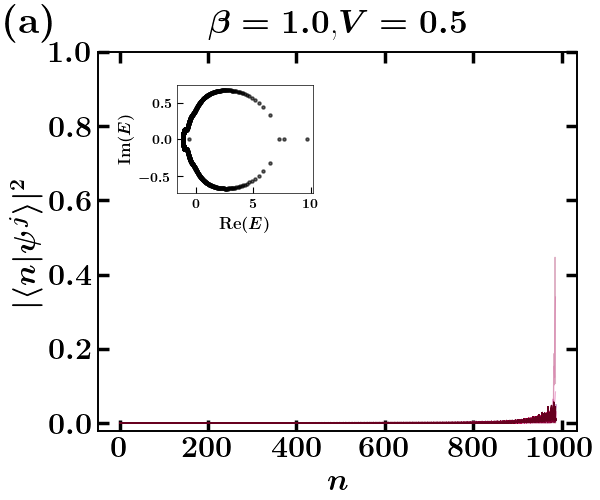}\hspace{-0.25cm}
           \includegraphics[width=0.245\textwidth,height=0.2\textwidth]{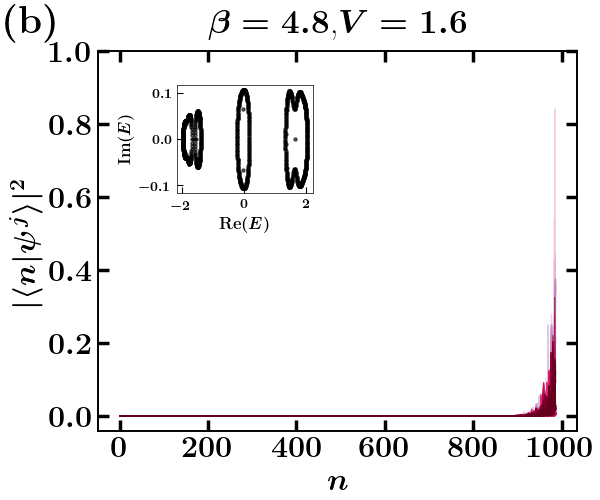}\\
           \includegraphics[width=0.245\textwidth,height=0.2\textwidth]{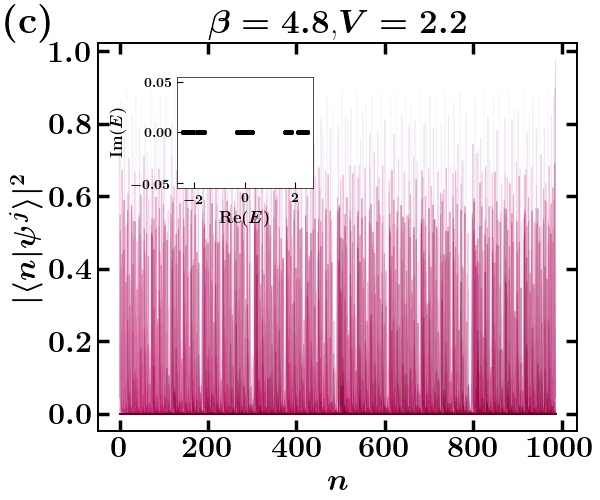}\hspace{-0.25cm}
           \includegraphics[width=0.245\textwidth,height=0.2\textwidth]{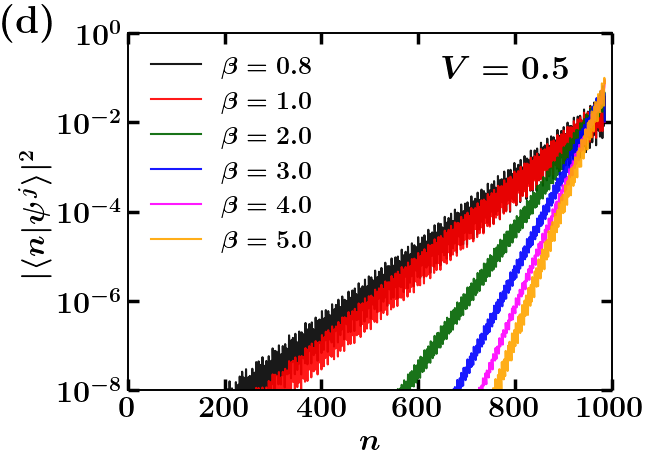}
		\caption{Demonstration of the existence or absence of the single particle SE. The parameters are chosen from Fig.~\ref{Fig:Fig_A1}, i.e., (a) $\beta=1.0,V=0.5$ (ergodic under PBC) , (b) $\beta=4.8,V=1.6$ (ergodic under PBC), (c) $\beta=4.8,V=2.2$ (localized under PBC). (d) Manifestation of the degree of localization of the $800$th single-particle eigenstate with varying $\beta$ at $V=0.5$. All these figures have been presented under OBC.}
		\label{Fig:Fig_A2}
		\end{tabular}
	\end{figure}

        From the past literature, it is well known that the ergodic phase under PBC gives rise to a skin phase under the OBC. To verify this, we consider three distinct combinations of $\beta$ and $V$, and plot the probability distribution of all the eigenstates under OBC. From Figs.~\ref{Fig:Fig_A2}(a) and ~\ref{Fig:Fig_A2}(b), it is evident that all the ergodic states turn into skin modes with maximum weightage towards the right, since $J_R>J_L$. On the other hand, the localized states are insensitive to the change in boundary condition as clear from Fig.~\ref{Fig:Fig_A2}(c). Furthermore, as evident from Fig.~\ref{Fig:Fig_A2}(d), the degree of exponential localization of these skin states (the decay profile of only one, i.e., the 800th eigenstate, has been demonstrated for clarity) towards the right end of the lattice decreases with an increase in the range of hopping (decrease in $\beta$), similar to that reported in Ref.~\cite{Xianlong}.
        
        \begin{figure}
        \begin{tabular}{p{\linewidth}c}
        \centering
            \includegraphics[width=0.245\textwidth,height=0.2\textwidth]{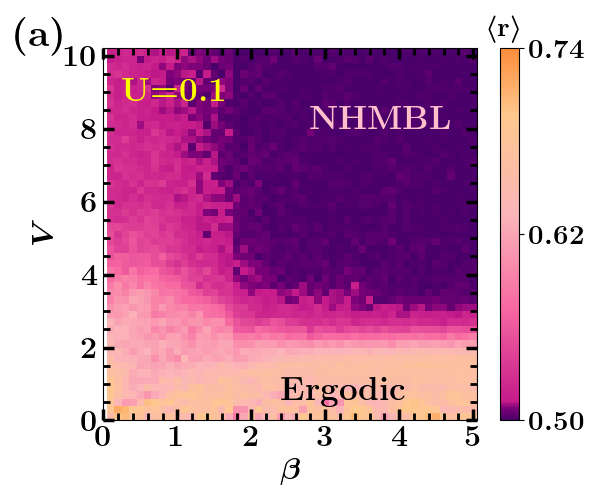}\hspace{-0.25cm}
            \includegraphics[width=0.245\textwidth,height=0.2\textwidth]{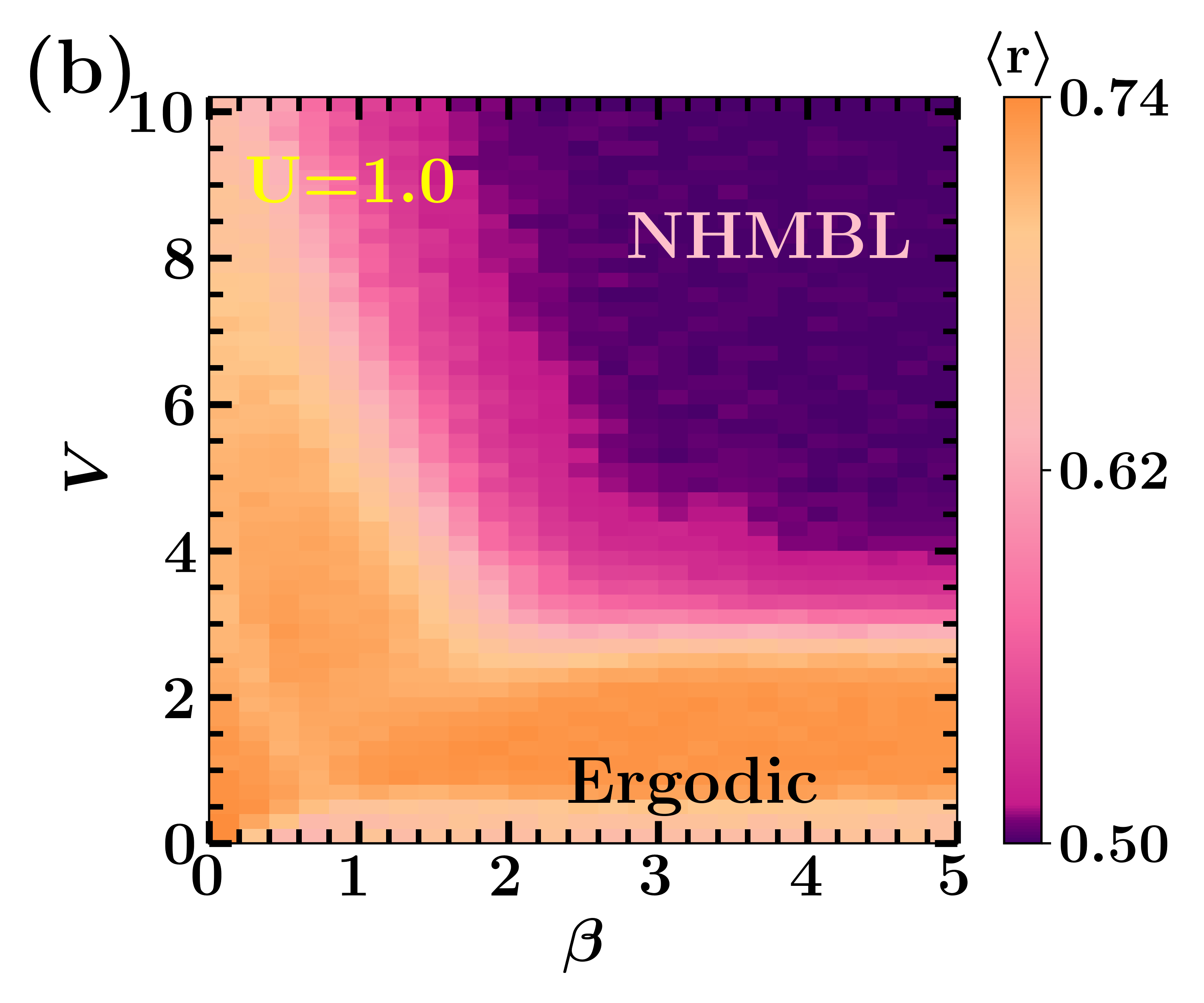}
        \caption{Demonstration of $\left\langle r \right \rangle$ in the non-Hermitian Hamiltonian considered in our work using Eq.~\ref{Eq:r_value}. In (a), we have considered $U=0.1$, whereas in (b) we have considered $U=1.0$, corresponding to Figs.~\ref{Fig:Fig_1}(a) and (b) in the main text respectively.}
        \label{Fig:Fig_A3}
        \end{tabular}
        \end{figure}
        
        \section{Verification of the distinct phases using $\left \langle r \right \rangle$}\label{App:r_avg}
        In the complex plane, the level spacing distribution is defined using a ratio of the complex distances given as \cite{Lucas},\\
        \begin{eqnarray}
         \Xi_n = \frac{z_n^{NN}-z_n}{z_n^{NNN}-z_n} = r_n e^{i\theta_n}.
         \label{Eq:r_value}
        \end{eqnarray}
        In Eq.~\ref{Eq:r_value}, $z_n^{NN}$ and  $z_n^{NNN}$ are the energies of the nearest and next-nearest neighbors respectively of the complex energy $z_n$ where the distances are estimated in $\mathbb{C}$. $r_n$ and $\theta_n$ give the magnitude and the direction of such a ratio respectively. This measure is reliable in non-Hermitian systems and is not sensitive to the unfolding procedure. To find out this ratio, we consider an energy window spanning $20\%$ of the entire energy range (i.e., we consider $\pm10\%$ span of the energy spectrum from the middle) and take realizations in $\phi$ such as the total number of eigenvalues after considering realizations $\sim 10^5$. $\left \langle r \right \rangle$ can characterize distinct phases and have different values following the Ginibre/Poisson ensembles according to the random matrix theory. While the chaotic systems in the thermalizing regime follows a Ginibre Orthogonal Ensemble (GinOE), the level statistics in the integrable (NHMBL) phase is Poisson in nature. Specifically, $\left \langle r \right \rangle \sim 0.738 (0.50)$ when the states follow the GinOE (1D Poisson) distribution.\\
        \indent 
        In Fig.~\ref{Fig:Fig_A3}(a)-(b), we have verified the nature of the eigenstates obtained in Figs.~\ref{Fig:Fig_1}(a) and (b) by estimating $\left \langle r \right \rangle$. It is evident that the ergodic states follow the GinOE, whereas in the NHMBL regime the states are in accordance to the 1D Poisson distribution as demonstrated in Ref.~\cite{Kulkarni}. We have additionally verified this for the other phase diagrams as well (which are not presented here for brevity), and establish that the level statistics reveal the same information regarding the nature of the many-body eigenstates.

          \begin{figure}
        \begin{tabular}{p{\linewidth}c}
        \centering
            \includegraphics[width=0.23\textwidth,height=0.22\textwidth]{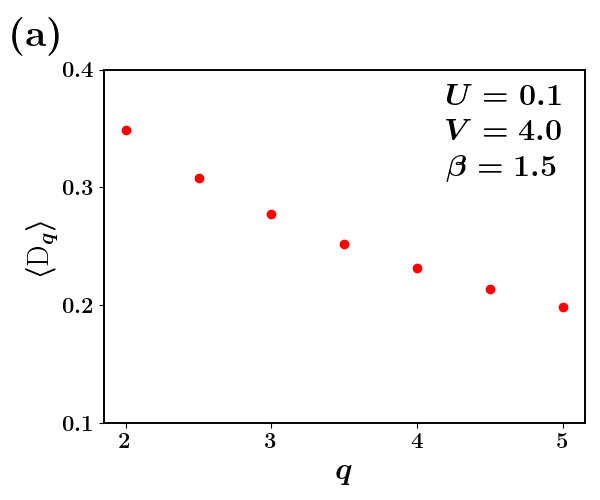}
            \includegraphics[width=0.23\textwidth,height=0.22\textwidth]{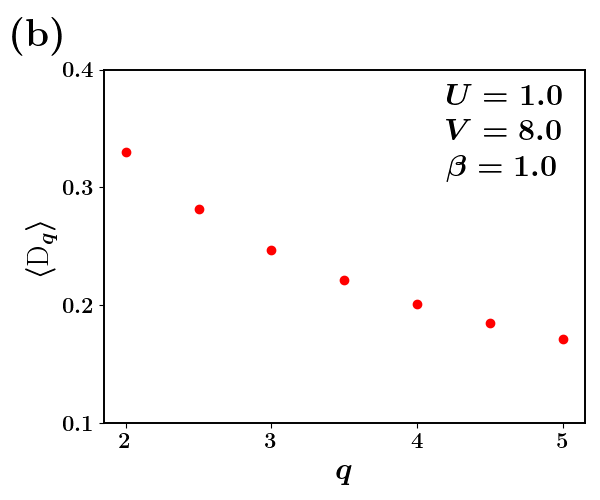}\\
            \includegraphics[width=0.23\textwidth,height=0.22\textwidth]{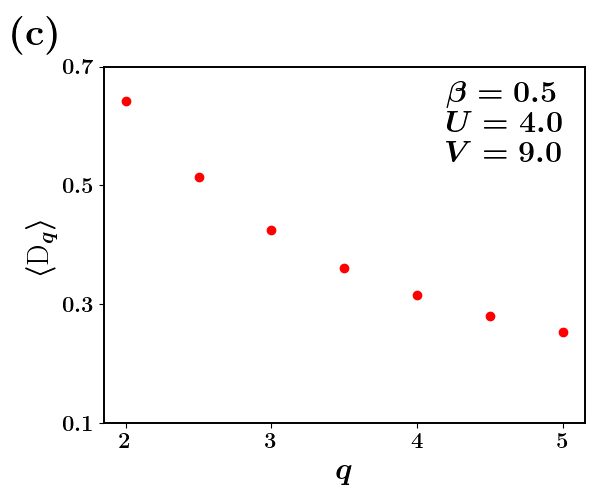}
            \includegraphics[width=0.23\textwidth,height=0.22\textwidth]{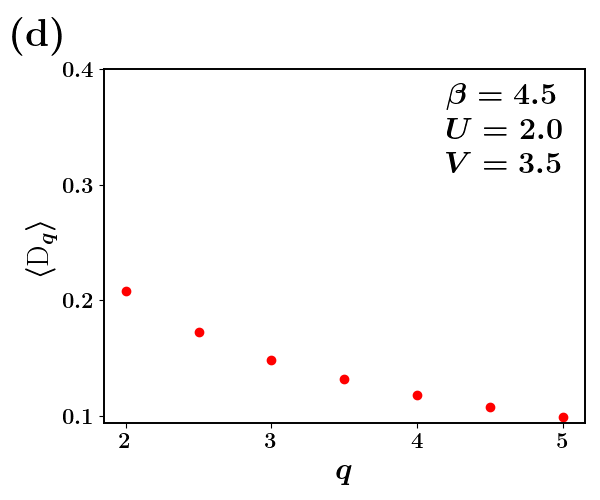}
        \caption{Illustration of $\left \langle \text{D}_q\right \rangle$ as a function of $q$ for: (a) $U=0.1, V=4.0, \beta=1.5$, (b) $U=1.0, V=8.0, \beta=1.0$, (c) $\beta=0.5, U=4.0, V=9.0$, and (d) $\beta=4.5, U=2.0, V=3.5$, which correspond to the intermediate regimes as shown in Figs.~\ref{Fig:Fig_1}(c)-(d) and Fig.~\ref{Fig:Fig_5}(c)-(d) in the manuscript.}
        \label{Fig:Fig_A4}
        \end{tabular}
        \end{figure}
        
        \section{Characterization of the fractal dimension $D_q$}\label{App:D_q}
        The fractal dimension $D_2$ alone as given in Eq.~\ref{Eq:D_2} cannot ascertain the multifractal nature of an eigenstate, although it is a good measure to identify that the states might be multifractal in nature. In this regard, the generalized IPR for different moments $q$ has been defined in Hermitian systems \cite{Aoki,Mirlin} and can be extended to non-Hermitian systems as,
        \begin{eqnarray}
            BIPR_j^{(q)} = \frac{\displaystyle\sum_{m=1}^{\mathcal{D}} |\psi_{mL}^{j}\psi_{mR}^{j}|^{q}}{\Big(\displaystyle\sum_{m=1}^{\mathcal{D}} |{\psi_{mL}^j\psi_{mR}^{j}}|\Big)^{q}},
            \label{Eq:Gen_BIPR}
        \end{eqnarray}
        where $BIPR_j=BIPR_j^{(2)}$. It is known that $BIPR_j^{(q)}\sim \mathcal{D}^{-\tau_q}$, where $\tau_q$ is the mass dimension and the fractal dimension for $q$th moment is given as $D_q=\tau_q/(q-1)$. $D_q=1 (0)$ for the ergodic (localized) states respectively, whereas if the states are multifractal in nature, they are characterized by $0<D_q<1$, for all $q$. We cross-check our interpretation about the multifractal behavior presented in Figs.~\ref{Fig:Fig_1}(c)-(d) and Fig.~\ref{Fig:Fig_5}(c)-(d) using the obtained values of $\left \langle \text{D}_q\right \rangle$, averaged over all eigenstates, for varying $q$ in Figs.~\ref{Fig:Fig_A4}(a)-(d), which suggests that the states in the intermediate regime depicted using the second moment $D_2$ are indeed mostly multifractal in nature.

        \begin{figure}
		\begin{tabular}{p{\linewidth}c}
			\centering
		\includegraphics[width=0.22\textwidth,height=0.2\textwidth]{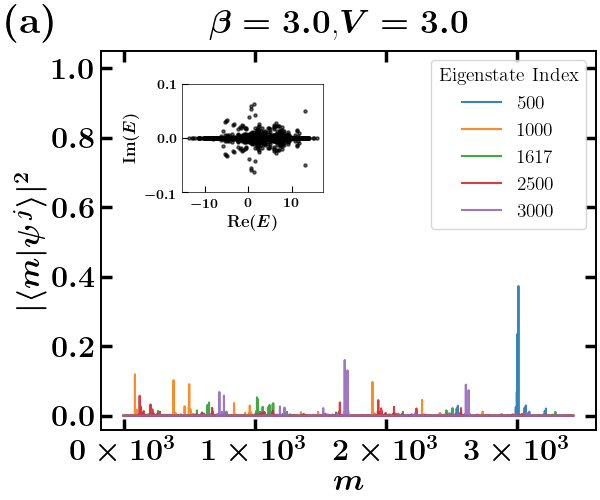}           \includegraphics[width=0.22\textwidth,height=0.2\textwidth]{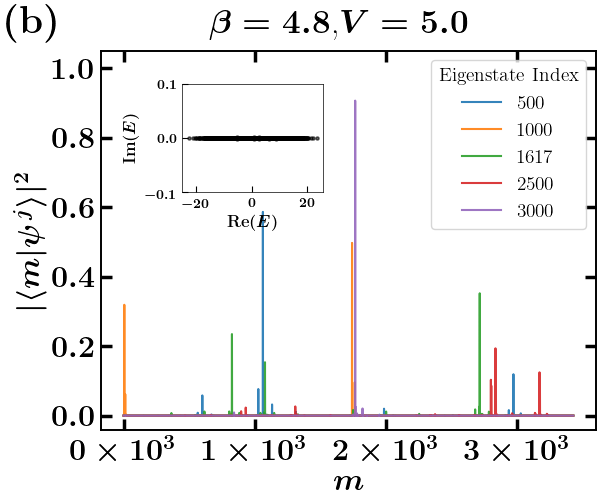}\\
		\caption{The Fock-space wave-function probabilities of the selected eigenstates corresponding to Fig.~\ref{Fig:Fig_2}(b) at $U=1.0$. The other parameters are chosen to select the parts of intermediate regime, i.e., (a) $\beta=3.0,V=3.0$ and (b) $\beta=4.8,V=5.0$.}
		\label{Fig:Fig_A5}
		\end{tabular}
	\end{figure}
        
        \section{Many-body spectrum and the Fock-space wavefunction probability distribution in the intermediate phase at $U=1.0$}\label{App:WF_analysis_U1.0}
        To validate one of the main findings in this work, which is the appearance of intermediate states with both complex and fully real eigenenergies, in this section, we demonstrate the wavefunction probabilities of five Fock-space eigenstates in Fig.~\ref{Fig:Fig_A5} at $U=1.0$, alike Fig.~\ref{Fig:Fig_4} demonstrated at $U=0.1$. We clearly verify the appearance of the intermediate states with small weightage at many nearby Fock-space configurations that can exist with both complex or completely real eigenenergies as illustrated in Fig.~\ref{Fig:Fig_A5}(a) and Fig.~\ref{Fig:Fig_A5}(b) respectively.

        \section{System-size dependence of the entanglement entropy in the observed phases}\label{App:EE_system_sizes}
\begin{figure}
		\begin{tabular}{p{\linewidth}c}
			\centering
	          \includegraphics[width=0.22\textwidth,height=0.2\textwidth]{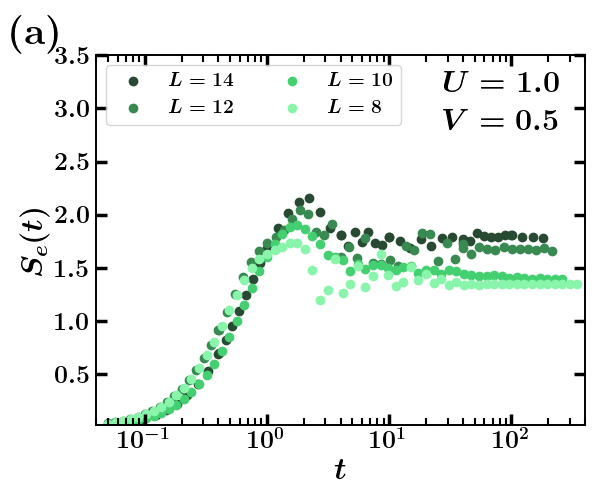}
                \includegraphics[width=0.22\textwidth,height=0.2\textwidth]{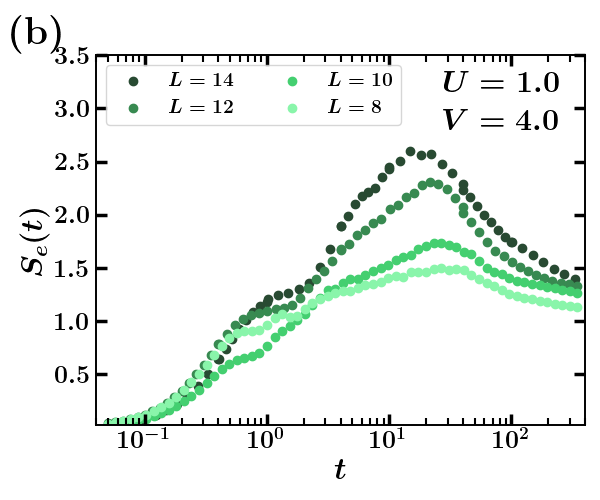}\\
                \includegraphics[width=0.22\textwidth,height=0.2\textwidth]{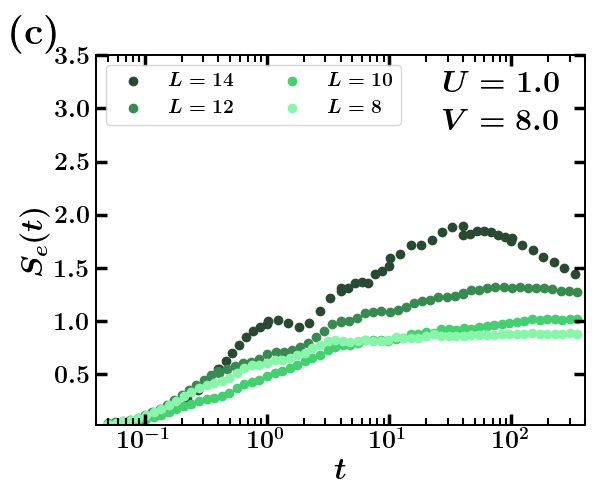}
                \includegraphics[width=0.22\textwidth,height=0.2\textwidth]{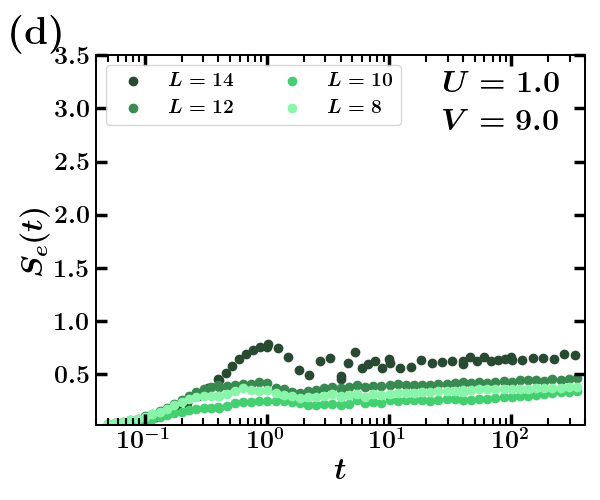}
		\caption{$S_e(t)$ vs. $t$ at different system sizes ($L=8,10,12,14$) and at different regimes given by the parameters: (a) $U=1.0,V=0.5,\beta=1.5$ (ergodic regime), (b) $U=1.0,V=4.0,\beta=1.5$ (intermediate regime), (c) $U=1.0,V=8.0,\beta=1.5$ (intermediate regime) and (d) $U=1.0,V=9.0,\beta=4.5$ (NHMBL regime) respectively.}
		\label{Fig:Fig_A6}
		\end{tabular}
	\end{figure}

         A reliable tracking of the ergodic-MBL transition has also been achieved using the scaling of the entanglement entropy by varying the system sizes. The localized states are restricted by the `area-law' scaling, i.e., $S_e\sim l^{d-1}$, with $l$ is the size of the subsystem and $d$ is the spatial dimension. On the other hand, the ergodic eigenstates exhibit $S_e\sim l^{d}$ behavior and follow the `volume law' scaling \cite{Nayak,Huse_2013}.
         Therefore, in this section, we try to understand the behavior of the entanglement entropy as a function of different system sizes as depicted in Fig.~\ref{Fig:Fig_A6}. To unfold the nature, we select the parameters as given in Figs.~\ref{Fig:Fig_7}(a)-(c) at $\beta=1.5$ and vary the number of lattice sites from $L=8$ to $L=14$ (a lighter to darker shade in green indicates an increase in $L$). In Fig.~\ref{Fig:Fig_A6}(a), we demonstrate the systematic scaling behavior in the ergodic regime where the saturation value of entanglement entropy has an increasing trend with the system-size and saturates to a large value. In the intermediate regimes as shown in Figs.~\ref{Fig:Fig_A6}(b)-(c), we find nearly similar scaling along with the non-monotonic behavior with time as already explained in the main text, for all the lattice sizes considered. 
         Finally, in the localized regime as demonstrated in Fig.~\ref{Fig:Fig_A6}(d), one finds a fairly low value in the saturation entanglement entropy which is a characteristic of NHMBL, along with the nearly `area-law' behavior (the slight change in saturation value is expected to be due to the finitie-size effects). Our present computational facilities did not permit us to have numerical estimates of the entanglement entropy beyond $L=14$ and 200 realizations within a reasonable time. Please note that the aim of this section is not to demonstrate the volume or area law behavior, but to validate that our result is not a finite-size effect and is not particular to $L=14$ as considered in our analyses. 

        \begin{figure}
		\begin{tabular}{p{\linewidth}c}
			\centering
	          \includegraphics[width=0.22\textwidth,height=0.2\textwidth]{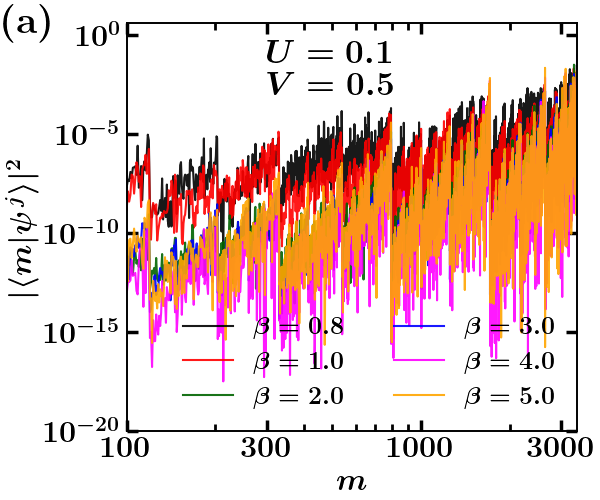}
                \includegraphics[width=0.22\textwidth,height=0.2\textwidth]{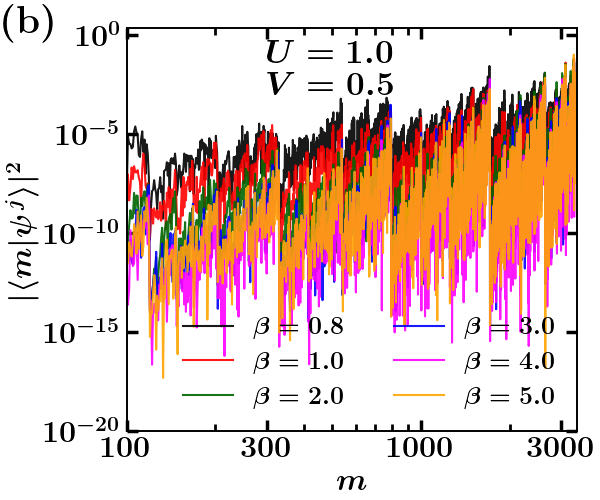}\\
		\caption{Degree of localization of the $2500$th eigenstate at $V=0.5$ and interaction strength: (a) $U=0.1$ and (b) $U=1.0$ with varying values of $\beta$ under the OBC.}
		\label{Fig:Fig_A7}
		\end{tabular}
	\end{figure}
         
        \section{Dependence of degree of MBSE on $\beta$}\label{App:SE_beta}
        Similar to Fig.~\ref{Fig:Fig_A2}(d), in this section, we try to understand the degree of localization of the skin states with varying hopping ranges in the presence of interaction. In contrary to the non-interacting counterpart, it is clear from Figs.~\ref{Fig:Fig_A7}(a) and (b) that although a discernible decay in the wave-function appears with an increase in $\beta$, at higher values of $\beta$ the decay becomes visibly non-systematic. One can therefore roughly think that in the presence of interaction, the decay profile of the skin states with a very long-range hopping is quite slow as compared to the profile with almost nearest-neighbor hopping. In addition, the localization is clearly much slower and not exponential as observed in the systems without inter-particle interaction.
  
	\bibliography{reference.bib}

\begin{thebibliography}{85}%
\makeatletter
\providecommand \@ifxundefined [1]{%
 \@ifx{#1\undefined}
}%
\providecommand \@ifnum [1]{%
 \ifnum #1\expandafter \@firstoftwo
 \else \expandafter \@secondoftwo
 \fi
}%
\providecommand \@ifx [1]{%
 \ifx #1\expandafter \@firstoftwo
 \else \expandafter \@secondoftwo
 \fi
}%
\providecommand \natexlab [1]{#1}%
\providecommand \enquote  [1]{``#1''}%
\providecommand \bibnamefont  [1]{#1}%
\providecommand \bibfnamefont [1]{#1}%
\providecommand \citenamefont [1]{#1}%
\providecommand \href@noop [0]{\@secondoftwo}%
\providecommand \href [0]{\begingroup \@sanitize@url \@href}%
\providecommand \@href[1]{\@@startlink{#1}\@@href}%
\providecommand \@@href[1]{\endgroup#1\@@endlink}%
\providecommand \@sanitize@url [0]{\catcode `\\12\catcode `\$12\catcode
  `\&12\catcode `\#12\catcode `\^12\catcode `\_12\catcode `\%12\relax}%
\providecommand \@@startlink[1]{}%
\providecommand \@@endlink[0]{}%
\providecommand \url  [0]{\begingroup\@sanitize@url \@url }%
\providecommand \@url [1]{\endgroup\@href {#1}{\urlprefix }}%
\providecommand \urlprefix  [0]{URL }%
\providecommand \Eprint [0]{\href }%
\providecommand \doibase [0]{https://doi.org/}%
\providecommand \selectlanguage [0]{\@gobble}%
\providecommand \bibinfo  [0]{\@secondoftwo}%
\providecommand \bibfield  [0]{\@secondoftwo}%
\providecommand \translation [1]{[#1]}%
\providecommand \BibitemOpen [0]{}%
\providecommand \bibitemStop [0]{}%
\providecommand \bibitemNoStop [0]{.\EOS\space}%
\providecommand \EOS [0]{\spacefactor3000\relax}%
\providecommand \BibitemShut  [1]{\csname bibitem#1\endcsname}%
\let\auto@bib@innerbib\@empty
\bibitem [{\citenamefont {Anderson}(1958)}]{Anderson}%
  \BibitemOpen
  \bibfield  {author} {\bibinfo {author} {\bibfnamefont {P.~W.}\ \bibnamefont
  {Anderson}},\ }\bibfield  {title} {\bibinfo {title} {Absence of diffusion in
  certain random lattices},\ }\href {https://doi.org/10.1103/PhysRev.109.1492}
  {\bibfield  {journal} {\bibinfo  {journal} {Phys. Rev.}\ }\textbf {\bibinfo
  {volume} {109}},\ \bibinfo {pages} {1492} (\bibinfo {year}
  {1958})}\BibitemShut {NoStop}%
\bibitem [{\citenamefont {Gornyi}\ \emph {et~al.}(2005)\citenamefont {Gornyi},
  \citenamefont {Mirlin},\ and\ \citenamefont {Polyakov}}]{Gornyi}%
  \BibitemOpen
  \bibfield  {author} {\bibinfo {author} {\bibfnamefont {I.~V.}\ \bibnamefont
  {Gornyi}}, \bibinfo {author} {\bibfnamefont {A.~D.}\ \bibnamefont {Mirlin}},\
  and\ \bibinfo {author} {\bibfnamefont {D.~G.}\ \bibnamefont {Polyakov}},\
  }\bibfield  {title} {\bibinfo {title} {Interacting electrons in disordered
  wires: {Anderson} localization and low-{T} transport},\ }\href
  {https://doi.org/10.1103/PhysRevLett.95.206603} {\bibfield  {journal}
  {\bibinfo  {journal} {Phys. Rev. Lett.}\ }\textbf {\bibinfo {volume} {95}},\
  \bibinfo {pages} {206603} (\bibinfo {year} {2005})}\BibitemShut {NoStop}%
\bibitem [{\citenamefont {Basko}\ \emph {et~al.}(2006)\citenamefont {Basko},
  \citenamefont {Aleiner},\ and\ \citenamefont {Altshuler}}]{Basko}%
  \BibitemOpen
  \bibfield  {author} {\bibinfo {author} {\bibfnamefont {D.}~\bibnamefont
  {Basko}}, \bibinfo {author} {\bibfnamefont {I.}~\bibnamefont {Aleiner}},\
  and\ \bibinfo {author} {\bibfnamefont {B.}~\bibnamefont {Altshuler}},\
  }\bibfield  {title} {\bibinfo {title} {Metal–insulator transition in a
  weakly interacting many-electron system with localized single-particle
  states},\ }\href {https://doi.org/https://doi.org/10.1016/j.aop.2005.11.014}
  {\bibfield  {journal} {\bibinfo  {journal} {Annals of Physics}\ }\textbf
  {\bibinfo {volume} {321}},\ \bibinfo {pages} {1126} (\bibinfo {year}
  {2006})}\BibitemShut {NoStop}%
\bibitem [{\citenamefont {Nandkishore}\ and\ \citenamefont
  {Huse}(2015)}]{Nandkishore}%
  \BibitemOpen
  \bibfield  {author} {\bibinfo {author} {\bibfnamefont {R.}~\bibnamefont
  {Nandkishore}}\ and\ \bibinfo {author} {\bibfnamefont {D.~A.}\ \bibnamefont
  {Huse}},\ }\bibfield  {title} {\bibinfo {title} {Many-body localization and
  thermalization in quantum statistical mechanics},\ }\href
  {https://doi.org/https://doi.org/10.1146/annurev-conmatphys-031214-014726}
  {\bibfield  {journal} {\bibinfo  {journal} {Annual Review of Condensed Matter
  Physics}\ }\textbf {\bibinfo {volume} {6}},\ \bibinfo {pages} {15} (\bibinfo
  {year} {2015})}\BibitemShut {NoStop}%
\bibitem [{\citenamefont {Abanin}\ and\ \citenamefont {Papić}(2017)}]{Zlatko}%
  \BibitemOpen
  \bibfield  {author} {\bibinfo {author} {\bibfnamefont {D.~A.}\ \bibnamefont
  {Abanin}}\ and\ \bibinfo {author} {\bibfnamefont {Z.}~\bibnamefont
  {Papić}},\ }\bibfield  {title} {\bibinfo {title} {Recent progress in
  many-body localization},\ }\href
  {https://doi.org/https://doi.org/10.1002/andp.201700169} {\bibfield
  {journal} {\bibinfo  {journal} {Annalen der Physik}\ }\textbf {\bibinfo
  {volume} {529}},\ \bibinfo {pages} {1700169} (\bibinfo {year}
  {2017})}\BibitemShut {NoStop}%
\bibitem [{\citenamefont {Abanin}\ \emph {et~al.}(2019)\citenamefont {Abanin},
  \citenamefont {Altman}, \citenamefont {Bloch},\ and\ \citenamefont
  {Serbyn}}]{Maksym}%
  \BibitemOpen
  \bibfield  {author} {\bibinfo {author} {\bibfnamefont {D.~A.}\ \bibnamefont
  {Abanin}}, \bibinfo {author} {\bibfnamefont {E.}~\bibnamefont {Altman}},
  \bibinfo {author} {\bibfnamefont {I.}~\bibnamefont {Bloch}},\ and\ \bibinfo
  {author} {\bibfnamefont {M.}~\bibnamefont {Serbyn}},\ }\bibfield  {title}
  {\bibinfo {title} {Colloquium: Many-body localization, thermalization, and
  entanglement},\ }\href {https://doi.org/10.1103/RevModPhys.91.021001}
  {\bibfield  {journal} {\bibinfo  {journal} {Rev. Mod. Phys.}\ }\textbf
  {\bibinfo {volume} {91}},\ \bibinfo {pages} {021001} (\bibinfo {year}
  {2019})}\BibitemShut {NoStop}%
\bibitem [{\citenamefont {Pal}\ and\ \citenamefont {Huse}(2010)}]{Pal}%
  \BibitemOpen
  \bibfield  {author} {\bibinfo {author} {\bibfnamefont {A.}~\bibnamefont
  {Pal}}\ and\ \bibinfo {author} {\bibfnamefont {D.~A.}\ \bibnamefont {Huse}},\
  }\bibfield  {title} {\bibinfo {title} {Many-body localization phase
  transition},\ }\href {https://doi.org/10.1103/PhysRevB.82.174411} {\bibfield
  {journal} {\bibinfo  {journal} {Phys. Rev. B}\ }\textbf {\bibinfo {volume}
  {82}},\ \bibinfo {pages} {174411} (\bibinfo {year} {2010})}\BibitemShut
  {NoStop}%
\bibitem [{\citenamefont {Kj\"all}\ \emph {et~al.}(2014)\citenamefont
  {Kj\"all}, \citenamefont {Bardarson},\ and\ \citenamefont
  {Pollmann}}]{Frank}%
  \BibitemOpen
  \bibfield  {author} {\bibinfo {author} {\bibfnamefont {J.~A.}\ \bibnamefont
  {Kj\"all}}, \bibinfo {author} {\bibfnamefont {J.~H.}\ \bibnamefont
  {Bardarson}},\ and\ \bibinfo {author} {\bibfnamefont {F.}~\bibnamefont
  {Pollmann}},\ }\bibfield  {title} {\bibinfo {title} {Many-body localization
  in a disordered quantum {I}sing chain},\ }\href
  {https://doi.org/10.1103/PhysRevLett.113.107204} {\bibfield  {journal}
  {\bibinfo  {journal} {Phys. Rev. Lett.}\ }\textbf {\bibinfo {volume} {113}},\
  \bibinfo {pages} {107204} (\bibinfo {year} {2014})}\BibitemShut {NoStop}%
\bibitem [{\citenamefont {Bera}\ \emph {et~al.}(2015)\citenamefont {Bera},
  \citenamefont {Schomerus}, \citenamefont {Heidrich-Meisner},\ and\
  \citenamefont {Bardarson}}]{Bardarson}%
  \BibitemOpen
  \bibfield  {author} {\bibinfo {author} {\bibfnamefont {S.}~\bibnamefont
  {Bera}}, \bibinfo {author} {\bibfnamefont {H.}~\bibnamefont {Schomerus}},
  \bibinfo {author} {\bibfnamefont {F.}~\bibnamefont {Heidrich-Meisner}},\ and\
  \bibinfo {author} {\bibfnamefont {J.~H.}\ \bibnamefont {Bardarson}},\
  }\bibfield  {title} {\bibinfo {title} {Many-body localization characterized
  from a one-particle perspective},\ }\href
  {https://doi.org/10.1103/PhysRevLett.115.046603} {\bibfield  {journal}
  {\bibinfo  {journal} {Phys. Rev. Lett.}\ }\textbf {\bibinfo {volume} {115}},\
  \bibinfo {pages} {046603} (\bibinfo {year} {2015})}\BibitemShut {NoStop}%
\bibitem [{\citenamefont {Sierant}\ \emph {et~al.}(2017)\citenamefont
  {Sierant}, \citenamefont {Delande},\ and\ \citenamefont
  {Zakrzewski}}]{Sierant}%
  \BibitemOpen
  \bibfield  {author} {\bibinfo {author} {\bibfnamefont {P.}~\bibnamefont
  {Sierant}}, \bibinfo {author} {\bibfnamefont {D.}~\bibnamefont {Delande}},\
  and\ \bibinfo {author} {\bibfnamefont {J.}~\bibnamefont {Zakrzewski}},\
  }\bibfield  {title} {\bibinfo {title} {Many-body localization for randomly
  interacting bosons},\ }\href {https://doi.org/10.12693/aphyspola.132.1707}
  {\bibfield  {journal} {\bibinfo  {journal} {Acta Physica Polonica A}\
  }\textbf {\bibinfo {volume} {132}},\ \bibinfo {pages} {1707–1712} (\bibinfo
  {year} {2017})}\BibitemShut {NoStop}%
\bibitem [{\citenamefont {Alet}\ and\ \citenamefont
  {Laflorencie}(2018)}]{Alet}%
  \BibitemOpen
  \bibfield  {author} {\bibinfo {author} {\bibfnamefont {F.}~\bibnamefont
  {Alet}}\ and\ \bibinfo {author} {\bibfnamefont {N.}~\bibnamefont
  {Laflorencie}},\ }\bibfield  {title} {\bibinfo {title} {Many-body
  localization: An introduction and selected topics},\ }\href
  {https://doi.org/10.1016/j.crhy.2018.03.003} {\bibfield  {journal} {\bibinfo
  {journal} {Comptes Rendus. Physique}\ }\textbf {\bibinfo {volume} {19}},\
  \bibinfo {pages} {498} (\bibinfo {year} {2018})}\BibitemShut {NoStop}%
\bibitem [{\citenamefont {Kondov}\ \emph {et~al.}(2015)\citenamefont {Kondov},
  \citenamefont {McGehee}, \citenamefont {Xu},\ and\ \citenamefont
  {DeMarco}}]{Kondov}%
  \BibitemOpen
  \bibfield  {author} {\bibinfo {author} {\bibfnamefont {S.~S.}\ \bibnamefont
  {Kondov}}, \bibinfo {author} {\bibfnamefont {W.~R.}\ \bibnamefont {McGehee}},
  \bibinfo {author} {\bibfnamefont {W.}~\bibnamefont {Xu}},\ and\ \bibinfo
  {author} {\bibfnamefont {B.}~\bibnamefont {DeMarco}},\ }\bibfield  {title}
  {\bibinfo {title} {Disorder-induced localization in a strongly correlated
  atomic hubbard gas},\ }\href {https://doi.org/10.1103/PhysRevLett.114.083002}
  {\bibfield  {journal} {\bibinfo  {journal} {Phys. Rev. Lett.}\ }\textbf
  {\bibinfo {volume} {114}},\ \bibinfo {pages} {083002} (\bibinfo {year}
  {2015})}\BibitemShut {NoStop}%
\bibitem [{\citenamefont {yoon Choi}\ \emph {et~al.}(2016)\citenamefont {yoon
  Choi}, \citenamefont {Hild}, \citenamefont {Zeiher}, \citenamefont {Schauß},
  \citenamefont {Rubio-Abadal}, \citenamefont {Yefsah}, \citenamefont
  {Khemani}, \citenamefont {Huse}, \citenamefont {Bloch},\ and\ \citenamefont
  {Gross}}]{Choi}%
  \BibitemOpen
  \bibfield  {author} {\bibinfo {author} {\bibfnamefont {J.}~\bibnamefont {yoon
  Choi}}, \bibinfo {author} {\bibfnamefont {S.}~\bibnamefont {Hild}}, \bibinfo
  {author} {\bibfnamefont {J.}~\bibnamefont {Zeiher}}, \bibinfo {author}
  {\bibfnamefont {P.}~\bibnamefont {Schauß}}, \bibinfo {author} {\bibfnamefont
  {A.}~\bibnamefont {Rubio-Abadal}}, \bibinfo {author} {\bibfnamefont
  {T.}~\bibnamefont {Yefsah}}, \bibinfo {author} {\bibfnamefont
  {V.}~\bibnamefont {Khemani}}, \bibinfo {author} {\bibfnamefont {D.~A.}\
  \bibnamefont {Huse}}, \bibinfo {author} {\bibfnamefont {I.}~\bibnamefont
  {Bloch}},\ and\ \bibinfo {author} {\bibfnamefont {C.}~\bibnamefont {Gross}},\
  }\bibfield  {title} {\bibinfo {title} {Exploring the many-body localization
  transition in two dimensions},\ }\href
  {https://doi.org/10.1126/science.aaf8834} {\bibfield  {journal} {\bibinfo
  {journal} {Science}\ }\textbf {\bibinfo {volume} {352}},\ \bibinfo {pages}
  {1547} (\bibinfo {year} {2016})}\BibitemShut {NoStop}%
\bibitem [{\citenamefont {Smith}\ \emph {et~al.}(2016)\citenamefont {Smith},
  \citenamefont {Lee}, \citenamefont {Richerme}, \citenamefont {Neyenhuis},
  \citenamefont {Hess}, \citenamefont {Hauke}, \citenamefont {Heyl},
  \citenamefont {Huse},\ and\ \citenamefont {Monroe}}]{Smith}%
  \BibitemOpen
  \bibfield  {author} {\bibinfo {author} {\bibfnamefont {J.}~\bibnamefont
  {Smith}}, \bibinfo {author} {\bibfnamefont {A.}~\bibnamefont {Lee}}, \bibinfo
  {author} {\bibfnamefont {P.}~\bibnamefont {Richerme}}, \bibinfo {author}
  {\bibfnamefont {B.}~\bibnamefont {Neyenhuis}}, \bibinfo {author}
  {\bibfnamefont {P.~W.}\ \bibnamefont {Hess}}, \bibinfo {author}
  {\bibfnamefont {P.}~\bibnamefont {Hauke}}, \bibinfo {author} {\bibfnamefont
  {M.}~\bibnamefont {Heyl}}, \bibinfo {author} {\bibfnamefont {D.~A.}\
  \bibnamefont {Huse}},\ and\ \bibinfo {author} {\bibfnamefont
  {C.}~\bibnamefont {Monroe}},\ }\bibfield  {title} {\bibinfo {title}
  {Many-body localization in a quantum simulator with programmable random
  disorder},\ }\href {https://doi.org/https://doi.org/10.1038/nphys3783}
  {\bibfield  {journal} {\bibinfo  {journal} {Nature Physics}\ }\textbf
  {\bibinfo {volume} {12}},\ \bibinfo {pages} {907} (\bibinfo {year}
  {2016})}\BibitemShut {NoStop}%
\bibitem [{\citenamefont {Zhang}\ \emph {et~al.}(2017)\citenamefont {Zhang},
  \citenamefont {Hess}, \citenamefont {Kyprianidis}, \citenamefont {Becker},
  \citenamefont {Lee}, \citenamefont {Smith}, \citenamefont {Pagano},
  \citenamefont {Potirniche}, \citenamefont {Potter}, \citenamefont
  {Vishwanath}, \citenamefont {Yao},\ and\ \citenamefont {Monroe}}]{Hess}%
  \BibitemOpen
  \bibfield  {author} {\bibinfo {author} {\bibfnamefont {J.}~\bibnamefont
  {Zhang}}, \bibinfo {author} {\bibfnamefont {P.~W.}\ \bibnamefont {Hess}},
  \bibinfo {author} {\bibfnamefont {A.}~\bibnamefont {Kyprianidis}}, \bibinfo
  {author} {\bibfnamefont {P.}~\bibnamefont {Becker}}, \bibinfo {author}
  {\bibfnamefont {A.}~\bibnamefont {Lee}}, \bibinfo {author} {\bibfnamefont
  {J.}~\bibnamefont {Smith}}, \bibinfo {author} {\bibfnamefont
  {G.}~\bibnamefont {Pagano}}, \bibinfo {author} {\bibfnamefont {I.-D.}\
  \bibnamefont {Potirniche}}, \bibinfo {author} {\bibfnamefont {A.~C.}\
  \bibnamefont {Potter}}, \bibinfo {author} {\bibfnamefont {A.}~\bibnamefont
  {Vishwanath}}, \bibinfo {author} {\bibfnamefont {N.~Y.}\ \bibnamefont
  {Yao}},\ and\ \bibinfo {author} {\bibfnamefont {C.}~\bibnamefont {Monroe}},\
  }\bibfield  {title} {\bibinfo {title} {Observation of a discrete time
  crystal},\ }\href {https://doi.org/https://doi.org/10.1038/nature21413}
  {\bibfield  {journal} {\bibinfo  {journal} {Nature}\ }\textbf {\bibinfo
  {volume} {543}},\ \bibinfo {pages} {217} (\bibinfo {year}
  {2017})}\BibitemShut {NoStop}%
\bibitem [{\citenamefont {Deutsch}(1991)}]{Deutsch}%
  \BibitemOpen
  \bibfield  {author} {\bibinfo {author} {\bibfnamefont {J.~M.}\ \bibnamefont
  {Deutsch}},\ }\bibfield  {title} {\bibinfo {title} {Quantum statistical
  mechanics in a closed system},\ }\href
  {https://doi.org/10.1103/PhysRevA.43.2046} {\bibfield  {journal} {\bibinfo
  {journal} {Phys. Rev. A}\ }\textbf {\bibinfo {volume} {43}},\ \bibinfo
  {pages} {2046} (\bibinfo {year} {1991})}\BibitemShut {NoStop}%
\bibitem [{\citenamefont {Srednicki}(1994)}]{Srednicki}%
  \BibitemOpen
  \bibfield  {author} {\bibinfo {author} {\bibfnamefont {M.}~\bibnamefont
  {Srednicki}},\ }\bibfield  {title} {\bibinfo {title} {Chaos and quantum
  thermalization},\ }\href {https://doi.org/10.1103/PhysRevE.50.888} {\bibfield
   {journal} {\bibinfo  {journal} {Phys. Rev. E}\ }\textbf {\bibinfo {volume}
  {50}},\ \bibinfo {pages} {888} (\bibinfo {year} {1994})}\BibitemShut
  {NoStop}%
\bibitem [{\citenamefont {\ifmmode \check{Z}\else
  \v{Z}\fi{}nidari\ifmmode~\check{c}\else \v{c}\fi{}}\ \emph
  {et~al.}(2008)\citenamefont {\ifmmode \check{Z}\else
  \v{Z}\fi{}nidari\ifmmode~\check{c}\else \v{c}\fi{}}, \citenamefont {Prosen},\
  and\ \citenamefont {Prelov\ifmmode~\check{s}\else \v{s}\fi{}ek}}]{Znidaric}%
  \BibitemOpen
  \bibfield  {author} {\bibinfo {author} {\bibfnamefont {M.}~\bibnamefont
  {\ifmmode \check{Z}\else \v{Z}\fi{}nidari\ifmmode~\check{c}\else
  \v{c}\fi{}}}, \bibinfo {author} {\bibfnamefont {T.}~\bibnamefont {Prosen}},\
  and\ \bibinfo {author} {\bibfnamefont {P.}~\bibnamefont
  {Prelov\ifmmode~\check{s}\else \v{s}\fi{}ek}},\ }\bibfield  {title} {\bibinfo
  {title} {Many-body localization in the {H}eisenberg {XXZ} magnet in a random
  field},\ }\href {https://doi.org/10.1103/PhysRevB.77.064426} {\bibfield
  {journal} {\bibinfo  {journal} {Phys. Rev. B}\ }\textbf {\bibinfo {volume}
  {77}},\ \bibinfo {pages} {064426} (\bibinfo {year} {2008})}\BibitemShut
  {NoStop}%
\bibitem [{\citenamefont {Bardarson}\ \emph {et~al.}(2012)\citenamefont
  {Bardarson}, \citenamefont {Pollmann},\ and\ \citenamefont {Moore}}]{Moore}%
  \BibitemOpen
  \bibfield  {author} {\bibinfo {author} {\bibfnamefont {J.~H.}\ \bibnamefont
  {Bardarson}}, \bibinfo {author} {\bibfnamefont {F.}~\bibnamefont
  {Pollmann}},\ and\ \bibinfo {author} {\bibfnamefont {J.~E.}\ \bibnamefont
  {Moore}},\ }\bibfield  {title} {\bibinfo {title} {Unbounded growth of
  entanglement in models of many-body localization},\ }\href
  {https://doi.org/10.1103/PhysRevLett.109.017202} {\bibfield  {journal}
  {\bibinfo  {journal} {Phys. Rev. Lett.}\ }\textbf {\bibinfo {volume} {109}},\
  \bibinfo {pages} {017202} (\bibinfo {year} {2012})}\BibitemShut {NoStop}%
\bibitem [{\citenamefont {Serbyn}\ \emph {et~al.}(2013)\citenamefont {Serbyn},
  \citenamefont {Papi\ifmmode~\acute{c}\else \'{c}\fi{}},\ and\ \citenamefont
  {Abanin}}]{Abanin}%
  \BibitemOpen
  \bibfield  {author} {\bibinfo {author} {\bibfnamefont {M.}~\bibnamefont
  {Serbyn}}, \bibinfo {author} {\bibfnamefont {Z.}~\bibnamefont
  {Papi\ifmmode~\acute{c}\else \'{c}\fi{}}},\ and\ \bibinfo {author}
  {\bibfnamefont {D.~A.}\ \bibnamefont {Abanin}},\ }\bibfield  {title}
  {\bibinfo {title} {Local conservation laws and the structure of the many-body
  localized states},\ }\href {https://doi.org/10.1103/PhysRevLett.111.127201}
  {\bibfield  {journal} {\bibinfo  {journal} {Phys. Rev. Lett.}\ }\textbf
  {\bibinfo {volume} {111}},\ \bibinfo {pages} {127201} (\bibinfo {year}
  {2013})}\BibitemShut {NoStop}%
\bibitem [{\citenamefont {Ros}\ \emph {et~al.}(2015)\citenamefont {Ros},
  \citenamefont {Müller},\ and\ \citenamefont {Scardicchio}}]{Scardicchio}%
  \BibitemOpen
  \bibfield  {author} {\bibinfo {author} {\bibfnamefont {V.}~\bibnamefont
  {Ros}}, \bibinfo {author} {\bibfnamefont {M.}~\bibnamefont {Müller}},\ and\
  \bibinfo {author} {\bibfnamefont {A.}~\bibnamefont {Scardicchio}},\
  }\bibfield  {title} {\bibinfo {title} {Integrals of motion in the many-body
  localized phase},\ }\href
  {https://doi.org/https://doi.org/10.1016/j.nuclphysb.2014.12.014} {\bibfield
  {journal} {\bibinfo  {journal} {Nuclear Physics B}\ }\textbf {\bibinfo
  {volume} {891}},\ \bibinfo {pages} {420} (\bibinfo {year}
  {2015})}\BibitemShut {NoStop}%
\bibitem [{\citenamefont {Huse}\ \emph {et~al.}(2014)\citenamefont {Huse},
  \citenamefont {Nandkishore},\ and\ \citenamefont {Oganesyan}}]{Organesyan}%
  \BibitemOpen
  \bibfield  {author} {\bibinfo {author} {\bibfnamefont {D.~A.}\ \bibnamefont
  {Huse}}, \bibinfo {author} {\bibfnamefont {R.}~\bibnamefont {Nandkishore}},\
  and\ \bibinfo {author} {\bibfnamefont {V.}~\bibnamefont {Oganesyan}},\
  }\bibfield  {title} {\bibinfo {title} {Phenomenology of fully
  many-body-localized systems},\ }\href
  {https://doi.org/10.1103/PhysRevB.90.174202} {\bibfield  {journal} {\bibinfo
  {journal} {Phys. Rev. B}\ }\textbf {\bibinfo {volume} {90}},\ \bibinfo
  {pages} {174202} (\bibinfo {year} {2014})}\BibitemShut {NoStop}%
\bibitem [{\citenamefont {Iyer}\ \emph {et~al.}(2013)\citenamefont {Iyer},
  \citenamefont {Oganesyan}, \citenamefont {Refael},\ and\ \citenamefont
  {Huse}}]{Huse_2013}%
  \BibitemOpen
  \bibfield  {author} {\bibinfo {author} {\bibfnamefont {S.}~\bibnamefont
  {Iyer}}, \bibinfo {author} {\bibfnamefont {V.}~\bibnamefont {Oganesyan}},
  \bibinfo {author} {\bibfnamefont {G.}~\bibnamefont {Refael}},\ and\ \bibinfo
  {author} {\bibfnamefont {D.~A.}\ \bibnamefont {Huse}},\ }\bibfield  {title}
  {\bibinfo {title} {Many-body localization in a quasiperiodic system},\ }\href
  {https://doi.org/10.1103/PhysRevB.87.134202} {\bibfield  {journal} {\bibinfo
  {journal} {Phys. Rev. B}\ }\textbf {\bibinfo {volume} {87}},\ \bibinfo
  {pages} {134202} (\bibinfo {year} {2013})}\BibitemShut {NoStop}%
\bibitem [{\citenamefont {Michal}\ \emph {et~al.}(2014)\citenamefont {Michal},
  \citenamefont {Altshuler},\ and\ \citenamefont {Shlyapnikov}}]{Shylapnikov}%
  \BibitemOpen
  \bibfield  {author} {\bibinfo {author} {\bibfnamefont {V.~P.}\ \bibnamefont
  {Michal}}, \bibinfo {author} {\bibfnamefont {B.~L.}\ \bibnamefont
  {Altshuler}},\ and\ \bibinfo {author} {\bibfnamefont {G.~V.}\ \bibnamefont
  {Shlyapnikov}},\ }\bibfield  {title} {\bibinfo {title} {Delocalization of
  weakly interacting bosons in a {1D} quasiperiodic potential},\ }\href
  {https://doi.org/10.1103/PhysRevLett.113.045304} {\bibfield  {journal}
  {\bibinfo  {journal} {Phys. Rev. Lett.}\ }\textbf {\bibinfo {volume} {113}},\
  \bibinfo {pages} {045304} (\bibinfo {year} {2014})}\BibitemShut {NoStop}%
\bibitem [{\citenamefont {Li}\ \emph {et~al.}(2015)\citenamefont {Li},
  \citenamefont {Ganeshan}, \citenamefont {Pixley},\ and\ \citenamefont
  {Das~Sarma}}]{DasSarma}%
  \BibitemOpen
  \bibfield  {author} {\bibinfo {author} {\bibfnamefont {X.}~\bibnamefont
  {Li}}, \bibinfo {author} {\bibfnamefont {S.}~\bibnamefont {Ganeshan}},
  \bibinfo {author} {\bibfnamefont {J.~H.}\ \bibnamefont {Pixley}},\ and\
  \bibinfo {author} {\bibfnamefont {S.}~\bibnamefont {Das~Sarma}},\ }\bibfield
  {title} {\bibinfo {title} {Many-body localization and quantum nonergodicity
  in a model with a single-particle mobility edge},\ }\href
  {https://doi.org/10.1103/PhysRevLett.115.186601} {\bibfield  {journal}
  {\bibinfo  {journal} {Phys. Rev. Lett.}\ }\textbf {\bibinfo {volume} {115}},\
  \bibinfo {pages} {186601} (\bibinfo {year} {2015})}\BibitemShut {NoStop}%
\bibitem [{\citenamefont {Zhang}\ and\ \citenamefont {Yao}(2018)}]{Hong}%
  \BibitemOpen
  \bibfield  {author} {\bibinfo {author} {\bibfnamefont {S.-X.}\ \bibnamefont
  {Zhang}}\ and\ \bibinfo {author} {\bibfnamefont {H.}~\bibnamefont {Yao}},\
  }\bibfield  {title} {\bibinfo {title} {Universal properties of many-body
  localization transitions in quasiperiodic systems},\ }\href
  {https://doi.org/10.1103/PhysRevLett.121.206601} {\bibfield  {journal}
  {\bibinfo  {journal} {Phys. Rev. Lett.}\ }\textbf {\bibinfo {volume} {121}},\
  \bibinfo {pages} {206601} (\bibinfo {year} {2018})}\BibitemShut {NoStop}%
\bibitem [{\citenamefont {Xu}\ \emph {et~al.}(2019)\citenamefont {Xu},
  \citenamefont {Li}, \citenamefont {Hsu}, \citenamefont {Swingle},\ and\
  \citenamefont {Das~Sarma}}]{Xu}%
  \BibitemOpen
  \bibfield  {author} {\bibinfo {author} {\bibfnamefont {S.}~\bibnamefont
  {Xu}}, \bibinfo {author} {\bibfnamefont {X.}~\bibnamefont {Li}}, \bibinfo
  {author} {\bibfnamefont {Y.-T.}\ \bibnamefont {Hsu}}, \bibinfo {author}
  {\bibfnamefont {B.}~\bibnamefont {Swingle}},\ and\ \bibinfo {author}
  {\bibfnamefont {S.}~\bibnamefont {Das~Sarma}},\ }\bibfield  {title} {\bibinfo
  {title} {Butterfly effect in interacting {Aubry-Andre} model: Thermalization,
  slow scrambling, and many-body localization},\ }\href
  {https://doi.org/10.1103/PhysRevResearch.1.032039} {\bibfield  {journal}
  {\bibinfo  {journal} {Phys. Rev. Res.}\ }\textbf {\bibinfo {volume} {1}},\
  \bibinfo {pages} {032039} (\bibinfo {year} {2019})}\BibitemShut {NoStop}%
\bibitem [{\citenamefont {Aubry}\ and\ \citenamefont
  {Andr\'{e}}(1980)}]{Aubry}%
  \BibitemOpen
  \bibfield  {author} {\bibinfo {author} {\bibfnamefont {S.}~\bibnamefont
  {Aubry}}\ and\ \bibinfo {author} {\bibfnamefont {G.}~\bibnamefont
  {Andr\'{e}}},\ }\bibfield  {title} {\bibinfo {title} {Analyticity breaking
  and {Anderson} localization in incommensurate lattices},\ }\href
  {https://chaos.if.uj.edu.pl/~delande/Lectures/files/An.Is.Phys.Soc.pdf}
  {\bibfield  {journal} {\bibinfo  {journal} {Ann. Israel. Phys. Soc.}\
  }\textbf {\bibinfo {volume} {3}},\ \bibinfo {pages} {133} (\bibinfo {year}
  {1980})}\BibitemShut {NoStop}%
\bibitem [{\citenamefont {Khemani}\ \emph {et~al.}(2017)\citenamefont
  {Khemani}, \citenamefont {Sheng},\ and\ \citenamefont {Huse}}]{Khemani}%
  \BibitemOpen
  \bibfield  {author} {\bibinfo {author} {\bibfnamefont {V.}~\bibnamefont
  {Khemani}}, \bibinfo {author} {\bibfnamefont {D.~N.}\ \bibnamefont {Sheng}},\
  and\ \bibinfo {author} {\bibfnamefont {D.~A.}\ \bibnamefont {Huse}},\
  }\bibfield  {title} {\bibinfo {title} {Two universality classes for the
  many-body localization transition},\ }\href
  {https://doi.org/10.1103/PhysRevLett.119.075702} {\bibfield  {journal}
  {\bibinfo  {journal} {Phys. Rev. Lett.}\ }\textbf {\bibinfo {volume} {119}},\
  \bibinfo {pages} {075702} (\bibinfo {year} {2017})}\BibitemShut {NoStop}%
\bibitem [{\citenamefont {Prasad}\ and\ \citenamefont
  {Garg}(2024)}]{Garg_2024}%
  \BibitemOpen
  \bibfield  {author} {\bibinfo {author} {\bibfnamefont {Y.}~\bibnamefont
  {Prasad}}\ and\ \bibinfo {author} {\bibfnamefont {A.}~\bibnamefont {Garg}},\
  }\bibfield  {title} {\bibinfo {title} {Single-particle excitations across the
  localization and many-body localization transition in quasiperiodic
  systems},\ }\href {https://doi.org/10.1103/PhysRevB.109.094204} {\bibfield
  {journal} {\bibinfo  {journal} {Phys. Rev. B}\ }\textbf {\bibinfo {volume}
  {109}},\ \bibinfo {pages} {094204} (\bibinfo {year} {2024})}\BibitemShut
  {NoStop}%
\bibitem [{\citenamefont {Schreiber}\ \emph {et~al.}(2015)\citenamefont
  {Schreiber}, \citenamefont {Hodgman}, \citenamefont {Bordia}, \citenamefont
  {Lüschen}, \citenamefont {Fischer}, \citenamefont {Vosk}, \citenamefont
  {Altman}, \citenamefont {Schneider},\ and\ \citenamefont
  {Bloch}}]{Schreiber}%
  \BibitemOpen
  \bibfield  {author} {\bibinfo {author} {\bibfnamefont {M.}~\bibnamefont
  {Schreiber}}, \bibinfo {author} {\bibfnamefont {S.~S.}\ \bibnamefont
  {Hodgman}}, \bibinfo {author} {\bibfnamefont {P.}~\bibnamefont {Bordia}},
  \bibinfo {author} {\bibfnamefont {H.~P.}\ \bibnamefont {Lüschen}}, \bibinfo
  {author} {\bibfnamefont {M.~H.}\ \bibnamefont {Fischer}}, \bibinfo {author}
  {\bibfnamefont {R.}~\bibnamefont {Vosk}}, \bibinfo {author} {\bibfnamefont
  {E.}~\bibnamefont {Altman}}, \bibinfo {author} {\bibfnamefont
  {U.}~\bibnamefont {Schneider}},\ and\ \bibinfo {author} {\bibfnamefont
  {I.}~\bibnamefont {Bloch}},\ }\bibfield  {title} {\bibinfo {title}
  {Observation of many-body localization of interacting fermions in a
  quasirandom optical lattice},\ }\href
  {https://doi.org/10.1126/science.aaa7432} {\bibfield  {journal} {\bibinfo
  {journal} {Science}\ }\textbf {\bibinfo {volume} {349}},\ \bibinfo {pages}
  {842} (\bibinfo {year} {2015})}\BibitemShut {NoStop}%
\bibitem [{\citenamefont {Bordia}\ \emph {et~al.}(2016)\citenamefont {Bordia},
  \citenamefont {L\"uschen}, \citenamefont {Hodgman}, \citenamefont
  {Schreiber}, \citenamefont {Bloch},\ and\ \citenamefont
  {Schneider}}]{Henrik}%
  \BibitemOpen
  \bibfield  {author} {\bibinfo {author} {\bibfnamefont {P.}~\bibnamefont
  {Bordia}}, \bibinfo {author} {\bibfnamefont {H.~P.}\ \bibnamefont
  {L\"uschen}}, \bibinfo {author} {\bibfnamefont {S.~S.}\ \bibnamefont
  {Hodgman}}, \bibinfo {author} {\bibfnamefont {M.}~\bibnamefont {Schreiber}},
  \bibinfo {author} {\bibfnamefont {I.}~\bibnamefont {Bloch}},\ and\ \bibinfo
  {author} {\bibfnamefont {U.}~\bibnamefont {Schneider}},\ }\bibfield  {title}
  {\bibinfo {title} {Coupling identical one-dimensional many-body localized
  systems},\ }\href {https://doi.org/10.1103/PhysRevLett.116.140401} {\bibfield
   {journal} {\bibinfo  {journal} {Phys. Rev. Lett.}\ }\textbf {\bibinfo
  {volume} {116}},\ \bibinfo {pages} {140401} (\bibinfo {year}
  {2016})}\BibitemShut {NoStop}%
\bibitem [{\citenamefont {L\"uschen}\ \emph
  {et~al.}(2017{\natexlab{a}})\citenamefont {L\"uschen}, \citenamefont
  {Bordia}, \citenamefont {Scherg}, \citenamefont {Alet}, \citenamefont
  {Altman}, \citenamefont {Schneider},\ and\ \citenamefont {Bloch}}]{Scherg}%
  \BibitemOpen
  \bibfield  {author} {\bibinfo {author} {\bibfnamefont {H.~P.}\ \bibnamefont
  {L\"uschen}}, \bibinfo {author} {\bibfnamefont {P.}~\bibnamefont {Bordia}},
  \bibinfo {author} {\bibfnamefont {S.}~\bibnamefont {Scherg}}, \bibinfo
  {author} {\bibfnamefont {F.}~\bibnamefont {Alet}}, \bibinfo {author}
  {\bibfnamefont {E.}~\bibnamefont {Altman}}, \bibinfo {author} {\bibfnamefont
  {U.}~\bibnamefont {Schneider}},\ and\ \bibinfo {author} {\bibfnamefont
  {I.}~\bibnamefont {Bloch}},\ }\bibfield  {title} {\bibinfo {title}
  {Observation of slow dynamics near the many-body localization transition in
  one-dimensional quasiperiodic systems},\ }\href
  {https://doi.org/10.1103/PhysRevLett.119.260401} {\bibfield  {journal}
  {\bibinfo  {journal} {Phys. Rev. Lett.}\ }\textbf {\bibinfo {volume} {119}},\
  \bibinfo {pages} {260401} (\bibinfo {year} {2017}{\natexlab{a}})}\BibitemShut
  {NoStop}%
\bibitem [{\citenamefont {Bordia}\ \emph {et~al.}(2017)\citenamefont {Bordia},
  \citenamefont {L\"uschen}, \citenamefont {Scherg}, \citenamefont
  {Gopalakrishnan}, \citenamefont {Knap}, \citenamefont {Schneider},\ and\
  \citenamefont {Bloch}}]{Bordia}%
  \BibitemOpen
  \bibfield  {author} {\bibinfo {author} {\bibfnamefont {P.}~\bibnamefont
  {Bordia}}, \bibinfo {author} {\bibfnamefont {H.}~\bibnamefont {L\"uschen}},
  \bibinfo {author} {\bibfnamefont {S.}~\bibnamefont {Scherg}}, \bibinfo
  {author} {\bibfnamefont {S.}~\bibnamefont {Gopalakrishnan}}, \bibinfo
  {author} {\bibfnamefont {M.}~\bibnamefont {Knap}}, \bibinfo {author}
  {\bibfnamefont {U.}~\bibnamefont {Schneider}},\ and\ \bibinfo {author}
  {\bibfnamefont {I.}~\bibnamefont {Bloch}},\ }\bibfield  {title} {\bibinfo
  {title} {Probing slow relaxation and many-body localization in
  two-dimensional quasiperiodic systems},\ }\href
  {https://doi.org/10.1103/PhysRevX.7.041047} {\bibfield  {journal} {\bibinfo
  {journal} {Phys. Rev. X}\ }\textbf {\bibinfo {volume} {7}},\ \bibinfo {pages}
  {041047} (\bibinfo {year} {2017})}\BibitemShut {NoStop}%
\bibitem [{\citenamefont {Kohlert}\ \emph {et~al.}(2019)\citenamefont
  {Kohlert}, \citenamefont {Scherg}, \citenamefont {Li}, \citenamefont
  {L\"uschen}, \citenamefont {Das~Sarma}, \citenamefont {Bloch},\ and\
  \citenamefont {Aidelsburger}}]{Kohlert}%
  \BibitemOpen
  \bibfield  {author} {\bibinfo {author} {\bibfnamefont {T.}~\bibnamefont
  {Kohlert}}, \bibinfo {author} {\bibfnamefont {S.}~\bibnamefont {Scherg}},
  \bibinfo {author} {\bibfnamefont {X.}~\bibnamefont {Li}}, \bibinfo {author}
  {\bibfnamefont {H.~P.}\ \bibnamefont {L\"uschen}}, \bibinfo {author}
  {\bibfnamefont {S.}~\bibnamefont {Das~Sarma}}, \bibinfo {author}
  {\bibfnamefont {I.}~\bibnamefont {Bloch}},\ and\ \bibinfo {author}
  {\bibfnamefont {M.}~\bibnamefont {Aidelsburger}},\ }\bibfield  {title}
  {\bibinfo {title} {Observation of many-body localization in a one-dimensional
  system with a single-particle mobility edge},\ }\href
  {https://doi.org/10.1103/PhysRevLett.122.170403} {\bibfield  {journal}
  {\bibinfo  {journal} {Phys. Rev. Lett.}\ }\textbf {\bibinfo {volume} {122}},\
  \bibinfo {pages} {170403} (\bibinfo {year} {2019})}\BibitemShut {NoStop}%
\bibitem [{\citenamefont {L\"uschen}\ \emph
  {et~al.}(2017{\natexlab{b}})\citenamefont {L\"uschen}, \citenamefont
  {Bordia}, \citenamefont {Hodgman}, \citenamefont {Schreiber}, \citenamefont
  {Sarkar}, \citenamefont {Daley}, \citenamefont {Fischer}, \citenamefont
  {Altman}, \citenamefont {Bloch},\ and\ \citenamefont {Schneider}}]{Ehud}%
  \BibitemOpen
  \bibfield  {author} {\bibinfo {author} {\bibfnamefont {H.~P.}\ \bibnamefont
  {L\"uschen}}, \bibinfo {author} {\bibfnamefont {P.}~\bibnamefont {Bordia}},
  \bibinfo {author} {\bibfnamefont {S.~S.}\ \bibnamefont {Hodgman}}, \bibinfo
  {author} {\bibfnamefont {M.}~\bibnamefont {Schreiber}}, \bibinfo {author}
  {\bibfnamefont {S.}~\bibnamefont {Sarkar}}, \bibinfo {author} {\bibfnamefont
  {A.~J.}\ \bibnamefont {Daley}}, \bibinfo {author} {\bibfnamefont {M.~H.}\
  \bibnamefont {Fischer}}, \bibinfo {author} {\bibfnamefont {E.}~\bibnamefont
  {Altman}}, \bibinfo {author} {\bibfnamefont {I.}~\bibnamefont {Bloch}},\ and\
  \bibinfo {author} {\bibfnamefont {U.}~\bibnamefont {Schneider}},\ }\bibfield
  {title} {\bibinfo {title} {Signatures of many-body localization in a
  controlled open quantum system},\ }\href
  {https://doi.org/10.1103/PhysRevX.7.011034} {\bibfield  {journal} {\bibinfo
  {journal} {Phys. Rev. X}\ }\textbf {\bibinfo {volume} {7}},\ \bibinfo {pages}
  {011034} (\bibinfo {year} {2017}{\natexlab{b}})}\BibitemShut {NoStop}%
\bibitem [{\citenamefont {Hatano}\ and\ \citenamefont
  {Nelson}(1996)}]{HatanoNelson1996}%
  \BibitemOpen
  \bibfield  {author} {\bibinfo {author} {\bibfnamefont {N.}~\bibnamefont
  {Hatano}}\ and\ \bibinfo {author} {\bibfnamefont {D.~R.}\ \bibnamefont
  {Nelson}},\ }\bibfield  {title} {\bibinfo {title} {Localization transitions
  in non-{Hermitian} quantum mechanics},\ }\href
  {https://doi.org/10.1103/PhysRevLett.77.570} {\bibfield  {journal} {\bibinfo
  {journal} {Phys. Rev. Lett.}\ }\textbf {\bibinfo {volume} {77}},\ \bibinfo
  {pages} {570} (\bibinfo {year} {1996})}\BibitemShut {NoStop}%
\bibitem [{\citenamefont {Hatano}\ and\ \citenamefont
  {Nelson}(1998)}]{HatanoNelson1998}%
  \BibitemOpen
  \bibfield  {author} {\bibinfo {author} {\bibfnamefont {N.}~\bibnamefont
  {Hatano}}\ and\ \bibinfo {author} {\bibfnamefont {D.~R.}\ \bibnamefont
  {Nelson}},\ }\bibfield  {title} {\bibinfo {title} {Non-{Hermitian}
  delocalization and eigenfunctions},\ }\href
  {https://doi.org/10.1103/PhysRevB.58.8384} {\bibfield  {journal} {\bibinfo
  {journal} {Phys. Rev. B}\ }\textbf {\bibinfo {volume} {58}},\ \bibinfo
  {pages} {8384} (\bibinfo {year} {1998})}\BibitemShut {NoStop}%
\bibitem [{\citenamefont {Hamazaki}\ \emph {et~al.}(2019)\citenamefont
  {Hamazaki}, \citenamefont {Kawabata},\ and\ \citenamefont {Ueda}}]{Kawabata}%
  \BibitemOpen
  \bibfield  {author} {\bibinfo {author} {\bibfnamefont {R.}~\bibnamefont
  {Hamazaki}}, \bibinfo {author} {\bibfnamefont {K.}~\bibnamefont {Kawabata}},\
  and\ \bibinfo {author} {\bibfnamefont {M.}~\bibnamefont {Ueda}},\ }\bibfield
  {title} {\bibinfo {title} {Non-{Hermitian} many-body localization},\ }\href
  {https://doi.org/10.1103/PhysRevLett.123.090603} {\bibfield  {journal}
  {\bibinfo  {journal} {Phys. Rev. Lett.}\ }\textbf {\bibinfo {volume} {123}},\
  \bibinfo {pages} {090603} (\bibinfo {year} {2019})}\BibitemShut {NoStop}%
\bibitem [{\citenamefont {Panda}\ and\ \citenamefont
  {Banerjee}(2020)}]{Banerjee}%
  \BibitemOpen
  \bibfield  {author} {\bibinfo {author} {\bibfnamefont {A.}~\bibnamefont
  {Panda}}\ and\ \bibinfo {author} {\bibfnamefont {S.}~\bibnamefont
  {Banerjee}},\ }\bibfield  {title} {\bibinfo {title} {Entanglement in
  nonequilibrium steady states and many-body localization breakdown in a
  current-driven system},\ }\href {https://doi.org/10.1103/PhysRevB.101.184201}
  {\bibfield  {journal} {\bibinfo  {journal} {Phys. Rev. B}\ }\textbf {\bibinfo
  {volume} {101}},\ \bibinfo {pages} {184201} (\bibinfo {year}
  {2020})}\BibitemShut {NoStop}%
\bibitem [{\citenamefont {Zhai}\ \emph {et~al.}(2020)\citenamefont {Zhai},
  \citenamefont {Yin},\ and\ \citenamefont {Huang}}]{Huang}%
  \BibitemOpen
  \bibfield  {author} {\bibinfo {author} {\bibfnamefont {L.-J.}\ \bibnamefont
  {Zhai}}, \bibinfo {author} {\bibfnamefont {S.}~\bibnamefont {Yin}},\ and\
  \bibinfo {author} {\bibfnamefont {G.-Y.}\ \bibnamefont {Huang}},\ }\bibfield
  {title} {\bibinfo {title} {Many-body localization in a non-{Hermitian}
  quasiperiodic system},\ }\href {https://doi.org/10.1103/PhysRevB.102.064206}
  {\bibfield  {journal} {\bibinfo  {journal} {Phys. Rev. B}\ }\textbf {\bibinfo
  {volume} {102}},\ \bibinfo {pages} {064206} (\bibinfo {year}
  {2020})}\BibitemShut {NoStop}%
\bibitem [{\citenamefont {Tang}\ \emph {et~al.}(2021)\citenamefont {Tang},
  \citenamefont {Zhang}, \citenamefont {Zhang},\ and\ \citenamefont
  {Zhang}}]{Dan-Wei-PRA21}%
  \BibitemOpen
  \bibfield  {author} {\bibinfo {author} {\bibfnamefont {L.-Z.}\ \bibnamefont
  {Tang}}, \bibinfo {author} {\bibfnamefont {G.-Q.}\ \bibnamefont {Zhang}},
  \bibinfo {author} {\bibfnamefont {L.-F.}\ \bibnamefont {Zhang}},\ and\
  \bibinfo {author} {\bibfnamefont {D.-W.}\ \bibnamefont {Zhang}},\ }\bibfield
  {title} {\bibinfo {title} {Localization and topological transitions in
  non-{H}ermitian quasiperiodic lattices},\ }\href
  {https://doi.org/10.1103/PhysRevA.103.033325} {\bibfield  {journal} {\bibinfo
   {journal} {Phys. Rev. A}\ }\textbf {\bibinfo {volume} {103}},\ \bibinfo
  {pages} {033325} (\bibinfo {year} {2021})}\BibitemShut {NoStop}%
\bibitem [{\citenamefont {Orito}\ and\ \citenamefont
  {Imura}(2022)}]{Imura_2022}%
  \BibitemOpen
  \bibfield  {author} {\bibinfo {author} {\bibfnamefont {T.}~\bibnamefont
  {Orito}}\ and\ \bibinfo {author} {\bibfnamefont {K.-I.}\ \bibnamefont
  {Imura}},\ }\bibfield  {title} {\bibinfo {title} {Unusual wave-packet
  spreading and entanglement dynamics in non-{Hermitian} disordered many-body
  systems},\ }\href {https://doi.org/10.1103/PhysRevB.105.024303} {\bibfield
  {journal} {\bibinfo  {journal} {Phys. Rev. B}\ }\textbf {\bibinfo {volume}
  {105}},\ \bibinfo {pages} {024303} (\bibinfo {year} {2022})}\BibitemShut
  {NoStop}%
\bibitem [{\citenamefont {Orito}\ and\ \citenamefont
  {Imura}(2023)}]{Imura_2023}%
  \BibitemOpen
  \bibfield  {author} {\bibinfo {author} {\bibfnamefont {T.}~\bibnamefont
  {Orito}}\ and\ \bibinfo {author} {\bibfnamefont {K.-I.}\ \bibnamefont
  {Imura}},\ }\bibfield  {title} {\bibinfo {title} {Entanglement dynamics in
  the many-body {Hatano-Nelson} model},\ }\href
  {https://doi.org/10.1103/PhysRevB.108.214308} {\bibfield  {journal} {\bibinfo
   {journal} {Phys. Rev. B}\ }\textbf {\bibinfo {volume} {108}},\ \bibinfo
  {pages} {214308} (\bibinfo {year} {2023})}\BibitemShut {NoStop}%
\bibitem [{\citenamefont {Lee}\ \emph {et~al.}(2020)\citenamefont {Lee},
  \citenamefont {Lee},\ and\ \citenamefont {Yang}}]{Lee_2020}%
  \BibitemOpen
  \bibfield  {author} {\bibinfo {author} {\bibfnamefont {E.}~\bibnamefont
  {Lee}}, \bibinfo {author} {\bibfnamefont {H.}~\bibnamefont {Lee}},\ and\
  \bibinfo {author} {\bibfnamefont {B.-J.}\ \bibnamefont {Yang}},\ }\bibfield
  {title} {\bibinfo {title} {Many-body approach to non-{Hermitian} physics in
  fermionic systems},\ }\href {https://doi.org/10.1103/PhysRevB.101.121109}
  {\bibfield  {journal} {\bibinfo  {journal} {Phys. Rev. B}\ }\textbf {\bibinfo
  {volume} {101}},\ \bibinfo {pages} {121109} (\bibinfo {year}
  {2020})}\BibitemShut {NoStop}%
\bibitem [{\citenamefont {Suthar}\ \emph {et~al.}(2022)\citenamefont {Suthar},
  \citenamefont {Wang}, \citenamefont {Huang}, \citenamefont {Jen},\ and\
  \citenamefont {You}}]{Suthar}%
  \BibitemOpen
  \bibfield  {author} {\bibinfo {author} {\bibfnamefont {K.}~\bibnamefont
  {Suthar}}, \bibinfo {author} {\bibfnamefont {Y.-C.}\ \bibnamefont {Wang}},
  \bibinfo {author} {\bibfnamefont {Y.-P.}\ \bibnamefont {Huang}}, \bibinfo
  {author} {\bibfnamefont {H.~H.}\ \bibnamefont {Jen}},\ and\ \bibinfo {author}
  {\bibfnamefont {J.-S.}\ \bibnamefont {You}},\ }\bibfield  {title} {\bibinfo
  {title} {Non-{Hermitian} many-body localization with open boundaries},\
  }\href {https://doi.org/10.1103/PhysRevB.106.064208} {\bibfield  {journal}
  {\bibinfo  {journal} {Phys. Rev. B}\ }\textbf {\bibinfo {volume} {106}},\
  \bibinfo {pages} {064208} (\bibinfo {year} {2022})}\BibitemShut {NoStop}%
\bibitem [{\citenamefont {Kawabata}\ \emph {et~al.}(2023)\citenamefont
  {Kawabata}, \citenamefont {Numasawa},\ and\ \citenamefont {Ryu}}]{Numasawa}%
  \BibitemOpen
  \bibfield  {author} {\bibinfo {author} {\bibfnamefont {K.}~\bibnamefont
  {Kawabata}}, \bibinfo {author} {\bibfnamefont {T.}~\bibnamefont {Numasawa}},\
  and\ \bibinfo {author} {\bibfnamefont {S.}~\bibnamefont {Ryu}},\ }\bibfield
  {title} {\bibinfo {title} {Entanglement phase transition induced by the
  non-{Hermitian} skin effect},\ }\href
  {https://doi.org/10.1103/PhysRevX.13.021007} {\bibfield  {journal} {\bibinfo
  {journal} {Phys. Rev. X}\ }\textbf {\bibinfo {volume} {13}},\ \bibinfo
  {pages} {021007} (\bibinfo {year} {2023})}\BibitemShut {NoStop}%
\bibitem [{\citenamefont {Li}\ \emph {et~al.}(2023)\citenamefont {Li},
  \citenamefont {Liu},\ and\ \citenamefont {Xu}}]{Li}%
  \BibitemOpen
  \bibfield  {author} {\bibinfo {author} {\bibfnamefont {K.}~\bibnamefont
  {Li}}, \bibinfo {author} {\bibfnamefont {Z.-C.}\ \bibnamefont {Liu}},\ and\
  \bibinfo {author} {\bibfnamefont {Y.}~\bibnamefont {Xu}},\ }\href
  {https://arxiv.org/abs/2305.12342} {\bibinfo {title} {Disorder-induced
  entanglement phase transitions in non-{Hermitian} systems with skin effects}}
  (\bibinfo {year} {2023}),\ \Eprint {https://arxiv.org/abs/2305.12342}
  {arXiv:2305.12342 [quant-ph]} \BibitemShut {NoStop}%
\bibitem [{\citenamefont {Li}\ \emph {et~al.}(2024)\citenamefont {Li},
  \citenamefont {Yu},\ and\ \citenamefont {Li}}]{Zhi}%
  \BibitemOpen
  \bibfield  {author} {\bibinfo {author} {\bibfnamefont {S.-Z.}\ \bibnamefont
  {Li}}, \bibinfo {author} {\bibfnamefont {X.-J.}\ \bibnamefont {Yu}},\ and\
  \bibinfo {author} {\bibfnamefont {Z.}~\bibnamefont {Li}},\ }\bibfield
  {title} {\bibinfo {title} {Emergent entanglement phase transitions in
  non-{Hermitian} {Aubry-Andr\'e-Harper} chains},\ }\href
  {https://doi.org/10.1103/PhysRevB.109.024306} {\bibfield  {journal} {\bibinfo
   {journal} {Phys. Rev. B}\ }\textbf {\bibinfo {volume} {109}},\ \bibinfo
  {pages} {024306} (\bibinfo {year} {2024})}\BibitemShut {NoStop}%
\bibitem [{\citenamefont {Lee}(2016)}]{Lee}%
  \BibitemOpen
  \bibfield  {author} {\bibinfo {author} {\bibfnamefont {T.~E.}\ \bibnamefont
  {Lee}},\ }\bibfield  {title} {\bibinfo {title} {Anomalous edge state in a
  non-{Hermitian} lattice},\ }\href
  {https://doi.org/10.1103/PhysRevLett.116.133903} {\bibfield  {journal}
  {\bibinfo  {journal} {Phys. Rev. Lett.}\ }\textbf {\bibinfo {volume} {116}},\
  \bibinfo {pages} {133903} (\bibinfo {year} {2016})}\BibitemShut {NoStop}%
\bibitem [{\citenamefont {Zeuner}\ \emph {et~al.}(2015)\citenamefont {Zeuner},
  \citenamefont {Rechtsman}, \citenamefont {Plotnik}, \citenamefont {Lumer},
  \citenamefont {Nolte}, \citenamefont {Rudner}, \citenamefont {Segev},\ and\
  \citenamefont {Szameit}}]{Zeuner}%
  \BibitemOpen
  \bibfield  {author} {\bibinfo {author} {\bibfnamefont {J.~M.}\ \bibnamefont
  {Zeuner}}, \bibinfo {author} {\bibfnamefont {M.~C.}\ \bibnamefont
  {Rechtsman}}, \bibinfo {author} {\bibfnamefont {Y.}~\bibnamefont {Plotnik}},
  \bibinfo {author} {\bibfnamefont {Y.}~\bibnamefont {Lumer}}, \bibinfo
  {author} {\bibfnamefont {S.}~\bibnamefont {Nolte}}, \bibinfo {author}
  {\bibfnamefont {M.~S.}\ \bibnamefont {Rudner}}, \bibinfo {author}
  {\bibfnamefont {M.}~\bibnamefont {Segev}},\ and\ \bibinfo {author}
  {\bibfnamefont {A.}~\bibnamefont {Szameit}},\ }\bibfield  {title} {\bibinfo
  {title} {Observation of a topological transition in the bulk of a
  non-{Hermitian} system},\ }\href
  {https://doi.org/10.1103/PhysRevLett.115.040402} {\bibfield  {journal}
  {\bibinfo  {journal} {Phys. Rev. Lett.}\ }\textbf {\bibinfo {volume} {115}},\
  \bibinfo {pages} {040402} (\bibinfo {year} {2015})}\BibitemShut {NoStop}%
\bibitem [{\citenamefont {Yao}\ and\ \citenamefont {Wang}(2018)}]{Yao}%
  \BibitemOpen
  \bibfield  {author} {\bibinfo {author} {\bibfnamefont {S.}~\bibnamefont
  {Yao}}\ and\ \bibinfo {author} {\bibfnamefont {Z.}~\bibnamefont {Wang}},\
  }\bibfield  {title} {\bibinfo {title} {Edge states and topological invariants
  of non-{Hermitian} systems},\ }\href
  {https://doi.org/10.1103/PhysRevLett.121.086803} {\bibfield  {journal}
  {\bibinfo  {journal} {Phys. Rev. Lett.}\ }\textbf {\bibinfo {volume} {121}},\
  \bibinfo {pages} {086803} (\bibinfo {year} {2018})}\BibitemShut {NoStop}%
\bibitem [{\citenamefont {Gong}\ \emph {et~al.}(2018)\citenamefont {Gong},
  \citenamefont {Ashida}, \citenamefont {Kawabata}, \citenamefont {Takasan},
  \citenamefont {Higashikawa},\ and\ \citenamefont {Ueda}}]{Gong}%
  \BibitemOpen
  \bibfield  {author} {\bibinfo {author} {\bibfnamefont {Z.}~\bibnamefont
  {Gong}}, \bibinfo {author} {\bibfnamefont {Y.}~\bibnamefont {Ashida}},
  \bibinfo {author} {\bibfnamefont {K.}~\bibnamefont {Kawabata}}, \bibinfo
  {author} {\bibfnamefont {K.}~\bibnamefont {Takasan}}, \bibinfo {author}
  {\bibfnamefont {S.}~\bibnamefont {Higashikawa}},\ and\ \bibinfo {author}
  {\bibfnamefont {M.}~\bibnamefont {Ueda}},\ }\bibfield  {title} {\bibinfo
  {title} {Topological phases of non-{Hermitian} systems},\ }\href
  {https://doi.org/10.1103/PhysRevX.8.031079} {\bibfield  {journal} {\bibinfo
  {journal} {Phys. Rev. X}\ }\textbf {\bibinfo {volume} {8}},\ \bibinfo {pages}
  {031079} (\bibinfo {year} {2018})}\BibitemShut {NoStop}%
\bibitem [{\citenamefont {Kawabata}\ \emph {et~al.}(2019)\citenamefont
  {Kawabata}, \citenamefont {Shiozaki}, \citenamefont {Ueda},\ and\
  \citenamefont {Sato}}]{Kawabata_2019}%
  \BibitemOpen
  \bibfield  {author} {\bibinfo {author} {\bibfnamefont {K.}~\bibnamefont
  {Kawabata}}, \bibinfo {author} {\bibfnamefont {K.}~\bibnamefont {Shiozaki}},
  \bibinfo {author} {\bibfnamefont {M.}~\bibnamefont {Ueda}},\ and\ \bibinfo
  {author} {\bibfnamefont {M.}~\bibnamefont {Sato}},\ }\bibfield  {title}
  {\bibinfo {title} {Symmetry and topology in non-{Hermitian} physics},\ }\href
  {https://doi.org/10.1103/PhysRevX.9.041015} {\bibfield  {journal} {\bibinfo
  {journal} {Phys. Rev. X}\ }\textbf {\bibinfo {volume} {9}},\ \bibinfo {pages}
  {041015} (\bibinfo {year} {2019})}\BibitemShut {NoStop}%
\bibitem [{\citenamefont {Chiu}\ \emph {et~al.}(2016)\citenamefont {Chiu},
  \citenamefont {Teo}, \citenamefont {Schnyder},\ and\ \citenamefont
  {Ryu}}]{Ryu}%
  \BibitemOpen
  \bibfield  {author} {\bibinfo {author} {\bibfnamefont {C.-K.}\ \bibnamefont
  {Chiu}}, \bibinfo {author} {\bibfnamefont {J.~C.~Y.}\ \bibnamefont {Teo}},
  \bibinfo {author} {\bibfnamefont {A.~P.}\ \bibnamefont {Schnyder}},\ and\
  \bibinfo {author} {\bibfnamefont {S.}~\bibnamefont {Ryu}},\ }\bibfield
  {title} {\bibinfo {title} {Classification of topological quantum matter with
  symmetries},\ }\href {https://doi.org/10.1103/RevModPhys.88.035005}
  {\bibfield  {journal} {\bibinfo  {journal} {Rev. Mod. Phys.}\ }\textbf
  {\bibinfo {volume} {88}},\ \bibinfo {pages} {035005} (\bibinfo {year}
  {2016})}\BibitemShut {NoStop}%
\bibitem [{\citenamefont {Lapa}\ \emph {et~al.}(2016)\citenamefont {Lapa},
  \citenamefont {Teo},\ and\ \citenamefont {Hughes}}]{Hughes}%
  \BibitemOpen
  \bibfield  {author} {\bibinfo {author} {\bibfnamefont {M.~F.}\ \bibnamefont
  {Lapa}}, \bibinfo {author} {\bibfnamefont {J.~C.~Y.}\ \bibnamefont {Teo}},\
  and\ \bibinfo {author} {\bibfnamefont {T.~L.}\ \bibnamefont {Hughes}},\
  }\bibfield  {title} {\bibinfo {title} {Interaction-enabled topological
  crystalline phases},\ }\href {https://doi.org/10.1103/PhysRevB.93.115131}
  {\bibfield  {journal} {\bibinfo  {journal} {Phys. Rev. B}\ }\textbf {\bibinfo
  {volume} {93}},\ \bibinfo {pages} {115131} (\bibinfo {year}
  {2016})}\BibitemShut {NoStop}%
\bibitem [{\citenamefont {Zhang}\ \emph {et~al.}(2020)\citenamefont {Zhang},
  \citenamefont {Yang},\ and\ \citenamefont {Fang}}]{Fang}%
  \BibitemOpen
  \bibfield  {author} {\bibinfo {author} {\bibfnamefont {K.}~\bibnamefont
  {Zhang}}, \bibinfo {author} {\bibfnamefont {Z.}~\bibnamefont {Yang}},\ and\
  \bibinfo {author} {\bibfnamefont {C.}~\bibnamefont {Fang}},\ }\bibfield
  {title} {\bibinfo {title} {Correspondence between winding numbers and skin
  modes in non-{Hermitian} systems},\ }\href
  {https://doi.org/10.1103/PhysRevLett.125.126402} {\bibfield  {journal}
  {\bibinfo  {journal} {Phys. Rev. Lett.}\ }\textbf {\bibinfo {volume} {125}},\
  \bibinfo {pages} {126402} (\bibinfo {year} {2020})}\BibitemShut {NoStop}%
\bibitem [{\citenamefont {Okuma}\ \emph {et~al.}(2020)\citenamefont {Okuma},
  \citenamefont {Kawabata}, \citenamefont {Shiozaki},\ and\ \citenamefont
  {Sato}}]{Sato}%
  \BibitemOpen
  \bibfield  {author} {\bibinfo {author} {\bibfnamefont {N.}~\bibnamefont
  {Okuma}}, \bibinfo {author} {\bibfnamefont {K.}~\bibnamefont {Kawabata}},
  \bibinfo {author} {\bibfnamefont {K.}~\bibnamefont {Shiozaki}},\ and\
  \bibinfo {author} {\bibfnamefont {M.}~\bibnamefont {Sato}},\ }\bibfield
  {title} {\bibinfo {title} {Topological origin of non-{Hermitian} skin
  effects},\ }\href {https://doi.org/10.1103/PhysRevLett.124.086801} {\bibfield
   {journal} {\bibinfo  {journal} {Phys. Rev. Lett.}\ }\textbf {\bibinfo
  {volume} {124}},\ \bibinfo {pages} {086801} (\bibinfo {year}
  {2020})}\BibitemShut {NoStop}%
\bibitem [{\citenamefont {Claes}\ and\ \citenamefont {Hughes}(2021)}]{Claes}%
  \BibitemOpen
  \bibfield  {author} {\bibinfo {author} {\bibfnamefont {J.}~\bibnamefont
  {Claes}}\ and\ \bibinfo {author} {\bibfnamefont {T.~L.}\ \bibnamefont
  {Hughes}},\ }\bibfield  {title} {\bibinfo {title} {Skin effect and winding
  number in disordered non-{Hermitian} systems},\ }\href
  {https://doi.org/10.1103/PhysRevB.103.L140201} {\bibfield  {journal}
  {\bibinfo  {journal} {Phys. Rev. B}\ }\textbf {\bibinfo {volume} {103}},\
  \bibinfo {pages} {L140201} (\bibinfo {year} {2021})}\BibitemShut {NoStop}%
\bibitem [{\citenamefont {Khatami}\ \emph {et~al.}(2012)\citenamefont
  {Khatami}, \citenamefont {Rigol}, \citenamefont {Rela\~no},\ and\
  \citenamefont {Garc\'{\i}a-Garc\'{\i}a}}]{Khatami}%
  \BibitemOpen
  \bibfield  {author} {\bibinfo {author} {\bibfnamefont {E.}~\bibnamefont
  {Khatami}}, \bibinfo {author} {\bibfnamefont {M.}~\bibnamefont {Rigol}},
  \bibinfo {author} {\bibfnamefont {A.}~\bibnamefont {Rela\~no}},\ and\
  \bibinfo {author} {\bibfnamefont {A.~M.}\ \bibnamefont
  {Garc\'{\i}a-Garc\'{\i}a}},\ }\bibfield  {title} {\bibinfo {title} {Quantum
  quenches in disordered systems: Approach to thermal equilibrium without a
  typical relaxation time},\ }\href
  {https://doi.org/10.1103/PhysRevE.85.050102} {\bibfield  {journal} {\bibinfo
  {journal} {Phys. Rev. E}\ }\textbf {\bibinfo {volume} {85}},\ \bibinfo
  {pages} {050102} (\bibinfo {year} {2012})}\BibitemShut {NoStop}%
\bibitem [{\citenamefont {Deng}\ \emph {et~al.}(2019)\citenamefont {Deng},
  \citenamefont {Ray}, \citenamefont {Sinha}, \citenamefont {Shlyapnikov},\
  and\ \citenamefont {Santos}}]{Shlyapnikov}%
  \BibitemOpen
  \bibfield  {author} {\bibinfo {author} {\bibfnamefont {X.}~\bibnamefont
  {Deng}}, \bibinfo {author} {\bibfnamefont {S.}~\bibnamefont {Ray}}, \bibinfo
  {author} {\bibfnamefont {S.}~\bibnamefont {Sinha}}, \bibinfo {author}
  {\bibfnamefont {G.~V.}\ \bibnamefont {Shlyapnikov}},\ and\ \bibinfo {author}
  {\bibfnamefont {L.}~\bibnamefont {Santos}},\ }\bibfield  {title} {\bibinfo
  {title} {One-dimensional quasicrystals with power-law hopping},\ }\href
  {https://doi.org/10.1103/PhysRevLett.123.025301} {\bibfield  {journal}
  {\bibinfo  {journal} {Phys. Rev. Lett.}\ }\textbf {\bibinfo {volume} {123}},\
  \bibinfo {pages} {025301} (\bibinfo {year} {2019})}\BibitemShut {NoStop}%
\bibitem [{\citenamefont {Nag}\ and\ \citenamefont {Garg}(2019)}]{Sabyasachi}%
  \BibitemOpen
  \bibfield  {author} {\bibinfo {author} {\bibfnamefont {S.}~\bibnamefont
  {Nag}}\ and\ \bibinfo {author} {\bibfnamefont {A.}~\bibnamefont {Garg}},\
  }\bibfield  {title} {\bibinfo {title} {Many-body localization in the presence
  of long-range interactions and long-range hopping},\ }\href
  {https://doi.org/10.1103/PhysRevB.99.224203} {\bibfield  {journal} {\bibinfo
  {journal} {Phys. Rev. B}\ }\textbf {\bibinfo {volume} {99}},\ \bibinfo
  {pages} {224203} (\bibinfo {year} {2019})}\BibitemShut {NoStop}%
\bibitem [{\citenamefont {Burin}(2015)}]{Burin}%
  \BibitemOpen
  \bibfield  {author} {\bibinfo {author} {\bibfnamefont {A.~L.}\ \bibnamefont
  {Burin}},\ }\bibfield  {title} {\bibinfo {title} {Many-body delocalization in
  a strongly disordered system with long-range interactions: Finite-size
  scaling},\ }\href {https://doi.org/10.1103/PhysRevB.91.094202} {\bibfield
  {journal} {\bibinfo  {journal} {Phys. Rev. B}\ }\textbf {\bibinfo {volume}
  {91}},\ \bibinfo {pages} {094202} (\bibinfo {year} {2015})}\BibitemShut
  {NoStop}%
\bibitem [{\citenamefont {Nag}\ and\ \citenamefont {Garg}(2017)}]{Garg}%
  \BibitemOpen
  \bibfield  {author} {\bibinfo {author} {\bibfnamefont {S.}~\bibnamefont
  {Nag}}\ and\ \bibinfo {author} {\bibfnamefont {A.}~\bibnamefont {Garg}},\
  }\bibfield  {title} {\bibinfo {title} {Many-body mobility edges in a
  one-dimensional system of interacting fermions},\ }\href
  {https://doi.org/10.1103/PhysRevB.96.060203} {\bibfield  {journal} {\bibinfo
  {journal} {Phys. Rev. B}\ }\textbf {\bibinfo {volume} {96}},\ \bibinfo
  {pages} {060203} (\bibinfo {year} {2017})}\BibitemShut {NoStop}%
\bibitem [{\citenamefont {Xu}\ \emph {et~al.}(2021)\citenamefont {Xu},
  \citenamefont {Xia},\ and\ \citenamefont {Chen}}]{Chen}%
  \BibitemOpen
  \bibfield  {author} {\bibinfo {author} {\bibfnamefont {Z.}~\bibnamefont
  {Xu}}, \bibinfo {author} {\bibfnamefont {X.}~\bibnamefont {Xia}},\ and\
  \bibinfo {author} {\bibfnamefont {S.}~\bibnamefont {Chen}},\ }\bibfield
  {title} {\bibinfo {title} {Non-{Hermitian} {Aubry-Andr\'e} model with
  power-law hopping},\ }\href {https://doi.org/10.1103/PhysRevB.104.224204}
  {\bibfield  {journal} {\bibinfo  {journal} {Phys. Rev. B}\ }\textbf {\bibinfo
  {volume} {104}},\ \bibinfo {pages} {224204} (\bibinfo {year}
  {2021})}\BibitemShut {NoStop}%
\bibitem [{\citenamefont {Liu}\ \emph {et~al.}(2024)\citenamefont {Liu},
  \citenamefont {Zhang}, \citenamefont {Li},\ and\ \citenamefont {Li}}]{Juan}%
  \BibitemOpen
  \bibfield  {author} {\bibinfo {author} {\bibfnamefont {G.-J.}\ \bibnamefont
  {Liu}}, \bibinfo {author} {\bibfnamefont {J.-M.}\ \bibnamefont {Zhang}},
  \bibinfo {author} {\bibfnamefont {S.-Z.}\ \bibnamefont {Li}},\ and\ \bibinfo
  {author} {\bibfnamefont {Z.}~\bibnamefont {Li}},\ }\bibfield  {title}
  {\bibinfo {title} {Emergent strength-dependent scale-free mobility edge in a
  nonreciprocal long-range {Aubry-Andr\'e-Harper} model},\ }\href
  {https://doi.org/10.1103/PhysRevA.110.012222} {\bibfield  {journal} {\bibinfo
   {journal} {Phys. Rev. A}\ }\textbf {\bibinfo {volume} {110}},\ \bibinfo
  {pages} {012222} (\bibinfo {year} {2024})}\BibitemShut {NoStop}%
\bibitem [{\citenamefont {Peng}\ \emph {et~al.}(2025)\citenamefont {Peng},
  \citenamefont {Cheng},\ and\ \citenamefont {Xianlong}}]{Xianlong}%
  \BibitemOpen
  \bibfield  {author} {\bibinfo {author} {\bibfnamefont {D.}~\bibnamefont
  {Peng}}, \bibinfo {author} {\bibfnamefont {S.}~\bibnamefont {Cheng}},\ and\
  \bibinfo {author} {\bibfnamefont {G.}~\bibnamefont {Xianlong}},\ }\bibfield
  {title} {\bibinfo {title} {Long-range hopping in a quasiperiodic potential
  weakens the non-{Hermitian} skin effect},\ }\href
  {https://doi.org/10.1103/PhysRevB.111.094204} {\bibfield  {journal} {\bibinfo
   {journal} {Phys. Rev. B}\ }\textbf {\bibinfo {volume} {111}},\ \bibinfo
  {pages} {094204} (\bibinfo {year} {2025})}\BibitemShut {NoStop}%
\bibitem [{\citenamefont {Lazarides}\ \emph {et~al.}(2014)\citenamefont
  {Lazarides}, \citenamefont {Das},\ and\ \citenamefont {Moessner}}]{Moessner}%
  \BibitemOpen
  \bibfield  {author} {\bibinfo {author} {\bibfnamefont {A.}~\bibnamefont
  {Lazarides}}, \bibinfo {author} {\bibfnamefont {A.}~\bibnamefont {Das}},\
  and\ \bibinfo {author} {\bibfnamefont {R.}~\bibnamefont {Moessner}},\
  }\bibfield  {title} {\bibinfo {title} {Equilibrium states of generic quantum
  systems subject to periodic driving},\ }\href
  {https://doi.org/10.1103/PhysRevE.90.012110} {\bibfield  {journal} {\bibinfo
  {journal} {Phys. Rev. E}\ }\textbf {\bibinfo {volume} {90}},\ \bibinfo
  {pages} {012110} (\bibinfo {year} {2014})}\BibitemShut {NoStop}%
\bibitem [{\citenamefont {Wang}\ \emph {et~al.}(2019)\citenamefont {Wang},
  \citenamefont {Jin},\ and\ \citenamefont {Song}}]{Wang_2019}%
  \BibitemOpen
  \bibfield  {author} {\bibinfo {author} {\bibfnamefont {P.}~\bibnamefont
  {Wang}}, \bibinfo {author} {\bibfnamefont {L.}~\bibnamefont {Jin}},\ and\
  \bibinfo {author} {\bibfnamefont {Z.}~\bibnamefont {Song}},\ }\bibfield
  {title} {\bibinfo {title} {Non-{Hermitian} phase transition and eigenstate
  localization induced by asymmetric coupling},\ }\href
  {https://doi.org/10.1103/PhysRevA.99.062112} {\bibfield  {journal} {\bibinfo
  {journal} {Phys. Rev. A}\ }\textbf {\bibinfo {volume} {99}},\ \bibinfo
  {pages} {062112} (\bibinfo {year} {2019})}\BibitemShut {NoStop}%
\bibitem [{\citenamefont {Fraxanet}\ \emph {et~al.}(2022)\citenamefont
  {Fraxanet}, \citenamefont {Bhattacharya}, \citenamefont {Grass},
  \citenamefont {Lewenstein},\ and\ \citenamefont {Dauphin}}]{Joana}%
  \BibitemOpen
  \bibfield  {author} {\bibinfo {author} {\bibfnamefont {J.}~\bibnamefont
  {Fraxanet}}, \bibinfo {author} {\bibfnamefont {U.}~\bibnamefont
  {Bhattacharya}}, \bibinfo {author} {\bibfnamefont {T.}~\bibnamefont {Grass}},
  \bibinfo {author} {\bibfnamefont {M.}~\bibnamefont {Lewenstein}},\ and\
  \bibinfo {author} {\bibfnamefont {A.}~\bibnamefont {Dauphin}},\ }\bibfield
  {title} {\bibinfo {title} {Localization and multifractal properties of the
  long-range {Kitaev} chain in the presence of an {Aubry-Andr\'e-Harper}
  modulation},\ }\href {https://doi.org/10.1103/PhysRevB.106.024204} {\bibfield
   {journal} {\bibinfo  {journal} {Phys. Rev. B}\ }\textbf {\bibinfo {volume}
  {106}},\ \bibinfo {pages} {024204} (\bibinfo {year} {2022})}\BibitemShut
  {NoStop}%
\bibitem [{\citenamefont {Kawabata}\ \emph {et~al.}(2022)\citenamefont
  {Kawabata}, \citenamefont {Shiozaki},\ and\ \citenamefont {Ryu}}]{Shiozaki}%
  \BibitemOpen
  \bibfield  {author} {\bibinfo {author} {\bibfnamefont {K.}~\bibnamefont
  {Kawabata}}, \bibinfo {author} {\bibfnamefont {K.}~\bibnamefont {Shiozaki}},\
  and\ \bibinfo {author} {\bibfnamefont {S.}~\bibnamefont {Ryu}},\ }\bibfield
  {title} {\bibinfo {title} {Many-body topology of non-{Hermitian} systems},\
  }\href {https://doi.org/10.1103/PhysRevB.105.165137} {\bibfield  {journal}
  {\bibinfo  {journal} {Phys. Rev. B}\ }\textbf {\bibinfo {volume} {105}},\
  \bibinfo {pages} {165137} (\bibinfo {year} {2022})}\BibitemShut {NoStop}%
\bibitem [{\citenamefont {Alsallom}\ \emph {et~al.}(2022)\citenamefont
  {Alsallom}, \citenamefont {Herviou}, \citenamefont {Yazyev},\ and\
  \citenamefont {Brzezi\ifmmode~\acute{n}\else \'{n}\fi{}ska}}]{Alsallom}%
  \BibitemOpen
  \bibfield  {author} {\bibinfo {author} {\bibfnamefont {F.}~\bibnamefont
  {Alsallom}}, \bibinfo {author} {\bibfnamefont {L.}~\bibnamefont {Herviou}},
  \bibinfo {author} {\bibfnamefont {O.~V.}\ \bibnamefont {Yazyev}},\ and\
  \bibinfo {author} {\bibfnamefont {M.}~\bibnamefont
  {Brzezi\ifmmode~\acute{n}\else \'{n}\fi{}ska}},\ }\bibfield  {title}
  {\bibinfo {title} {Fate of the non-{Hermitian} skin effect in many-body
  fermionic systems},\ }\href
  {https://doi.org/10.1103/PhysRevResearch.4.033122} {\bibfield  {journal}
  {\bibinfo  {journal} {Phys. Rev. Res.}\ }\textbf {\bibinfo {volume} {4}},\
  \bibinfo {pages} {033122} (\bibinfo {year} {2022})}\BibitemShut {NoStop}%
\bibitem [{\citenamefont {Peng}\ \emph {et~al.}(2023)\citenamefont {Peng},
  \citenamefont {Cheng},\ and\ \citenamefont {Xianlong}}]{Xianlong_2023}%
  \BibitemOpen
  \bibfield  {author} {\bibinfo {author} {\bibfnamefont {D.}~\bibnamefont
  {Peng}}, \bibinfo {author} {\bibfnamefont {S.}~\bibnamefont {Cheng}},\ and\
  \bibinfo {author} {\bibfnamefont {G.}~\bibnamefont {Xianlong}},\ }\bibfield
  {title} {\bibinfo {title} {Power law hopping of single particles in
  one-dimensional non-{Hermitian} quasicrystals},\ }\href
  {https://doi.org/10.1103/PhysRevB.107.174205} {\bibfield  {journal} {\bibinfo
   {journal} {Phys. Rev. B}\ }\textbf {\bibinfo {volume} {107}},\ \bibinfo
  {pages} {174205} (\bibinfo {year} {2023})}\BibitemShut {NoStop}%
\bibitem [{\citenamefont {Jiang}\ \emph {et~al.}(2019)\citenamefont {Jiang},
  \citenamefont {Lang}, \citenamefont {Yang}, \citenamefont {Zhu},\ and\
  \citenamefont {Chen}}]{Jiang}%
  \BibitemOpen
  \bibfield  {author} {\bibinfo {author} {\bibfnamefont {H.}~\bibnamefont
  {Jiang}}, \bibinfo {author} {\bibfnamefont {L.-J.}\ \bibnamefont {Lang}},
  \bibinfo {author} {\bibfnamefont {C.}~\bibnamefont {Yang}}, \bibinfo {author}
  {\bibfnamefont {S.-L.}\ \bibnamefont {Zhu}},\ and\ \bibinfo {author}
  {\bibfnamefont {S.}~\bibnamefont {Chen}},\ }\bibfield  {title} {\bibinfo
  {title} {Interplay of non-{Hermitian} skin effects and {Anderson}
  localization in nonreciprocal quasiperiodic lattices},\ }\href
  {https://doi.org/10.1103/PhysRevB.100.054301} {\bibfield  {journal} {\bibinfo
   {journal} {Phys. Rev. B}\ }\textbf {\bibinfo {volume} {100}},\ \bibinfo
  {pages} {054301} (\bibinfo {year} {2019})}\BibitemShut {NoStop}%
\bibitem [{\citenamefont {Bera}\ \emph {et~al.}(2017)\citenamefont {Bera},
  \citenamefont {Martynec}, \citenamefont {Schomerus}, \citenamefont
  {Heidrich-Meisner},\ and\ \citenamefont {Bardarson}}]{Bera}%
  \BibitemOpen
  \bibfield  {author} {\bibinfo {author} {\bibfnamefont {S.}~\bibnamefont
  {Bera}}, \bibinfo {author} {\bibfnamefont {T.}~\bibnamefont {Martynec}},
  \bibinfo {author} {\bibfnamefont {H.}~\bibnamefont {Schomerus}}, \bibinfo
  {author} {\bibfnamefont {F.}~\bibnamefont {Heidrich-Meisner}},\ and\ \bibinfo
  {author} {\bibfnamefont {J.~H.}\ \bibnamefont {Bardarson}},\ }\bibfield
  {title} {\bibinfo {title} {One-particle density matrix characterization of
  many-body localization},\ }\href
  {https://doi.org/https://doi.org/10.1002/andp.201600356} {\bibfield
  {journal} {\bibinfo  {journal} {Annalen der Physik}\ }\textbf {\bibinfo
  {volume} {529}},\ \bibinfo {pages} {1600356} (\bibinfo {year}
  {2017})}\BibitemShut {NoStop}%
\bibitem [{\citenamefont {Singh}\ \emph {et~al.}(2017)\citenamefont {Singh},
  \citenamefont {Moessner},\ and\ \citenamefont {Roy}}]{Roy}%
  \BibitemOpen
  \bibfield  {author} {\bibinfo {author} {\bibfnamefont {R.}~\bibnamefont
  {Singh}}, \bibinfo {author} {\bibfnamefont {R.}~\bibnamefont {Moessner}},\
  and\ \bibinfo {author} {\bibfnamefont {D.}~\bibnamefont {Roy}},\ }\bibfield
  {title} {\bibinfo {title} {Effect of long-range hopping and interactions on
  entanglement dynamics and many-body localization},\ }\href
  {https://doi.org/10.1103/PhysRevB.95.094205} {\bibfield  {journal} {\bibinfo
  {journal} {Phys. Rev. B}\ }\textbf {\bibinfo {volume} {95}},\ \bibinfo
  {pages} {094205} (\bibinfo {year} {2017})}\BibitemShut {NoStop}%
\bibitem [{\citenamefont {Modak}\ and\ \citenamefont {Nag}(2020)}]{Nag}%
  \BibitemOpen
  \bibfield  {author} {\bibinfo {author} {\bibfnamefont {R.}~\bibnamefont
  {Modak}}\ and\ \bibinfo {author} {\bibfnamefont {T.}~\bibnamefont {Nag}},\
  }\bibfield  {title} {\bibinfo {title} {Many-body dynamics in long-range
  hopping models in the presence of correlated and uncorrelated disorder},\
  }\href {https://doi.org/10.1103/PhysRevResearch.2.012074} {\bibfield
  {journal} {\bibinfo  {journal} {Phys. Rev. Res.}\ }\textbf {\bibinfo {volume}
  {2}},\ \bibinfo {pages} {012074} (\bibinfo {year} {2020})}\BibitemShut
  {NoStop}%
\bibitem [{\citenamefont {Page}(1993)}]{Page}%
  \BibitemOpen
  \bibfield  {author} {\bibinfo {author} {\bibfnamefont {D.~N.}\ \bibnamefont
  {Page}},\ }\bibfield  {title} {\bibinfo {title} {Average entropy of a
  subsystem},\ }\href {https://doi.org/10.1103/PhysRevLett.71.1291} {\bibfield
  {journal} {\bibinfo  {journal} {Phys. Rev. Lett.}\ }\textbf {\bibinfo
  {volume} {71}},\ \bibinfo {pages} {1291} (\bibinfo {year}
  {1993})}\BibitemShut {NoStop}%
\bibitem [{\citenamefont {Chakrabarty}\ and\ \citenamefont
  {Datta}(2023)}]{Chakrabarty}%
  \BibitemOpen
  \bibfield  {author} {\bibinfo {author} {\bibfnamefont {A.}~\bibnamefont
  {Chakrabarty}}\ and\ \bibinfo {author} {\bibfnamefont {S.}~\bibnamefont
  {Datta}},\ }\bibfield  {title} {\bibinfo {title} {Skin effect and dynamical
  delocalization in non-{H}ermitian quasicrystals with spin-orbit
  interaction},\ }\href {https://doi.org/10.1103/PhysRevB.107.064305}
  {\bibfield  {journal} {\bibinfo  {journal} {Phys. Rev. B}\ }\textbf {\bibinfo
  {volume} {107}},\ \bibinfo {pages} {064305} (\bibinfo {year}
  {2023})}\BibitemShut {NoStop}%
\bibitem [{\citenamefont {Weinberg}\ and\ \citenamefont
  {Bukov}(2019)}]{Weinberg_Quspin_II}%
  \BibitemOpen
  \bibfield  {author} {\bibinfo {author} {\bibfnamefont {P.}~\bibnamefont
  {Weinberg}}\ and\ \bibinfo {author} {\bibfnamefont {M.}~\bibnamefont
  {Bukov}},\ }\bibfield  {title} {\bibinfo {title} {{QuSpin: a Python package
  for dynamics and exact diagonalisation of quantum many body systems. Part II:
  bosons, fermions and higher spins}},\ }\href
  {https://doi.org/10.21468/SciPostPhys.7.2.020} {\bibfield  {journal}
  {\bibinfo  {journal} {SciPost Phys.}\ }\textbf {\bibinfo {volume} {7}},\
  \bibinfo {pages} {020} (\bibinfo {year} {2019})}\BibitemShut {NoStop}%
\bibitem [{\citenamefont {S\'a}\ \emph {et~al.}(2020)\citenamefont {S\'a},
  \citenamefont {Ribeiro},\ and\ \citenamefont {Prosen}}]{Lucas}%
  \BibitemOpen
  \bibfield  {author} {\bibinfo {author} {\bibfnamefont {L.}~\bibnamefont
  {S\'a}}, \bibinfo {author} {\bibfnamefont {P.}~\bibnamefont {Ribeiro}},\ and\
  \bibinfo {author} {\bibfnamefont {T.~c.~v.}\ \bibnamefont {Prosen}},\
  }\bibfield  {title} {\bibinfo {title} {Complex spacing ratios: A signature of
  dissipative quantum chaos},\ }\href
  {https://doi.org/10.1103/PhysRevX.10.021019} {\bibfield  {journal} {\bibinfo
  {journal} {Phys. Rev. X}\ }\textbf {\bibinfo {volume} {10}},\ \bibinfo
  {pages} {021019} (\bibinfo {year} {2020})}\BibitemShut {NoStop}%
\bibitem [{\citenamefont {Ghosh}\ \emph {et~al.}(2022)\citenamefont {Ghosh},
  \citenamefont {Gupta},\ and\ \citenamefont {Kulkarni}}]{Kulkarni}%
  \BibitemOpen
  \bibfield  {author} {\bibinfo {author} {\bibfnamefont {S.}~\bibnamefont
  {Ghosh}}, \bibinfo {author} {\bibfnamefont {S.}~\bibnamefont {Gupta}},\ and\
  \bibinfo {author} {\bibfnamefont {M.}~\bibnamefont {Kulkarni}},\ }\bibfield
  {title} {\bibinfo {title} {Spectral properties of disordered interacting
  non-{Hermitian} systems},\ }\href
  {https://doi.org/10.1103/PhysRevB.106.134202} {\bibfield  {journal} {\bibinfo
   {journal} {Phys. Rev. B}\ }\textbf {\bibinfo {volume} {106}},\ \bibinfo
  {pages} {134202} (\bibinfo {year} {2022})}\BibitemShut {NoStop}%
\bibitem [{\citenamefont {Aoki}(1986)}]{Aoki}%
  \BibitemOpen
  \bibfield  {author} {\bibinfo {author} {\bibfnamefont {H.}~\bibnamefont
  {Aoki}},\ }\bibfield  {title} {\bibinfo {title} {Fractal dimensionality of
  wave functions at the mobility edge: Quantum fractal in the landau levels},\
  }\href {https://doi.org/10.1103/PhysRevB.33.7310} {\bibfield  {journal}
  {\bibinfo  {journal} {Phys. Rev. B}\ }\textbf {\bibinfo {volume} {33}},\
  \bibinfo {pages} {7310} (\bibinfo {year} {1986})}\BibitemShut {NoStop}%
\bibitem [{\citenamefont {Mirlin}\ \emph {et~al.}(2006)\citenamefont {Mirlin},
  \citenamefont {Fyodorov}, \citenamefont {Mildenberger},\ and\ \citenamefont
  {Evers}}]{Mirlin}%
  \BibitemOpen
  \bibfield  {author} {\bibinfo {author} {\bibfnamefont {A.~D.}\ \bibnamefont
  {Mirlin}}, \bibinfo {author} {\bibfnamefont {Y.~V.}\ \bibnamefont
  {Fyodorov}}, \bibinfo {author} {\bibfnamefont {A.}~\bibnamefont
  {Mildenberger}},\ and\ \bibinfo {author} {\bibfnamefont {F.}~\bibnamefont
  {Evers}},\ }\bibfield  {title} {\bibinfo {title} {Exact relations between
  multifractal exponents at the anderson transition},\ }\href
  {https://doi.org/10.1103/PhysRevLett.97.046803} {\bibfield  {journal}
  {\bibinfo  {journal} {Phys. Rev. Lett.}\ }\textbf {\bibinfo {volume} {97}},\
  \bibinfo {pages} {046803} (\bibinfo {year} {2006})}\BibitemShut {NoStop}%
\bibitem [{\citenamefont {Bauer}\ and\ \citenamefont {Nayak}(2013)}]{Nayak}%
  \BibitemOpen
  \bibfield  {author} {\bibinfo {author} {\bibfnamefont {B.}~\bibnamefont
  {Bauer}}\ and\ \bibinfo {author} {\bibfnamefont {C.}~\bibnamefont {Nayak}},\
  }\bibfield  {title} {\bibinfo {title} {Area laws in a many-body localized
  state and its implications for topological order},\ }\href
  {https://doi.org/10.1088/1742-5468/2013/09/P09005} {\bibfield  {journal}
  {\bibinfo  {journal} {Journal of Statistical Mechanics: Theory and
  Experiment}\ }\textbf {\bibinfo {volume} {2013}},\ \bibinfo {pages} {P09005}
  (\bibinfo {year} {2013})}\BibitemShut {NoStop}%
\end{thebibliography}%
\end{document}